Enumeration of Spanning Trees Using Edge Exchange with Minimal Partitioning • Nasr Mohamed • 2014

جامعة الشارقة
UNIVERSITY OF SHARJAH

# Enumeration of Spanning Trees Using Edge Exchange with Minimal Partitioning

BY

**Nasr Mohamed**

A THESIS SUBMITTED IN PARTIAL FULFILMENT OF THE REQUIREMENTS FOR THE DEGREE OF MASTER OF SCIENCE

IN

COMPUTER SCIENCE

AT

DEPARTMENT OF COMPUTER SCIENCE

UNIVERSITY OF SHARJAH

SHARJAH, UNITED ARAB EMIRATES

MAY, 2014

UNIVERSITY OF SHARJAH

# Enumeration of Spanning Trees Using Edge Exchange with Minimal Partitioning

BY

**Nasr Mohamed**

A THESIS SUBMITTED IN PARTIAL FULFILMENT OF THE REQUIREMENTS FOR THE DEGREE OF MASTER OF SCIENCE

IN
COMPUTER SCIENCE

AT
DEPARTMENT OF COMPUTER SCIENCE
UNIVERSITY OF SHARJAH

SHARJAH, UNITED ARAB EMIRATES
MAY, 2014

UNIVERSITY OF SHARJAH

DEPARTMENT OF COMPUTER SCIENCE

The undersigned hereby certify that they have read and recommend to the College of Science for acceptance a thesis entitled "**Enumeration of Spanning Trees Using Edge Exchange with Minimal Partitioning**" by **Nasr Mohamed** in partial fulfilment of the requirements for the degree of **Master of Science**.

Date: May 26, 2014

Supervisor: ______________________

Prof. Ashraf Elnagar (Chair)

Examiners: ______________________

Dr. Saad Harous (External member)

______________________

Dr. Ahmed M. Khedr (Internal member)

UNIVERSITY OF SHARJAH

Author: **Nasr Mohamed**

Thesis Title: **Enumeration of Spanning Trees Using Edge Exchange with Minimal Partitioning**

Department: **Computer Science**

Degree: **Master of Science**

Defense date: **May 26, 2014**

Signature of Author

Dated: May 26, 2014

*To whom my success means*
*more than it means to me*
*To my parents*

# Contents

# List of Figures

## List of Algorithms

## List of Tables

# Acknowledgement

All praise is due to Allah, the Almighty, and the Most Merciful with whose favours all good can be accomplished. We thank You Allah for the countless blessings You bestowed upon us.

I would like to express my deepest appreciation to all those who contributed to the achievement of this work. Foremost, I am particularly indebted to the continuous assistance provided by my supervisor Prof. Ashraf Elnagar. His precious advice always enlightened my way during the whole course of research.

Additionally, I am grateful to the comments made by all reviewers, especially the valuable remarks by Dr. Saad Harous, which gave this thesis its final presentation.

I extend my appreciation to the family of Department of Computer Science and all professors, colleagues and friends who attended my defence session. They truly offered me the support I needed and made my defence a cheerful event.

Last but not least, my very special thanks to my family. Words cannot express how grateful I am to their prayers, sacrifice and help. That was my true incentive to strive toward my goal.

# Abstract

Spanning trees is the solution space of numerous problems in different areas of research. Many studies were conducted to detect their existence, list their occurrences and investigate their features. In this thesis, Minimal Partitioning (*MP*) algorithm, an innovative algorithm for enumerating all the spanning trees in an undirected graph is presented.

While MP algorithm uses a computational *tree graph* to traverse all possible spanning trees by the edge exchange technique, it has two unique properties compared to previous algorithms. In the first place, the algorithm maintains a state of minimal partition size in the spanning tree due to edge deletion. This is realized by swapping peripheral edges, more precisely leaf edges, in most of edge exchange operations. Consequently, the main structure of the spanning trees is preserved during the steps of the enumeration process. This extra constraint proves to be advantageous in many applications where the partition size is a factor in the solution cost. Secondly, we introduce, and utilize, the new concept of *edge promotion*: the exchanged edges always share one end. Practically, and as a result of this property, the interface between the two partitions of the spanning tree during edge exchange has to be maintained from one side only.

We prove the correctness of MP algorithm, offer detailed specifications for implementation and analyse time and space complexity of the algorithm.

For a graph $G(V,E)$, MP algorithm requires $O(log\ V + E/V)$ expected time and $O(V\ log\ V)$ worst case time for generating each spanning tree. MP algorithm requires a total expected space limit of $O(E\ log\ V)$ with worst case limit of $O(EV)$. Like all edge exchange algorithms, MP algorithm retains the advantage of compacted output of $O(1)$ per spanning tree by listing the relative differences only, i.e. the exchanged edges.

Three sample real-world applications of spanning trees enumeration are explored and the effects of using MP algorithm are studied. Namely: construction of nets of polyhedra, multi-robots spanning tree routing, and computing the electric current in edges of a network. The predicted benefits are highlighted as compared to other algorithms. We report that MP algorithm outperforms other algorithm by $O(V)$ time complexity.

# Abstract in Arabic

## سرد الأشجار الممتدة باستخدام تبديل الروابط مع الحفاظ على الحد الأدنى للتقسيم

**الخلاصة**:

تمثل الأشجار الممتدة بيئة الحل لكثيرٍ من المسائل في مجالات بحثية متعددة ولذلك كان الاهتمام بدراستها لتحديد وجودها، تعداد أفرادها، والتحقق من خواصها. في هذه الأطروحة نعرض خوارزمية التقسيم الأدنى وهي طريقة مبتكرة لسرد الأشجار الممتدة في الرسومات غير الموجهة.

على الرغم من أن خوارزمية التقسيم الأدنى تتبع أسلوب تبديل الروابط عبر الشجرة الموجهة التي تضم كافة الأشجار الممتدة للرسم، فإنها تتميز بميزتين فريدتين مقارنة بالخوارزميات الأخرى. في المقام الأول، تحافظ الخوارزمية على على حدٍ أدنى من التقسيم للشجرة الممتدة عند حذف الرابط ويتم ذلك عبر تبديل الروابط الهامشية (وبالأحرى الروابط الطرفية) في أكثر عمليات تبديل الروابط. كنتيجة لذلك تبقى البنية الأساسية للشجرة الممتدة سليمة خلال خطوات سرد الأشجار الممتدة. وقد أثبتت هذه الميزة أنها ذات فائدة كبيرة في العديد من التطبيقات التي يشكل حجم التقسيم فيها عاملاً في تحديد تكلفة الحل. ثانياً، قدمنا واستخدمنا مفهوماً جديداً للتبديل وهو ترقية الرابط حيث تشترك الروابط المستبدلة دائماً في طرفٍ واحد. عملياً، يعني ذلك أن الارتباط بين القسمين تتم صيانته من طرف واحد فقط أثناء تبديل الروابط.

في هذه الأطروحة أثبتنا صحة الحل وفصلنا طريقة تنفيذها ومن ثمّ قمنا بتحليل التكلفة الوقتية والتخزينية للخوارزمية.

بالنسبة للرسم G(V,E) تكون تكلفة الوقت المتوقع O(log V + E/V) وأسوأ الحالات O(V log V) لسرد شجرة ممتدة واحدة. أما تكلفة التخزين الكلية المتوقعة فهي O(E log V) وأسوأ الحالات O(EV). ومثل جميع خوارزميات تبديل الروابط تتميز خوارزمية التقسيم الأدنى بالناتج المختصر O(1) لكل شجرة حيث يتم عرض الفروق البينية فقط (الروابط المستبدلة).

كذلك تم استكشاف ثلاث تطبيقات عملية لسرد الأشجار الممتدة ومن ثمّ دراسة آثار استخدام خوارزمية التقسيم الأدنى عليها. تحديداً: إنشاء الشبكات للمجسمات ثلاثية الأبعاد، تسيير عدة رجال آليين عبر شجرة ممتدة، وحساب التيار الكهربي في روابط شبكة كهربائية. تم تحديد الفوائد المتوقعة من تطبيق الخوارزمية مقارنة بالخوارزميات الأخرى وأثبتنا أن خوارزمية التقسيم الأدنى تتفوق على الخورزميات الأخرى بـ O(V) في التكلفة الزمنية.

Chapter 1

# Introduction and Literature Review

Mathematicians have always been fascinated by the challenging nature of combinatorial problems. Listing of permutations, combinations, subsets, subsequences… are all types of enumeration problems that were extremely tedious, if not impossible, till the intervention of computers several decades ago. Graphs, being in the center of interest, have their notable share of these types of problems. Enumeration of special types of subgraphs such as cycles, cliques, $k$-trees, matchings... are types of combinatorial puzzles that have inspiring charm on passionate researchers, leave aside their important applications in many practical fields. The count of these special patterns is usually exponential in the size of the input graph. Enumeration of spanning trees in a graph belongs to this family of combinatorial graph problems.

In this introduction, before surveying previous research conducted on the problem, we shall shed some light on the context and history of spanning trees enumeration.

## 1.1 Background:

A spanning tree of any network is a communication sub-network that ensures connectivity between all nodes in the original network with minimum number of edges. This minimal edges condition contributes another interesting feature: Every pair of nodes in the network are interconnected via a unique path over a sequence of edges. In other words, a spanning tree guarantees existence and uniqueness of connection between any pair of nodes.

Spanning trees is one of the well-studied areas in graph theory with so many applications in diversified fields in engineering, computer science, bioinformatics, chemistry and other domains. Many problems in real life applications are modelled as spanning tree problem. To begin with, construction of many communication and supply networks can be abstracted as building a spanning tree over the underlying network (known as the *connector problem*)[1]. Spanning trees contribute to the solution of many interesting problems in computer sciences. To name a few: Taxonomy is one of the early applications of spanning trees [2] where the spanning tree is used to map a multidimensional proximity matrix and highlight inadequacies. The well-known travelling salesman problem can be approximated by processing spanning trees using Christofide's Algorithm [3]. Robotic path planning has several implementations using spanning trees [4]. In computer vision, spanning trees are employed for image registration [5], segmentation [6] and feature extraction [7]. Presently, in the social networks era, many new applications of spanning trees are re-studied to analyze and mine the massive amount of data beneath. Clustering analysis based on spanning trees was researched as it provides computationally low-cost algorithms for grouping the data [8,9]. Cluster hierarchies in social networks may also be identified using spanning trees [10].

Furthermore, spanning trees create a sparse subgraph that is easy to study and reflects a lot about the original graph [11]. Features related to path and flow can be studied much easier in the acyclic and unique-path configurations than in the original graphs.

Different spanning trees constructed over a network have many different features. Total cost of a spanning tree is one of the primary properties of interest with literature as early as 1926 [12]. The cost of a network link can be an interpretation of many physical properties that are desirable to be minimized such as distance travelled, time spent, resistance encountered, etc. The main algorithms to compute minimum cost spanning trees were introduced by Kruskal [13] and Prim [14]. Rectilinear [15] and Euclidean [16] minimum spanning trees are direct variations of the problem. The minimum bottleneck spanning tree [17] is the spanning tree having the minimum ceiling for all edge costs.

Other than the direct link-wise minimum spanning tree, more complicated properties of spanning trees are regarded as optimum communication trees. Another spanning tree property is quadratic minimum spanning trees [18] where the cost of edges is correlated with the use of other edges in the spanning trees. Degree constraint and degree preserving [19] spanning trees are instances of spanning trees with restriction on the degree of nodes, i.e. edges per node, in such trees.

## 1.2 Enumeration of spanning trees:

Enumeration of spanning trees is a combinatorial generation problem. The algorithms used are typically analyzed in terms of the output size rather than the input size [20]. A lower bound on the number of spanning trees in a graph $G(V,E)$ was computed [21] to be $\Omega(2^t)$ where $t$ is given by:

$$t = \left\lceil \frac{-1 + \sqrt{1 + 0.8(|E| - |V| + 1)}}{2} \right\rceil$$

In spite of this exponentially explosive lower bound of the problem, it remained a live topic of research for decades for many substantial reasons.

Primarily, the numerous aspects of comparison between spanning trees that may be constructed over a network and the absence of direct algorithms to calculate the optimum tree with respect to many of these comparisons, emphasizes the importance of enumerating all spanning trees. Several optimization problems on spanning trees have been proven to be NP-hard [22]. Dozens of other problems have no polynomial time solutions yet. Nevertheless, many applications require multiple properties to be optimized simultaneously [23]. In practice, this imposes that an exact solution can only be found by an exhaustive search [24].

Furthermore, some applications do require, explicitly, enumeration of all spanning trees in a graph. One important such application is the calculation of electric current in an electrical network. Introducing a voltage source to an electrical network causes different amount of electric current to run into edges of the network. In order to calculate the current flowing in each edge, every spanning tree of the electrical network is generated. A spanning tree will dictate a single path for the current to flow. The current of every edge is calculated as the normalized sum of current in that edge under all spanning trees [25].Another related application is found in chemistry, the calculation of ring-currents in molecules with closed cycles of atoms is determined by generating all the spanning trees of the graph which represents the molecule [26].

While many algorithms are proposed for the task of enumerating all spanning trees in a graph, MP solution is unique in two important behaviors:

1- The partition in the tree due to edge exchange is minimized. Essentially, most of the exchanges occur in leaf edges.
2- During the enumeration process, the edge exchange is always performed between 2 edges that share one end. This means that spanning trees are enumerated by single edge end exchange.

In the rest of this chapter, we briefly survey previous research in this topic. Besides, the application which initially inspired this novel solution is described. Subsequently, in chapter 2, we formulate and define MP solution. A conceptual description of the solution is provided. The correctness of the solution framework is established. Then, in chapter 3, details of the algorithm are specified. The algorithm is defined using implementation oriented pseudo-code that define the process details and the data structures used. The analysis of the solution is conducted in chapter 4. Time and space complexity of the algorithm is computed. More important, the fundamental objective of this algorithm, minimal partitioning of spanning trees, is proved. Finally, we show experimental results in chapter 5. The results verify the correctness of the algorithm and demonstrate the claimed tree partitioning limits. Additionally, we discuss some applications where the algorithm is more appropriate than previous algorithms.

## 1.3 Previous work:

Some of the early references to the topic of spanning trees enumeration in literature are Mayeda *et al* (1965)[27], Minty's (1965)[28] and Char's (1968)[29] contributions. Char's, conceptually simple approach, generates all combinations of $V$–1 edges of a graph $G(V,E)$, where $V$ is the set of nodes and $E$ the set of edges of $G$, and tests them by some tree testing algorithm to check whether the edges constitute a tree of $G$ or not. Jayakumar *et al* [30] found out the time complexity of Char's algorithm to be $O(E+V+V(T+N_{non_tree}))$, where $T$ and $N_{non_tree}$ denote the number of tree and non-tree sequences generated, for the graph $G$ [*].

Mayeda's algorithm uses a root spanning tree to generate all other spanning trees by exhaustive search and replace algorithm. For every edge in the root spanning tree, all possible replacements are tried recursively to generate new spanning trees. Minty's algorithm maintains a partial spanning tree, $T$, which is grown one edge at a time. After adding an edge, $e$, to $T$, all edges which would form a cycle in $T$ are removed from $G$. A recursive call then generates all spanning trees that include all edges in $T \cup \{e\}$. Then $e$ is removed from $T$ and all bridge edges in $G$, that reconnect $T$, are added one at a time. Read *et al* [21] analyzed the algorithm and determined its time complexity to be $O(VE+VET)$ and its space complexity $O(VE)$.

Ever since, many more elaborated ideas were introduced to solve the problem. The techniques used can be categorized into 2 main groups:

1- Edge exchange technique where all spanning trees are generated from a root tree. A single edge exchange is performed at each step to produce a new spanning tree.

2- Inclusion/exclusion technique to generate trees with particular edges of the graph being forcibly included or excluded.

Many different algorithms that use one of these techniques, or a hybrid of both, were developed.

A tree graph of a graph $G$ is defined as the graph with all spanning trees of $G$ as nodes. Every pair of nodes in the tree graph are connected by an edge if, and only if, they differ by 1 edge only (i.e. can be generated from each other by an edge exchange). Besides, a Hamiltonian circuit (or cycle) is a cycle in an undirected graph which visits each node exactly once and returns to the starting node.

* We shall use these notations in the rest of this thesis. Moreover, the same notation is used for the set and its cardinality as the intended meaning is always obvious from the context.

Cummins (1966)[31] proved that for any graph $G$ that has more than 2 spanning trees, the tree graph of $G$ has a Hamiltonian circuit. This interesting feature was employed by some algorithms to generate all spanning trees of $G$.

Kamae (1967)[32] generated the spanning trees by defining partitions over the tree graph. The partitions are devised using inclusion/exclusion rules. The trees in these partitions were generated sequentially. The algorithm then transfers to the next partition and generates all its spanning trees. The last tree generated in any partition always differ from the first tree generated in following partition by only one edge exchange. Kamae did not provide complexity analysis for his algorithm.

Berger (1967)[33] provided an algorithm that generates all spanning trees by growing smaller trees till the level of spanning trees. The algorithm start with an empty tree (0 edges) and goes through all the steps, recursively, till spanning trees ($V - 1$ edges). The paper did not analyze the complexity of the algorithm, but its time complexity is expected to be high as all the trees in lower level have to be generated, even though many of these trees do not generate spanning trees.

Kishi and Kajitani (1968)[34] used the Hamiltonian circuit feature to generate spanning trees. The algorithm partitions the graph into a set of complete subgraphs. The Hamiltonian circuit of the tree graph of a complete graph can be easily generated, thus, the spanning trees of subgraphs can be enumerated. The spanning trees of the whole graph are then assembled from all possible combinations of the subgraph spanning trees. Additionally, the algorithm uses an inclusion/exclusion list to avoid repetition of spanning trees.

A prominent algorithm for finding all spanning trees of directed and undirected graphs was presented by Gabow and Myers (1978)[35]. A spanning tree in a directed graph is defined with respect to a root node. Here the spanning tree should define a single path from the root to all nodes of the graph. According to the application being addressed, all the spanning trees with respect to a certain root or all the spanning trees with every node as root may be enumerated. The algorithm uses backtracking and a method for detecting bridges based on depth-first search. The time required is $O(V+E+ET)$ and the space is $O(V+E)$. If the graph is undirected, the time decreases to $O(V+E+VT)$, which is optimal within a constant factor. Gabow's algorithm was the fastest known algorithm when it was published. Moreover, the algorithm handles both directed and undirected graphs. Later, several algorithms that require constant amortized time, discussed below, were developed.

Winter (1986)[36] divided the original graph into smaller subgraphs by successive reductions and contractions. Trees of the original graph are then obtained from the trees of the sub graphs. The worst-case time and space complexities of the algorithm are $O(V+E+VT)$ and $O(V^2)$, respectively.

The algorithm proposed by Matsui (1993)[37] enumerates all spanning trees in a graph by starting from one spanning tree as a root and using it to generate all other trees. The basic observation behind the algorithm is that if we introduce an additional edge to a spanning tree we will obtain a cycle within the resultant graph. Then, any edge in this cycle can be removed to obtain a different spanning tree. This process is recursively repeated on all edges for all generated trees. An ordering on edges is used to eliminate duplications. The algorithm has time complexity of $O(V+E+T)$ and storage complexity of $O(V+E)$. Matsui (1997)[38] presented a modified algorithm that enumerates spanning trees of weighted graph in non-decreasing order in $O(V+E+VT)$ time and $O(E+VT)$ space.

Shioura, Tamura and Uno (1997)[39] developed an optimal algorithm based on a previous algorithm by Shioura and Tamura (1993)[40] to compact the space complexity to $O(V+E)$ while keeping the time complexity to $O(V+E+T)$. The algorithm envisions the tree graph of the given graph and develops a mechanism to traverse it using depth first to enumerate all spanning trees of the graph.

Kapoor and Ramesh (1995)[41] presented 3 algorithms for enumeration of spanning trees in undirected graphs using inclusion/exclusion technique. A computational tree is used to generate child trees from parent tree by introducing extra constraints over included and excluded edges at child tree. In each step a non-tree edge is added and an edge from a set of tree edges (forming cycles with the new edge) can be removed. The computational cost of the algorithm is $O(V+E+T)$. The space requirement can be limited to $O(V^2E)$ by considering all the cycles formed by the added edge in one step, thus, compacting the computational tree. The second algorithm uses the depth-first-search tree as the root of the computational tree. Tracking cycles during the process is simplified by tracking back edges with respect to the parent tree (by considering the tree edges to be directed from parent to child). The paper proves, for correctness of algorithm, that at any node of the computational tree, the spanning tree is the depth-first-tree of the node graph constrained by the inclusions and exclusions. Additionally, bridge edges that are required to connect sub-graphs are identified. They are added to inclusion list in order to reduce the process. The computation still remains $O(V+E+T)$ while the space cost is $O(VE)$. A third algorithm presented by the paper sorts the spanning trees (of a weighted graph) in increasing cost. The algorithm uses the same earlier algorithm with the minimum spanning tree as root and maintains priority queues in increasing cost order. The nodes in the computational tree are then traversed using an order dictated by the priority queues to output all the trees in sorted order. Naturally, the additional sorting operation invites an additional cost which makes the algorithm computation-wise $O(VE+T \log V)$ and space-wise $O(VE+T)$. The same authors extended their work to directed graphs (2000)[42] and introduced an algorithm to output all the spanning trees rooted at certain node of a directed graph by using the edge exchange technique. Unlike edge exchange in undirected graphs, only non-back edges (i.e. forward and cross edges) with respect to a tree can be used to produce another tree. When an edge is added to the tree, the edge in the tree with same destination as the new edge should be removed. The computational tree for enumerating all the spanning trees of a graph $G$ rooted at $v$ is a binary tree that has the depth-first-search tree $F$ at its root. The left child is a new tree formed by exchange of edge $e$ for $f$, where $f$ is the first edge in the list of possible candidates for exchange and $e$ is the only edge in $F$ that has same destination as $f$. The right child is the same as the parent tree $F$ but the edge $f$ introduced in the sibling tree is completely excluded from any further processing. That is, the left subtree enumerates all spanning trees including edge $f$ while the right subtree enumerates all subtrees excluding edge $f$. The union of those 2 subtrees is clearly the whole solution. The algorithm makes use of tree properties to eliminate avoidable calculations. The complexity of the algorithm for generating all the spanning trees with respect to a certain node as the root is $O(V^2+VT)$ with space requirement of $O(V^2)$.

Uno (1996)[43] introduced an algorithm for enumerating spanning trees in directed graphs by edge exchange. A root spanning tree is used and all possible exchanges with respect to a certain edge in it are performed recursively by considering all non-back edges not in the current spanning tree. The algorithm time and space complexity were not provided in input or output driven form. Later, Uno (1998) proposed an approach, trimming and balancing, for speeding up enumerations in graphs [44]. The approach removes unnecessary edges from the input and optimizes the enumeration of some special subgraphs such as spanning trees, matroids and matchings. For generating a single spanning tree the algorithm requires time of $O(log^2V)$.

Smith (1997)[45] modified the basic inclusion/exclusion algorithm to generate spanning trees in a Gray code [46] order. The edges of a graph are presented as bits where included edges in a spanning tree are presented by set bits. The modified algorithm enumerates trees in Gray code order. The paper analyzes the algorithm to be of time complexity of $O(V^2+T)$ and space complexity of $O(V^2)$.

Sörensen and Janssens (2005)[47] researched the area of enumerating all spanning trees of a graph in the order of increasing cost. The algorithm offered uses the minimum spanning tree (MST) as the root for generating other trees. The edges contained in the MST are used to partition the rest of the spanning trees by dictating included and excluded edges in each group of trees. Next, a slightly modified version of Kruskal's algorithm is used on each partition by initializing with included edges and finding the MST while avoiding the excluded edges. The partitioning process is repeated for the partition where the minimum of all partitions MST's is found. The process terminates any partition that does not contain a tree (i.e. the excluded edges result in a disconnected graph). The time complexity of this algorithm is $O(TE\ log\ E+T^2)$ and space complexity is $O(TE)$ which is higher than other unsorted enumeration algorithms.

Bhagat (2009)[48] proposed the use of the concept of loop domain to accelerate the enumeration of spanning trees. The original graph is transformed into a loop domain by eliminating all nodes of degree 2. The spanning trees in the transformed graph are enumerated. Subsequently, the spanning trees of the original graph are generated. The algorithm is optimized for graphs with significant number of nodes that have degree 2.

Chakraborty *et al* (2011)[49] proposed an algorithm that classifies the edges in a graph into 2 types: circuits and pendant (not part of any cycle). Pendant edges are part of any spanning tree and, therefore, are excluded from further processing. Circuits are then analyzed to obtain a structure of data about each circuit. Next a naïve approach is used to enumerate all possible sub-graphs based on the data extracted and non-tree sub-graphs are eliminated. The time complexity of the algorithm is high as non-tree graphs are generated and tested. The paper claims a complexity of $O((V+E)(T+N_{non_tree}))$ where $N_{non_tree}$ is the set of non-tree graphs generated and rejected.

## 1.4 Motivating application:

This research was motivated by an application in computational geometry where it was required to enumerate all the nets of a given polyhedron. The net of a polyhedron is a polygon that results from unfolding the facets of the polyhedron by cutting over edges. The facets are then rotated around the remaining common edge till all the facets are planar. The resultant facets should form a connected non-overlapping polygonal shape. In an earlier paper [50] we proved that this problem is equivalent to enumeration of spanning trees over a graph. The net adjacency graph is a spanning tree of the facet adjacency graph of the polyhedron. However, applying existing spanning trees enumeration algorithms proved to be inefficient. The main obstacle is that these algorithms assume that removing a tree edge and replacing it with another edge is $O(1)$ operation. Contrarily, in our application this operation is $O(V)$. Elimination of an edge from the adjacency graph is equivalent to cutting over an edge between 2 facets in the net. It partitions the net into two parts. Reconnecting the two parts using another edge involves geometrically transforming all the facets in one of the partitions to re-align it with the other partition (see Figure 1). If there is an algorithm to enumerate trees by exchanging leaf edges that connect only one facet to the net, the complexity of the algorithm can be reduced by $O(V)$. Any similar application where the edge exchange triggers a computation that depends on the size of partitions can benefit from this algorithm.

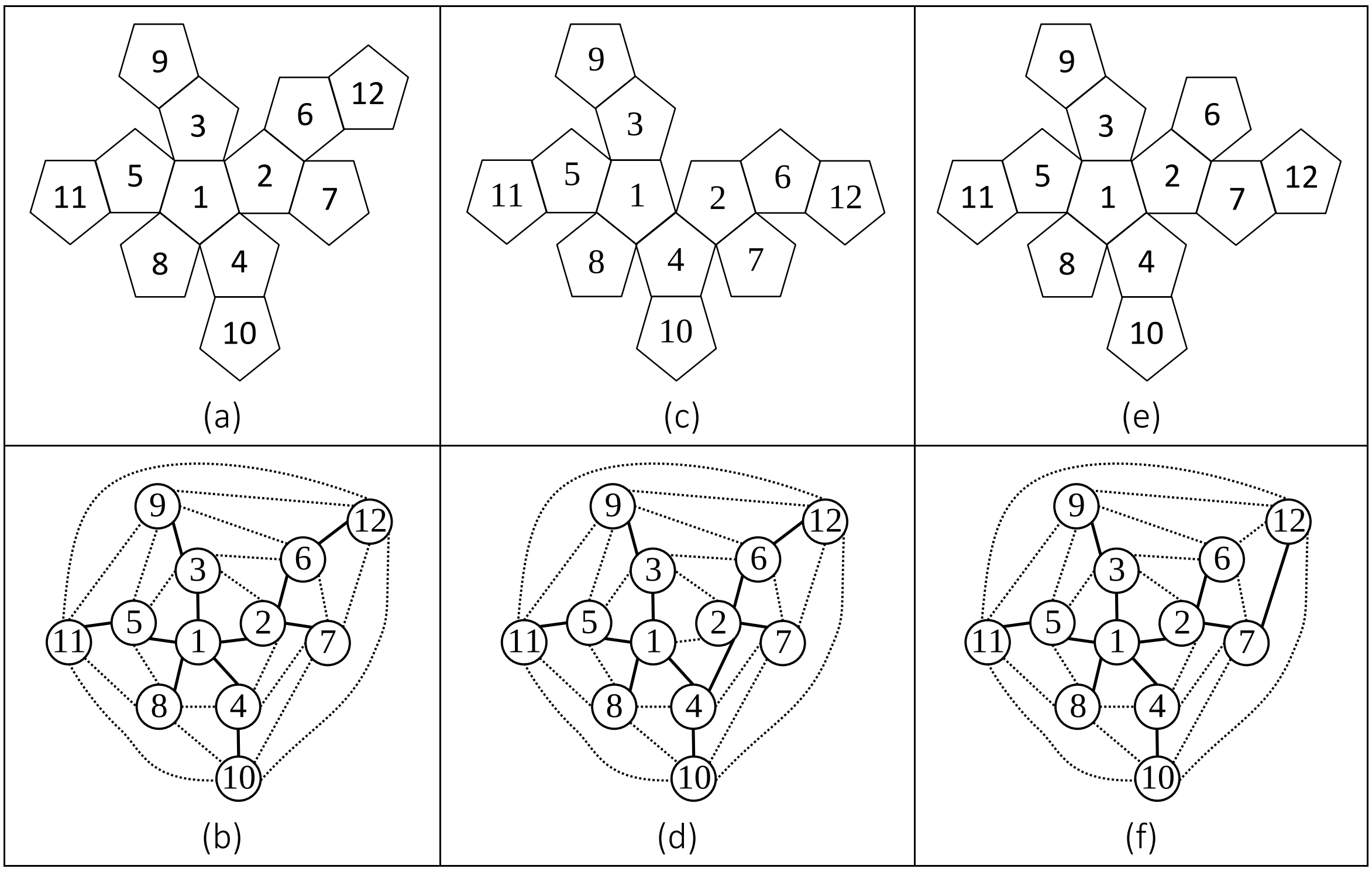

Figure 1: Enumerating nets of dodecahedron

In Figure 1, (a) is a possible net of the regular dodecahedron. (b) is the adjacency graph of the regular dodecahedron facets with the spanning tree representing the adjacency graph of net (a) in solid lines. (c) and (d) a net obtained from net (a) by exchanging 1 internal edge ($\overline{1,2}$ removed and $\overline{4,2}$ added) with the modified spanning tree. The partition containing facets 2,6,7 and 12 has to be transformed ($O(V)$ operation). (e) and (f) a net obtained from net (a) by exchanging 1 leaf edge ($\overline{6,12}$ removed and $\overline{7,12}$ added) with the modified spanning tree. Only facet 12 has to be transformed ($O(1)$ operation).

Chapter 2

# Problem Statement and Solution Framework

In this chapter, MP innovative method for enumerating all the spanning trees in a given graph is described. Initially, the problem shall be defined more formally then the terminology used for describing the solution is defined. The framework used for the solution is described and the theoretical proof of correctness is provided.

## 2.1 The problem statement:

Given a labelled connected undirected graph $G(V,E)$, define the set $S(G)$ as the set of all spanning trees over $G$. We shall assume that $G$ contains no self-edges or parallel edges. The problem is to:

1- Uniquely enumerate all members of $S(G)$ using edge exchange technique.

2- The edge exchanged in each step should partition the tree in such a way that the average size of the smaller partition is minimized (compared to $O(V)$ which is the average size for random partition size).

## 2.2 Terminology and definitions:

### 2.2.1. Node and edge order:

**Definition 1: Node order:**

Define the node order as a strictly increasing ordering function over all nodes $V$ of the graph $G$.

**Definition 2: Edge order:**

Define the edge order as a strictly increasing ordering function over all edges $E$ of the graph $G$.

We will denote the ordering function by *order*(). Besides, both ordering of nodes and edges will be represented using the same notation. The argument type will determine the intended function:

$$order: V \rightarrow \{0,2,\ldots,|V|-1\} \text{ such that } \forall\, v_i,v_j \in V\ order(v_i) = order(v_j) \Leftrightarrow v_i = v_j, \text{ and}$$

$order: E \rightarrow \{1,2,\ldots,|E|\}$ such that $\forall\ e_i,e_j \in E\ order(e_i) = order(e_j) \Leftrightarrow e_i = e_j$

A sufficient function for node ordering can be obtained if we take the index of a node in the adjacency list (or adjacency matrix) as node order. Likewise, a sufficient function for edge ordering can be obtained by the order of 1st appearance in the adjacency list (or adjacency matrix). Note that node ordering is 0-based to emphasize the concept of node zero explained shortly.

### 2.2.2. Edge-node pair:

While generating the initial spanning tree, using Prim's algorithm, every node is joined to the tree using a specific edge (except for the 1st node). A node together with its joining edge are considered to be an edge-node pair.

During the formation of spanning trees, an exchange of 2 edges to form a new tree is noted as an exchange of edge-node pair. This extends the concept of the edge-node pair to all spanning trees formed during the process. We shall see that exchanging an edge-node pair will always exchange an edge with another edge that shares one end (i.e. a node) with the original edge. Initially, the 1st node in the spanning tree is paired with NIL edge. Figure 2 demonstrates the concept of edge-node pair and shows how an exchange of a pair will generate a new spanning tree.

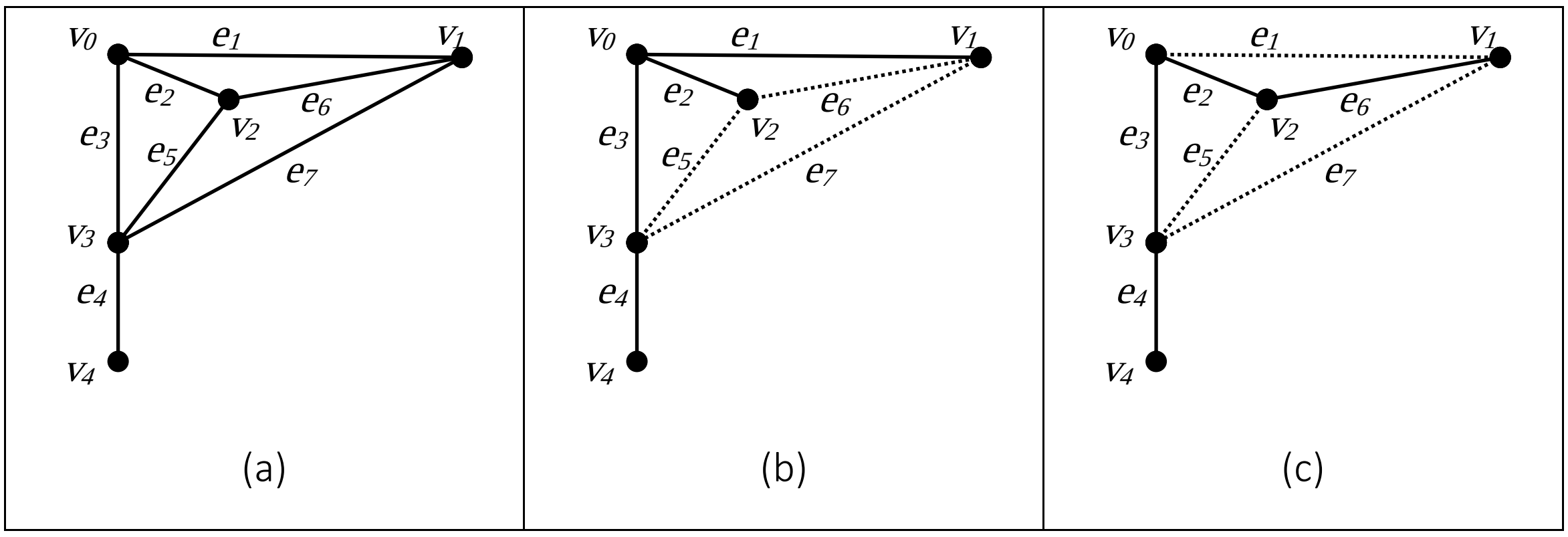

Figure 2: Exchange of spanning tree pairs

In Figure 2, (a) shows a graph $G(\{v_0,v_1,\ldots,v_4\},(e_1,e_2,\ldots,e_7\})$. (b) is a spanning tree constructed using $v_0$ as initial node and adding the pairs $(v_1,e_1)$, $(v_2,e_2)$, $(v_3,e_3)$, $(v_4,e_4)$. (c) is generated from (b) by exchanging the pair ($v_1$,$e_1$) with the pair ($v_1$,$e_6$). Spanning tree (b) has leaf pairs at ($v_1$,$e_1$), ($v_2$,$e_2$) and ($v_4$,$e_4$) while spanning tree (c) has leaf pairs at ($v_1$,$e_6$) and ($v_4$,$e_4$).

More formally:

**Definition 3: Node zero:**

This is the lowest order node in the graph. This node is used as a seed to generate the initial spanning tree using Prim's algorithm. We shall denote node zero by $v_0$.

Although node zero can be viewed as the root of the spanning tree, it is important to note that the concept of the root node here is only for computational purpose as the spanning trees generated are undirected and have no special node as root. (We shall avoid the term *root node* and keep to *node zero*). For the description of the solution, we shall use some terms that indicate a directed spanning tree, such as ancestors, descendants and subtree of a node. All these terms are defined for computational purpose only and always with respect to *node zero*.

**Definition 4: Edge-node pair:**

$\forall\ v \in V$ in $G$ where the path from $v$ to $v_0$ in spanning tree $T$ is $v,u,\ldots,v_0$ we define ($v$,$e$=($v$,$u$)) to be an edge-node pair.

That is, in any spanning tree $T$, a node $v$ with the edge $e$ incident to $v$ such that $e$ is the initial edge in the path from $v$ to $v_0$ is an edge-node pair (or simply a pair).

**Definition 5: Leaf pair:**

This is the same concept as edge-node pair applied to leaf nodes. Obviously, in the case of leaves, the partner edge is evident as each leaf has only 1 incident edge. These leaf pairs will be exchanged to generate most of the spanning trees (namely type 2 spanning trees, defined later) as we shall see. Figure 2 illustrates 2 spanning trees of a given graph and shows their leaf pairs.

We shall use the notations $pair(v)$ and $pair(e)$ to denote the pair that contains node $v$ and edge $e$. Conversely, $node(p)$ and $edge(p)$ shall be used to denote the node and the edge of the pair $p$.

### 2.2.3. Rank of edge-node pair:

We shall define a ranking function over edge-node pairs. This ranking function will be used to control exchanging of edges during the enumeration of spanning trees.

Definition 6: Rank of edge-node pair:

The rank of an edge-node pair (or a leaf pair) denoted by *rank*() is the order of the edge in the pair.

$\forall$pair $p = (v,e)$, $rank(p) = order(e)$

Definition 7: Minimal pair:

An edge-node pair (or leaf pair) is minimal within the node, or simply minimal, if the order of the associated edge is the minimum order of all edges incident to the node.

A pair $p = (v,e)$ is minimal $\Leftrightarrow \forall\ u \in adjacency\text{-}list(v)$ and $e \neq (v,u)$, $order(v,u) > order(e)$

Definition 8: Maximal pair:

An edge-node pair (or leaf pair) is maximal within the node, or simply maximal, if the order of the associated edge is the maximum order of all edges incident to the node.

A pair $p = (v,e)$ is maximal $\Leftrightarrow \forall\ u \in adjacency\text{-}list(v)$ and $e \neq (v,u)$, $order(v,u) < order(e)$

Note that if a node $v$ has only one incident edge $e$ in $G$ then its only pair $(v,e)$ is both minimal and maximal.

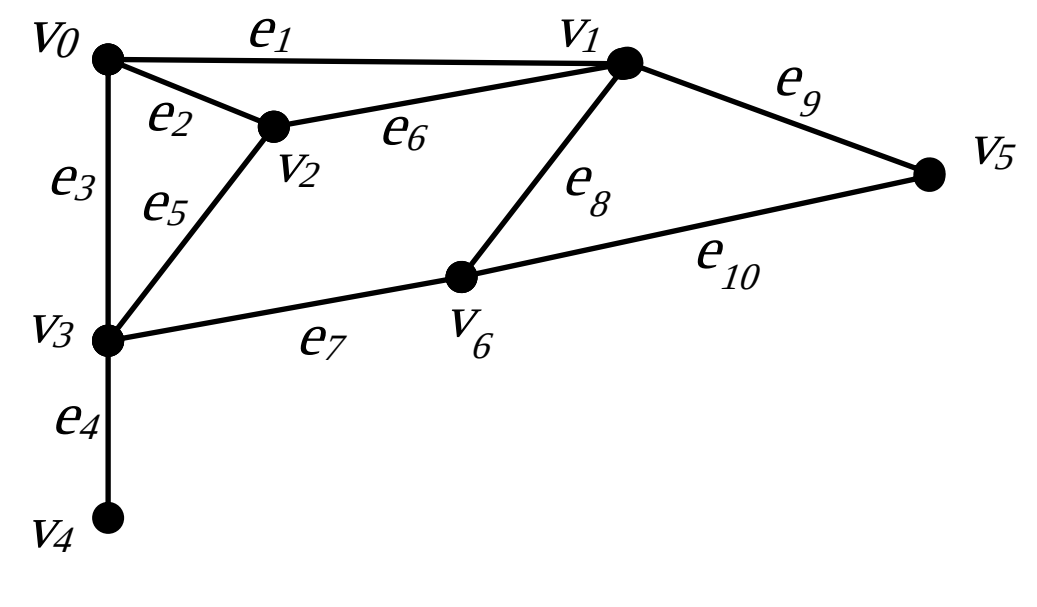

Figure 3: Rank of edge-node pairs

Figure 3 shows a graph $G(\{v_0,v_1,\ldots,v_6\},(e_1,e_2,\ldots,e_{10}\})$. Assuming that $order(e_1) < order(e_6) < order(e_8) < order(e_9)$ then the pair $(v_1,e_1)$ is minimal and the pair $(v_1,e_9)$ is maximal within node $v_1$. The pair $(v_4,e_4)$ is both minimal and maximal within node $v_4$.

### 2.2.4. Type 1 and type 2 spanning trees:

We shall define two disjoint and all inclusive sets of spanning trees for a graph:

#### Definition 9: Type 1 spanning trees:

These are spanning trees that have all their leaf pairs as minimal.

A spanning tree $T$ is type 1 $\Leftrightarrow \forall\ v \in V$: $pair(v)$ is minimal or $degree(v \text{ in } T) > 1$

#### Definition 10: Type 2 spanning trees:

These are spanning trees that have at least 1 non-minimal leaf pair.

A spanning tree $T$ is type 2 $\Leftrightarrow \exists\ v \in V$: $pair(v)$ is not minimal and $degree(v \text{ in } T) = 1$

Note that type 2 spanning trees are defined as the compliment of type 1 spanning trees which allows us to deduce the following observation:

#### Lemma 1:

Every spanning tree of a graph $G$ is either type 1 or type 2.

### 2.2.5. Pilot pair:

For each spanning tree, we define a single edge-node pair as the pilot pair. This concept plays an important role in the description of the solution. We shall denote the pilot pair of a spanning tree $T$ as $pilot(T)$. We start by defining the pilot pair for type 2 spanning tree as:

#### Definition 11: Type 2 spanning tree pilot pair:

The pilot pair of a type 2 spanning tree is the non-minimal leaf pair that has maximum node order. We know from Definition 10 that such a pair exists in a type 2 spanning tree.

For a spanning tree $T$ of type 2:

Let $L = \{(v,e): e \in T$ and $(v,e)$ is not minimal and $degree(v \text{ in } T) = 1\}$.

Then $pilot(T) = (v,e) \in L$ such that $\forall\ (u,f) \in L$ and $v \neq u$, $order(v) > order(u)$

In order to define the pilot pair for type 1 spanning trees, we need to explain the following concept:

Definition 12: Depth of a pair:

In order to support our principal objective of developing a solution that reduces the partition size of the edge exchange, we shall introduce the *depth* notion. The general concept is to place the edge-node pairs in a tree in layers starting from outermost layer, i.e. leaf pairs, and through the tree to the innermost pairs.

Using the above concept, the depth of a pair should be defined theoretically as the shortest distance from the node of the pair to any leaf node of the tree. The depth of any pair using this definition ranges from 0 (for leaf pairs) to $V/2 - 1$ (for the innermost pair when the tree degenerates to a Hamiltonian path).

Anyhow, for the sake of practicality, we shall adopt a relaxed definition for depth as:

$$depth(p=(v,e) \text{ in } T) = \begin{cases} 0 & \text{If } degree(v \text{ in } T) = 1 \\ |V| - \text{length of path from } v \text{ to } v_0 & \text{If } degree(v \text{ in } T) > 1 \end{cases} \quad \text{(Eq.1)}$$

Note that instead of defining the depth using the distance from a set of nodes, we redefine it using the distance from a single predefined node, which is node zero. Under this definition the depth function will range from 0 (for leaf pairs) to $V - 1$. Moreover, the depth of the leaf pairs remains 0 while the depth of the internal pairs provides a placement that does not follow the strict layering concept. This is sufficient as we are going to distinguish in the description and analysis of the solution between two ranges of depth: depth=0 and depth > 0.

In Figure 4, the depth of different pairs in a given spanning tree is presented as calculated using Equation 1. (a) shows a graph $G(\{v_0,v_1,\ldots,v_6\},(e_1,e_2,\ldots,e_{10}\})$, (b) is a spanning tree over $G$ with leaf pairs $(v_4,e_4)$, $(v_5,e_9)$ and $(v_6,e_8)$ of depth 0, the pair $(v_3,e_5)$ of depth $|V| - 2 = 5$, and pairs $(v_1,e_1)$ and $(v_2,e_2)$ of depth $|V| - 1 = 6$.

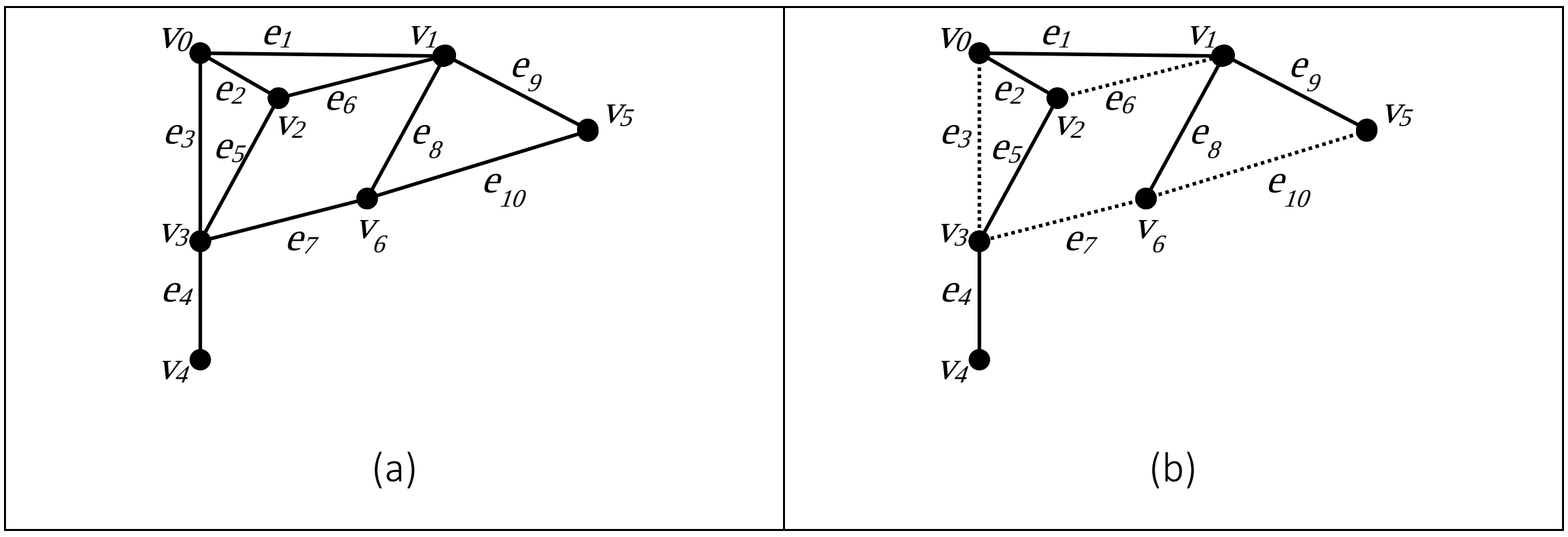

Figure 4: Depth of edge-node pairs

Definition 13: Type 1 spanning tree pilot pair:

The pilot pair of a type 1 spanning tree is the edge-node pair that satisfies the following conditions:

1- Is non-minimal,

2- Has depth less than or equal to all other non-minimal edge-node pairs,

3- Has node order higher than all non-minimal pairs with equal depth.

These conditions define a unique pair that is as far as possible from node zero to be used for the spanning tree generation.

For a spanning tree $T$ of type 1:

Let $L = \{(v,e): e \in T$ and $(v,e)$ is not minimal$\}$.

Then $pilot(T) = (v,e) \in L$ such that $\forall$ $(u,f) \in L$ and $v \neq u$, $depth(v) < depth(u)$ or $(depth(v) = depth(u)$ and $order(v) > order(u))$.

For completion, a type 1 spanning tree that has all its edge-node pairs as minimal has NIL pilot pair.

We notice that Definition 13 of type 1 pilot pair covers Definition 11 of type 2 as well. Obviously for type 2, the depth of pilot pair will be zero as it is a leaf pair, therefore, less than or equal to all other pairs. Anyhow, we select to distinguish between the two types for the sake of algorithm description and analysis.

Moreover, we shall also use the notion of *pilot node* and *pilot edge* to describe the node and edge of the pilot pair.

In Figure 5 we demonstrate the concept of pilot pair for both type 1 and type 2 spanning trees. (a) shows a graph $G(\{v_0,v_1,\ldots,v_6\},(e_1,e_2,\ldots,e_{10}\})$, (b) is a spanning tree over $G$ of type 1 and non-leaf pilot pair$(v_3,e_5)$, (c) is a spanning tree over $G$ of type 2 and leaf pilot pair $(v_6,e_8)$.

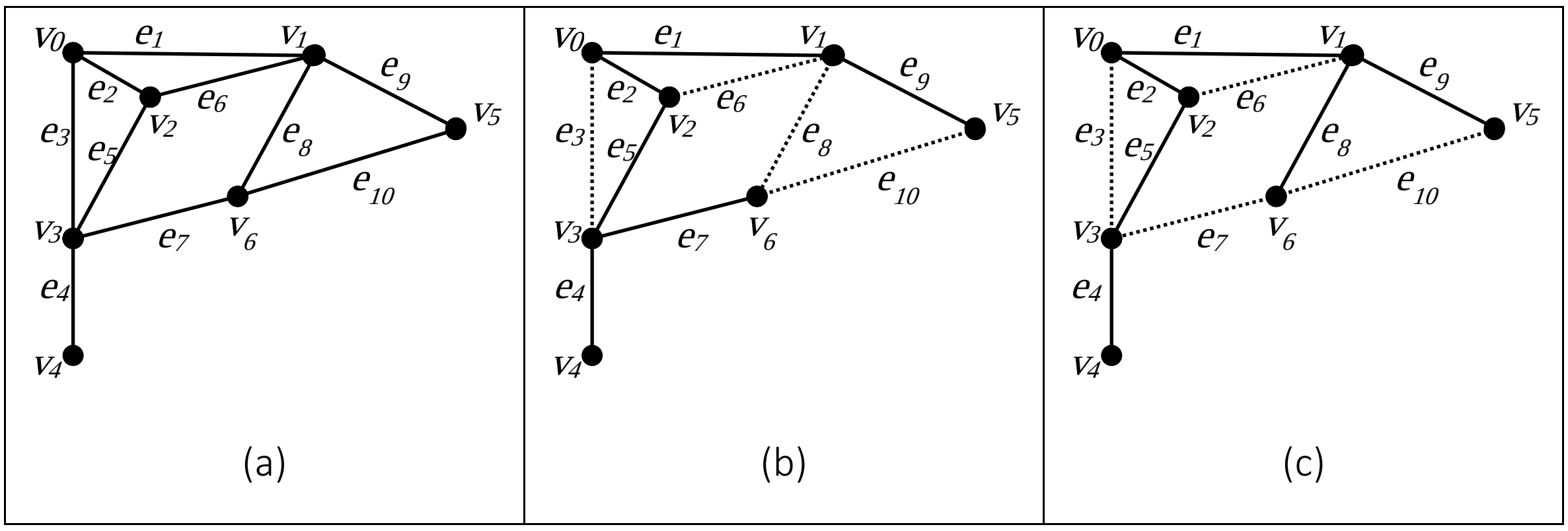

Figure 5: Pilot pair of type1 and type 2 spanning trees

Before we see how we are going to place the concept of pilot pair and pair depth in the solution framework, we shall state the following statements.

Lemma 2:

If the pilot pair of a spanning tree exists then it is unique.

Proof:

From Definition 11 and Definition 13, the pilot pair is defined as the pair that has maximum node order of all pairs satisfying the given conditions. As node ordering is distinct by Definition 1, then pilot pair is unique. ■

Lemma 3:

Edge-node pairs in the subtree rooted at a node $v$ (with respect to *node zero*) have depth lower than $pair(v)$.

Proof:

$\forall\ u \in subtree(v)$:

If $u$ is leaf pair then $depth(pair(u)) = 0 < depth(pair(v))$ as $v$ is internal.

If $u$ is internal pair then the path from $u$ to $v_0$ passes through $v$.

$\Rightarrow$ Length of path from $v$ to $v_0$ < length of path from $u$ to $v_0$

$\Rightarrow$ $depth(pair(v)) > depth(pair(u))$ By Definition 12. ■

Lemma 4:

All edge-node pairs in the subtree rooted at the pilot pair are minimal.

Proof:

By Lemma 3, all the edge-node pairs in the subtree of the pilot node $v$ have depth less than $v$.

By Definition 13, $\forall$ non-minimal pair $p \in T$: $depth(p) \geq depth(pair(v))$. ■

Lemma 5:

In any spanning tree, the pilot edge has an exchangeable edge of lower order that reconnects the spanning tree incident to the pilot node.

Proof:

We start the proof by type 2 spanning trees as it is straight forward:

As the pilot pair is non-minimal then there exists a lower order edge incident to the pilot node. But the pilot node is leaf and therefore any incident node will reconnect it to the tree.

For a type 1 spanning tree:

Let $v$ be the pilot node.

By Lemma 4, all pairs in the subtree rooted at $v$ are minimal.

But the edges incident to $v$ are the pilot edge (which is not the minimal edge of $v$ by Definition 13) and the edges from the subtree of $v$, which are all minimal edge of other nodes (which are not the minimal edge of $v$ by Lemma 6 below).

$\Rightarrow$ The minimal edge of the pilot node, let it be $(v,u)$, is not in the spanning tree.

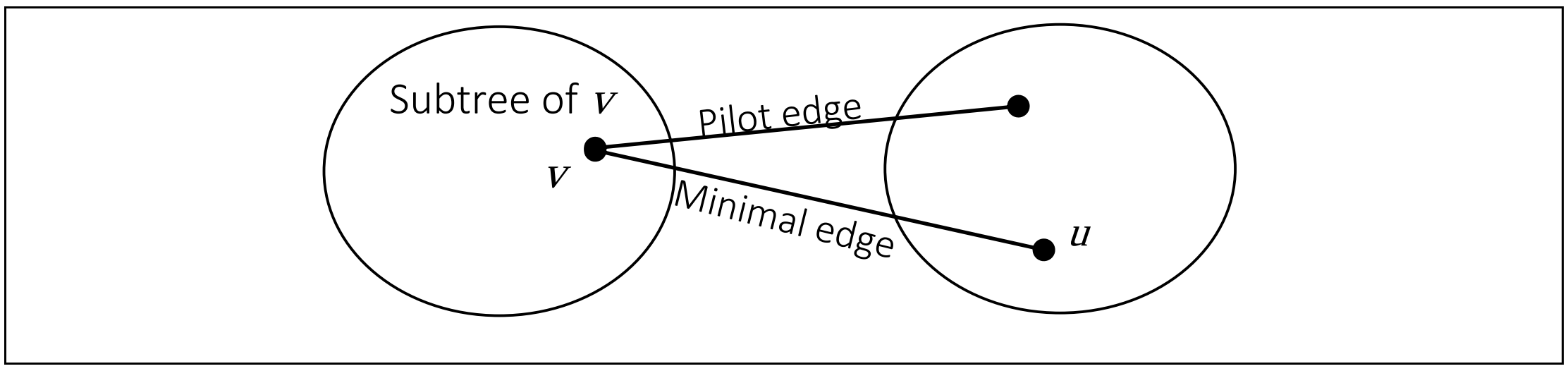

Figure 6: Existence of a lower order edge for pilot edge

Now, the subtree of $v$ is a subset of the subtree of $v$ in the minimal pairs spanning tree. This is due to that a pair can enter this subtree by promotion only. Then $u$ is not in the subtree of $v$.

⇒ The minimal edge of the pilot node, $(v,u)$, reconnects the spanning tree. ■

## 2.3 Parent/child relation of spanning trees:

The parent/child relation between spanning trees is the structural framework of the enumeration process. Once we have a well-defined parent/child relation that can be followed to reach any spanning tree then it only remains to recursively find all children of each spanning tree. Algorithm 1 depicts the abstract solution framework using the parent/child relation.

```
ENUMERATESPANNINGTREES(G)
  T0 ← a root spanning tree of G
  S(G) ← S(G) ∪ T0
  ENUMERATECHILDREN(T0)

ENUMERATECHILDREN (T)
  For each spanning tree TC in children of T
        S(G) ← S(G) ∪ TC
        ENUMERATECHILDREN (TC)
```

Algorithm 1: The theoretical solution framework of the spanning trees enumeration

We shall proceed by defining the parent spanning tree for both type 1 and type 2 spanning trees.

Definition 14: Parent of type 1 spanning tree:

The parent of a type 1 spanning tree is the spanning tree formed by exchanging the pilot pair of the tree with the edge-node pair that has the same node component paired with the edge that has maximum order less than pilot edge order while preserving the connectivity of the spanning tree.

Let $T$ be a type 1 spanning tree with pilot pair $(v,e = (v,u))$

Let $F = \{(v,t)\colon order((v,t)) < order(e) \text{ and } T - \{e\} \cup \{(v,t)\} \text{ is connected}\}$

Then:

$Parent(T) = T - \{e\} \cup \{e'\}$ where $e' \in F$ and $\forall$ $(v,t) \in F$, $e' = (v,t)$ or $order(e') > order((v,t))$.

Here we see the significance of Lemma 5. A pair that can be exchanged with the pilot pair always exists.

In Figure 7 a sample type 1 spanning tree with its parent is presented. (a) shows a graph $G(\{v_0,v_1,\ldots,v_6\},(e_1,e_2,\ldots,e_{10}\})$. (b) is a spanning tree over $G$ of type 1 and non-leaf pilot pair $(v_3,e_5)$. (c) is the parent of the spanning tree in (b) obtained by exchanging the pilot pair $(v_3,e_5)$ with the lower rank pair $(v_3,e_3)$.

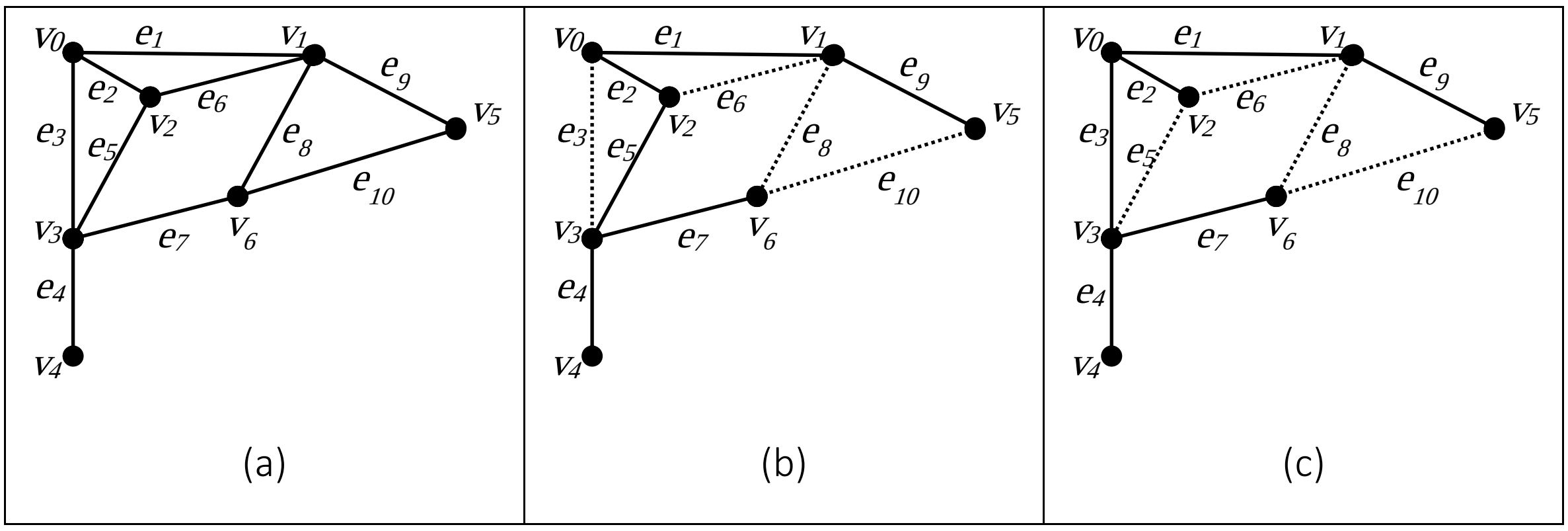

Figure 7: Parent of type 1 spanning tree

### Definition 15: Parent of type 2 spanning tree:

The parent of a type 2 spanning tree is the spanning tree formed by exchanging the leaf pilot pair of the tree with the leaf pair that has the same node component and with edge that has maximum order less than the pilot edge order.

Let $T$ be a type 2 spanning tree with pilot pair $(v,e = (v,u))$ where $v$ is leaf node. Let $F = \{(v,t): order((v,t)) < order(e)\}$

Then:

$Parent(T) = T - \{e\} \cup \{e'\}$ where $e' \in F$ and $\forall$ $(v,t) \in F$, $e' = (v,t)$ or $order(e') > order((v,t))$.

Again the concept of parent spanning tree for type 1 (Definition 14) can be generalized to type 2. Clearly, for type 2 trees there is no condition of connectivity as any edge incident to a leaf node will re-connect the graph.

In Figure 8 a sample type 2 spanning tree with its parent is presented. (a) shows a graph $G(\{v_0,v_1,\ldots,v_6\},(e_1,e_2,\ldots,e_{10}\})$. (b) is a spanning tree over $G$ of type 2 and leaf pilot pair $(v_6,e_8)$. (c) is the parent of the spanning tree in (b) obtained by exchanging the pilot pair $(v_6,e_8)$ with the lower rank pair $(v_6,e_7)$.

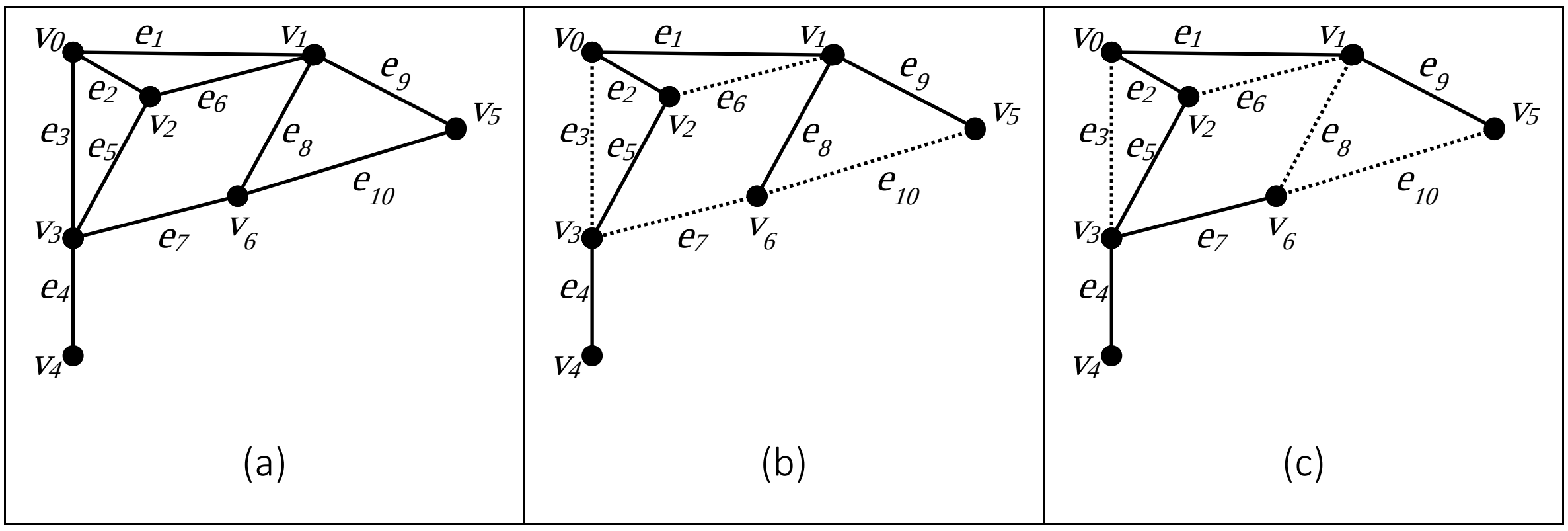

Figure 8: Parent of type 2 spanning tree

Definition 16: Pair promotion:

Define the promotion of pair $(v,e)$ as the exchange of the edge $e$ with the next in order edge incident to $v$ that retains the connectivity of the spanning tree. In case of non-leaf pair the connectivity of the spanning tree has to be verified as the exchange should not disconnect the tree. In case of leaf pair a promotion will not disturb the tree connectivity as a leaf node will remain connected by any incident edge.

Formally, a pair $(v_1, (v_1,v_2))$ in spanning tree $T$ is promoted by exchanging the edge $v_1,v_2$ with an edge $v_1,v_3$ such that:

1. $order(v_1,v_2) < order(v_1,v_3)$ and,
2. for any other edge $v_1,u$ either: (a) $order(v_1,u) < order(v_1,v_2)$, or (b) $order(v_1,u) > order(v_1,v_3)$, or (c) Exchanging $v_1,v_2$ with $v_1,u$ disconnects the tree.

We emphasize that in pair promotion one side of the edge is always fixed while exchanging the other side. This is a very important feature of MP solution framework. A spanning tree is generated from its parent by exchanging only one edge end. A great advantage over other enumeration algorithms based on edge exchange technique that has tangible impact on many spanning tree applications.

**Definition 17: Child spanning tree:**

Using the pair promotion notation (Definition 16), we can re-define the parent/child relation from the perspective of the child spanning tree for both type 1 and type 2 spanning trees as:

A child of a spanning tree is a spanning tree that is:

1- Generated by promoting a pair in the parent tree, and

2- The promoted pair becomes the pilot pair of the child tree.

Figure 9 shows a complete computational tree graph $\boldsymbol{C}(G)$ of a graph $G(\{v_0,v_1,v_2,v_3\},(e_1,e_2,e_3,e_4,e_5\})$ with parent/child relations. The nodes of $\boldsymbol{C}(G)$ are all the 8 spanning trees of $G$. Each arrow represents a parent/child relation tailed at the child. The annotation on each arrow presents the node of the promoted pair to generate the child. It can be seen how the parent/child relation makes every spanning tree reachable over the computational graph.

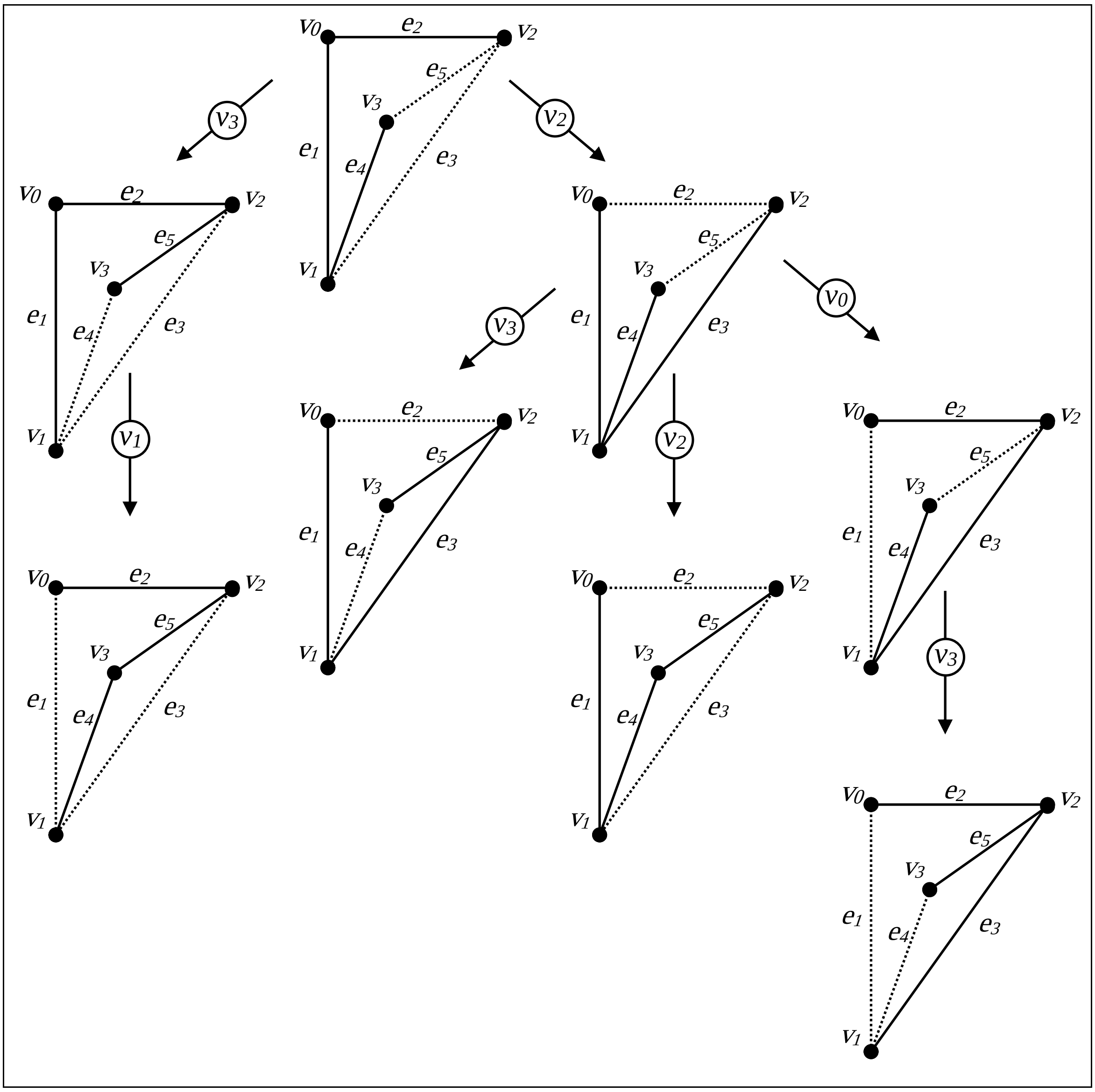

Figure 9: Computational tree graph with parent child relations.

### 2.3.1. Properties of parent/child relation:

After defining the parent/child relation, we state and prove 4 important properties of the parent/child relation we defined:

Property 1.1: Every spanning tree that has a pilot pair has a parent spanning tree.

Proof:

By Lemma 5, an exchangeable edge for the pilot pair with lower order always exists. The edge of maximum order of those edges is used to exchange the pilot edge and thus obtain the parent spanning tree. ■

**Property 1.2:** If a parent of a spanning tree exists then it is unique.

**Proof:**

By Lemma 2, the pilot pair is unique.

Let the pilot pair be $p$.

Let $F$ be the set of edges incident to the node of the pilot pair and have lower order than the edge pair.

$F = \{f: f$ is incident to $node(p)$ and $order(f) < order(edge(p))$ and $T - \{edge(p)\} \cup \{f\}$ is connected$\}$

Take $e = f$ such that $f \in F$ and $\forall f \in F: e = f$ or $order(e) > order(f)$

Now $e$ is unique and so is $T - \{edge(p)\} \cup \{e\}$ which is the parent of $T$. ■

**Property 1.3:** Parent/child relation is acyclic (a parent of a spanning tree cannot be its descendant).

**Proof:**

Define a function $\sigma(T)$ for any spanning tree $T$ as the sum of orders of all edges of a tree.

By Definition 16 of promotion, a child tree is generated by exchanging an edge $e_1$ in the parent tree with an edge $e_2$ in the child where $order(e_1) < order(e_2)$ (for both Type 1 and Type 2 trees).

$\Rightarrow \sigma(T_C) = \sigma(T_P) + order(e_2) - order(e_1) > \sigma(T_P)$ for any parent and child trees $T_P$ and $T_C$. Moreover $\sigma(T_A) < \sigma(T_D)$ for any ancestor and descendent trees $T_A$ and $T_D$.

$\Rightarrow$ If $T_1$ is the parent of $T_2$ then $\sigma(T_1) < \sigma(T_2)$ then $T_1$ cannot be a descendent of $T_2$. ■

### 2.3.2. Constructing a NIL pilot pair spanning tree:

The NIL pilot pair spanning tree plays an important role in MP enumeration algorithm. It is the root spanning tree which shall be used as the seed to generate all other spanning trees.

In order to construct a tree with NIL pilot pair, we need to join all the nodes of the graph to the spanning tree using the minimal rank pairs. An easy way to achieve this, while still using node and edge index as order, is to re-index the nodes of the spanning by the order of their appearance in a breadth (or depth)-first-search from node zero. Herewith, the all minimal pairs spanning tree will be generated using the minimum index edge for each node.

**Definition 18: Minimal-ranks-tree:**

The minimal-ranks-tree is the spanning tree generated by node zero and all minimal pairs of other nodes in a graph.

Generally, we can adjust the edge ordering function to ensure that any spanning tree constructed on the graph has NIL pilot pair by ordering the edge used to join any node before all other edges incident to that node. Whatever method we choose to generate the NIL pair spanning tree, an important observation can be realized from the construction of the NIL pilot spanning tree:

**Lemma 6:**

Every node other than node zero has a minimal order edge distinct than other nodes.

**Proof:**

Note that in the generation of *minimal-ranks-tree*, every node is joined to a node of lower order.

⇒ An edge $(u,v)$ can only be used to join the higher order node, out of $u$ and $v$, to the lower order node, but not both. ■

**Lemma 7:**

The *minimal-ranks-tree* can be constructed for any connected graph.

**Proof:**

The only requisite for constructing the *minimal-ranks-tree* is to order all nodes of the graph by *breadth-first-search* from node zero. This is possible for any connected graph. ■

### 2.3.3. Theorem 1: The root spanning tree

The only spanning tree that has NIL pilot pair is the *minimal-ranks-tree*.

Proof:

From Definition 13 of pilot pair, a spanning tree will have NIL pilot pair if no pair satisfies conditions 1-3. If any pair is non-minimal then all conditions will be met by one pair. ■

Property 1.4: The only spanning tree that has no parent is the *minimal-ranks-tree*.

Proof:

From Property 1.1, every spanning tree that has a pilot pair has a parent. But by Theorem 1 every spanning tree that has at least 1 non-minimal pair has a pilot pair.

⇒ The only spanning tree that has no parent is the spanning tree without non-minimal pair; which is the minimal-ranks-tree. ■

### 2.3.4. Theorem 2: Correctness of the solution framework

Given a connected undirected graph $G(V,E)$, consider the computational tree graph $\boldsymbol{C}(G)$ that has all spanning trees of $G$, $S(G)$, as nodes. The parent/child relation defined above constructs a directed spanning tree on $\boldsymbol{C}(G)$ rooted at the *minimal-ranks-tree*.

Proof:

The graph constructed using the child relation has the following properties:

1- Every node is reachable from the root (Property 1.1),

2- Every node has 1 path from root (Property 1.2).

3- Is acyclic (Property 1.3),

4- Has one single root at *minimal-ranks-tree* (Property 1.4).

⇒ The graph is a directed spanning tree over $\boldsymbol{C}(G)$ rooted at the *minimal-ranks-tree*.■

We conclude this chapter by the above theorem that proves the correctness of the presented solution framework. A recursive traversal of a directed spanning tree of a graph starting from the root, as in Algorithm 1, will cause each node to be visited exactly once.

Chapter 3

# Algorithm Description

In the previous chapter, we established the general structure of MP solution. Accordingly, and in order to enumerate all the spanning trees of $G$, we need to traverse the directed spanning tree generated over $\boldsymbol{C}(G)$ (as per Theorem 2) starting from the root. Taking into account that we want to distinguish between the 2 types of spanning trees, Algorithm 2 is a high level outline of the MP algorithm:

ENUMERATESPANNINGTREES($G$)
  $T_0 \leftarrow$ GENERATEMINIMUMRANKSSPANNINGTREE($G$)
  $S(G) \leftarrow S(G) \cup T_0$
  ENUMERATETYPE1ANDTYPE2CHILDREN($T_0$)

ENUMERATETYPE1ANDTYPE2CHILDREN($T$)
  **For each** spanning tree $T_C$ **in** children of $T$
      $S(G) \leftarrow S(G) \cup T_C$
      ENUMERATETYPE1ANDTYPE2CHILDREN($T_C$)

Algorithm 2: A conceptual level description of the enumeration process

We shall describe the method of initializing the algorithm by generating the root tree and go through the steps of enumeration, then the process of generating each type of tree. Subsequently, we shall study the data structures required for running these algorithms.

## 3.1 Generation of *minimal-ranks-tree*:

As we indicated earlier we shall adopt an ordering function for nodes and edges based on the index in the adjacency list (or matrix). Therefore, to generate the *minimal-ranks-tree*, we need to insure that every node, except for node zero, appears in the adjacency list (or the adjacency matrix) after at least one of its adjacent nodes. This will ensure that every node can be connected to the initial spanning tree using the minimal order edge incident to the node. In order to do that, we re-index the nodes of the graph using the order of appearance of each node in breadth-first-search (or possibly depth-first-search) starting from node zero. Then we proceed with generating the *minimal-ranks-tree.*

1. GENERATEMINIMUMRANKSSPANNINGTREE($G(V,E)$)
2. Traverse $G$ by *bfs* from $v_0$
3. Re-index $V$ by order of appearance in *bfs*
4. Sort nodes in each adjacency list
5. Use index of nodes as node order
6. Use order of appearance of edges as edge order
7. $T \leftarrow \Phi$
8. **For each** node $v$ **in** $V - \{v_0\}$
9. $T \leftarrow T \cup (v$, First element in *adjacency-list*$(v))$
10. **Return** $T$

Algorithm 3: Initialization step: generation of the root spanning tree: *minimal-ranks-tree*

Note that in Algorithm 3 above, every iteration in line 9 will add to $T$ a new node $v$ using the minimum order edge. Re-indexing the graph by *bfs* will ensure that the first element in *adjacency-list*$(v)$ is lower order than $v$ and, hence, already in the graph.

## 3.2 Properties of pair promotion:

In order to continue with the algorithm description, we need to further investigate the pair promotion action, Definition 16, as it is the technique used to generate all the spanning trees from the root spanning tree. We shall list and prove some important properties of pair promotion and its impact on the pairs and nodes of the tree after we define the following terms:

### Definition 19: Covered node and uncovered node by a promotion:

Consider a pair promotion that exchanges the edge $v_1,v_2$ with the edge $v_1,v_3$. The 3 nodes involved in the promotion are $v_1$, the node of the promoted pair, $v_2$, the uncovered node, and $v_3$ the covered node.

Figure 10 demonstrate the concept of covered and uncovered node for a promoted pair. (a) shows a graph $G(\{v_0,v_1,\ldots,v_6\},(e_1,e_2,\ldots,e_{10}\})$. (b) is the *minimal-ranks-tree* over $G$. (c) is generated from the spanning tree in (b) by promotion of pair $(v_3,e_3)$ to $(v_3,e_5)$. The covered node by the promotion is $v_2$ and the uncovered node is $v_0$.

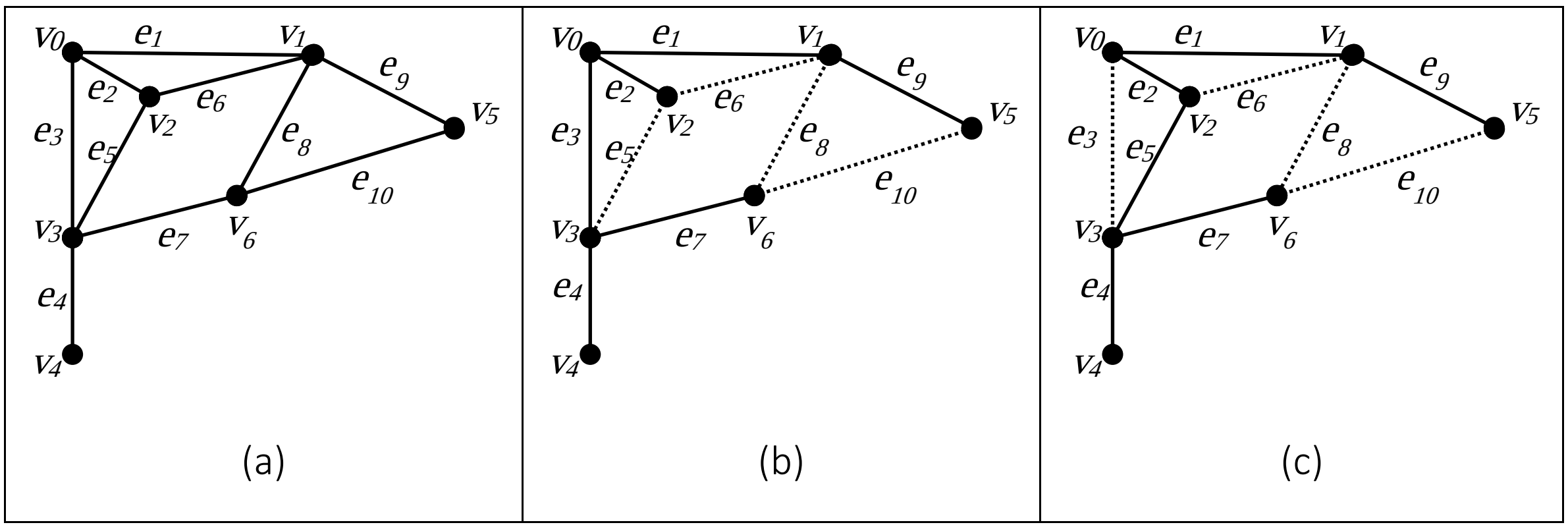

Figure 10: Covered node and uncovered node by promotion

**Property 2.1:** The rank of promoted pair is increased.

If promoted pair is minimal it becomes non-minimal and considered for the pilot pair. Promoted pair may become maximal (or saturated) and does not need to be considered for further promotion.

**Proof:**

Directly from Definition 16. ■

**Property 2.2:** The promoted pair will retain its state as leaf or non-leaf before and after promotion.

**Proof:**

By Definition 16: The number of edges incident to the node of the promoted pair is the same before and after the promotion. ■

**Property 2.3:** The degree of uncovered node in the tree is decreased. This may result in generating a new leaf of the tree at the uncovered node.

**Proof:**

By Definition 16: One of the edges incident to the uncovered node is removed by promotion. ■

**Property 2.4:** The degree of covered node in the tree is increased. This will eliminate the leaf of the tree at covered node (if exists).

**Proof:**

By Definition 16: An edge incident to the covered node is added by promotion. ■

**Property 2.5:** The depths of covered and uncovered nodes remain unchanged unless its leaf or non-leaf state is changed.

**Proof:**

As per Definition 12 the depth of a pair is based on 2 factors: Whether it is leaf and the path length between the node of the pair and node zero.

For covered and uncovered nodes the path to node zero remains the same.

$\Rightarrow$ Only the change of state from leaf to non-leaf will change its depth. ■

**Property 2.6:** The depth of promoted pair and all the pairs in its subtree may change.

**Proof:**

As per Definition 12, the depth of a pair is based on 2 factors: Whether it is leaf and the path length between the node of the pair and node zero.

As per Property 2.2, the state of being leaf or non-leaf will not change for the promoted node. On the other side, by Definition 4, the path from the promoted node to node zero before the promotion is through the uncovered node and after the promotion through covered node. For the promoted pair $p$, covered node $v_c$ and uncovered node $v_u$:

Length of path from $node(p)$ to $v_0$ (before promotion) = Length of path from $v_u$ to $v_0$ + 1

Length of path from $node(p)$ to $v_0$ (after promotion) = Length of path from $v_c$ to $v_0$ + 1 **(Eq.2)**

$\Rightarrow$ $depth(p)$ may change by promotion.

$\forall\ v \in subtree(node(p))$ path from $u$ to $v_0$ passes through $node(p)$

$\Rightarrow$ Length of path from $v$ to $v_0$ = length of path from $v$ to $node(p)$ + length of path from $node(p)$ to $v_0$ **(Eq.3)**

$\Rightarrow$ Length of path from $v$ to $v_0$ may change by promotion

$\Rightarrow$ $depth(pair(v))$ may change by promotion. ■

We notice that, as depths are calculated with respect to node zero, when node zero is promoted then all the paths of pairs to node zero (and subsequently the depths) have to be recomputed. Fortunately, there are some properties of node zero promotion, which we shall state and use (in Lemma 8) to eliminate this cost of node zero promotion.

**Property 2.7:** Node zero is promoted as a leaf pair only (not as internal node).

**Proof:**

As node zero has no associated edge (remember that in a spanning tree $E = V - 1$). Only when node zero becomes a leaf, we associate the only incident edge with it to consider the spanning tree to be the generated by promoting the leaf edge at node zero. ■

**Property 2.8:** No pair can be promoted to cover node zero.

**Proof:**

Promotion, By Definition 16, should exchange the edge of the pair with a higher order edge. The uncovered node should be lower order than the covered node. Which cannot occur when the covered node is node zero. ■

**Property 2.9:** When node zero becomes leaf node, it remains leaf node in all descendant trees.

**Proof:**

This follows directly from Property 2.8. ■

Using Property 2.7 - Property 2.9 we shall state the following important result which shall be used to enumerate all spanning trees generated by promotion of node zero.

**Lemma 8:**

All spanning trees where node zero is a leaf node with non-minimal edge can be generated from spanning trees where node zero is leaf edge with minimal edge by pair promotion.

**Proof:**

Given a spanning tree $T_1$ where node zero is leaf node with non-minimal edge, consider the spanning tree $T_2$ generated by exchanging the edge incident to node zero with the minimal edge of node zero.

$T_1$ can be generated from $T_2$ by promotion of node zero pair. ■

## 3.3 Different scenarios of spanning tree generation:

For generating spanning trees, we have 4 different scenarios based on the type of parent and child tree. We shall use the term *type-1-from-1* to indicate generating a type 1 child spanning tree from a type 1 parent spanning tree. Likewise, we shall use the terms *type-1-from-2*, *type-2-from-1* and *type-2-from-2* to indicate the corresponding parent and child types.

### 3.3.1. *Type-1-from-1*:

Lemma 9:

Type 1 tree can be generated as a child of another type 1 tree if (and only if):

1- A non-leaf pair is promoted and,

2- The promoted pair is the pilot of the child tree.

Proof:

Directly from Property 2.2, only non-leaf pairs can be promoted to type 1 tree. And from Definition 14 of parent of spanning tree condition 2 follows. ■

In Algorithm 4, we traverse all non-leaf pairs to generate possible type 1 child spanning trees. Anyhow, we shall see later, that by using suitable data structures, we do not need to test all non-leaf pairs.

```
1.  ENUMERATETYPE1CHILDRENOFTYPE1(T)
2.    For each p in non-leaf pairs of T
3.        T_C ← Promote p in T
4.        If T_C is connected and pilot(T_C) = p
5.            S(G) ← S(G) ∪ T_C
```

Algorithm 4: Enumeration of type 1 child spanning trees from a type 1 spanning tree parent

### 3.3.2. *Type-2-from-1*:

Lemma 10:

Type 2 spanning tree can be generated as a child of a type 1 spanning tree if (and only if):

1- A leaf pair is promoted and,

2- The promoted pair does not uncover the pilot pair of the child tree.

Proof:

Again condition 1 follows directly from Property 2.2.

For condition 2, refer to Definition 15 of parent of spanning tree, the promoted leaf should be the pilot of the child tree.

Note that all the leaf pairs of the parent are minimal (by Definition 9 of type 1 spanning tree). Therefore, to satisfy Definition 15, the promotion should not uncover a non-minimal leaf node higher in order than the promoted leaf. In other words, it should be checked that the node uncovered by the promotion is not the pilot node of the child tree; otherwise, the promoted pair is definitely the pilot of the child spanning tree as it is the only non-minimal leaf. ■

Algorithm 5 shows how enumeration of *type-2-from-1* is implemented.

```
1.  ENUMERATETYPE2CHILDRENOFTYPE1(T)
2.      For each p in leaf pairs of T that do not uncover a higher order non-minimal leaf
3.          T_C ← Promote p on T
4.          S(G) ← S(G) ∪ T_C
```

Algorithm 5: Enumeration of type 2 child spanning trees from a type 1 spanning tree parent

Later, we shall introduce data structures to ensure that every iteration in lines 3-4 in Algorithm 5 produces a new child tree.

### 3.3.3. *Type-1-from-2*:

Lemma 11:

Type 1 spanning tree can be generated as a child of a type 2 spanning tree if (and only if):

1. The parent tree has only 1 non-minimal leaf pair (which is the pilot pair) and,
2. A non-leaf pair is promoted and,
3. The promotion covers the pilot leaf pair and,
4. The promoted pair is the pilot of the child tree.

**Proof:**

Conditions 1, 2 and 3 can be deduced noticing that a type 1 tree has no non-minimal leaf pair. We know that a non-minimal pair is never exchanged by a minimal pair. This implies that the parent type 2 tree has only 1 non-minimal leaf and this leaf is covered, i.e. made non-leaf, by the promotion.

Condition 4 follows from Definition 14 of parent of type 1 spanning tree. ∎

Algorithm 6 shows how enumeration of *type-1-from-2* is implemented.

```
ENUMERATETYPE1CHILDRENOFTYPE2(T)
  If count(non-minimal leaves in T) = 1
        v_T ← pilot node of T
        For each v in adjacency-list(v_T)
              T_C ← Promote pair(v) on T
              If promotion of pair(v) covers v_T and pilot(T_C) = v
                    S(G) ← S(G) ∪ T_C
```

**Algorithm 6: Enumeration of type 1 child spanning trees from a type 2 spanning tree parent**

An important observation deduced from condition 3 Lemma 11, and utilized in Algorithm 6, is that only nodes adjacent to the pilot node of parent tree should be considered for this type of promotion.

### 3.3.4. *Type-2-from-2*:

**Lemma 12:**

A type 2 child tree can be generated from type 2 parent by promotion of pair $p$, if (and only if):

1. Pair $p$ is a leaf pair,
2. The node of the promoted pair:
    a. Is of order higher than or equal to the order of the node of the pilot pair in $T$ or,
    b. Covers the pilot pair and becomes the non-minimal leaf pair with highest node order.

3. The uncovered pair does not become the pilot pair (by uncovering a non-minimal leaf of higher node order)

Proof:

Condition 1 follows directly from Property 2.2. Type 2 spanning tree can be generated as a child of another type 2 spanning tree only by a leaf pair promotion.

For condition 2, note that a type 2 spanning tree already has non-minimal leaf pairs. In the parent, the non-minimal leaf pair with highest node order is the pilot pair. From Definition 15, and in order to promote a leaf pair that becomes the child pilot, this leaf pair should either have node order higher than or equal to the current pilot leaf node or should cover the pilot leaf node (i.e. make it non-leaf node) and become the highest non-minimal leaf pair.

Condition 3 follows by the same argument used to prove condition 2 in Lemma 10. ■

We shall call the first subtype, where the parent pilot node is not covered, as *type-2-from-2a* and the second subtype, where the parent pilot node is covered, as *type-2-from-2b*. It is clear that for *type-2-from-2b*, as in Algorithm 7, we need to consider only leaves in the adjacency list of the pilot node.

ENUMERATETYPE2CHILDRENOFTYPE2($T$)
    $v_T \leftarrow$ pilot node of $T$
    ▷ *type-2-from-2a*
    **For each** $v$ **in** leaf pairs of $T$ that do not uncover a higher order leaf and $order(v) \geq order(v_T)$
        $T_C \leftarrow$ Promote $v$ on $T$
        $S(G) \leftarrow S(G) \cup T_C$
    ▷ *type-2-from-2b*
    **For each** $v$ **in** *adjacency-list*($v_T$) and $v$ is leaf node of $T$
        $T_C \leftarrow$ Promote $pair(v)$ on $T$
        **If** promotion of $pair(v)$ covers $v_T$ **and** $pilot(T_C) = pair(v)$
            $S(G) \leftarrow S(G) \cup T_C$

Algorithm 7: Enumeration of type 2 child spanning trees from a type 2 spanning tree parent (2 subtypes are possible)

## 3.4 Data structures:

In order to run the algorithm described in the previous section, data structures are required to facilitate the enumeration process and perform the necessary tasks with minimum complexity. These data structures has to be initialized and maintained by the algorithm.

Initialization of data structures:

All the data structures have to be initialized together with the generation of *minimal-ranks-tree*. The data structures should reflect the status of the enumeration process at the root spanning tree.

Maintenance of data structures:

The data structures used by the algorithm have to be maintained whenever a new spanning tree is generated to reflect the current situation of process. As a new spanning tree is only generated by a promotion of a pair in the existing spanning tree, we can link the maintenance of the data structures with promotion only. All the changes that will occur in the data structures will be expressed in terms of the pair promotion being performed.

Now, if we examine the operations performed for generation of child spanning trees (Algorithm 4 - Algorithm 7) we shall find that 2 main operations are required:

1- Traversal of pairs that are candidates for generation of child trees.
2- Calculation of pilot for generated child tree.

We shall introduce 3 data structures that will support performing these operations.

### 3.4.1. Data structure 1: Augmenting the spanning tree:

The principal data structure used for enumeration is, naturally, the current spanning tree which is the parent tree from which we should fork all direct child trees. This spanning tree will be stored in the form of pairs where each pair contains the node and the edge of the pair. Additionally each pair in the spanning tree is augmented with extra attributes that facilitates different calculations in the process. The attributes added are:

1- **Degree in tree:** the count of incident edges to the node of the pair. This is useful to check leaf pairs.

2- **Distance from node zero:** the length of the path to node zero. This is used in depth calculation, and as a result in pilot appointment. The depth of any pair after a planned promotion can be calculated directly from the covered node (using Equation 2).

3- **Count of non-minimal descendants:** the count of non-minimal descendants for any leaf equals the count of non-minimal descendants of all its children plus its non-minimal children. This attribute is useful for testing Property 3.4 below.

We can see that by these extra attributes many tests required on pairs during the process are performed in constant time. This comes at the cost of maintaining these attributes with every promotion in the enumeration process.

Initialization:

The augmentation is part of the spanning tree and has to be initialized together with the generation of *minimal-ranks-tree*. The attributes calculated for each pair are:

1- The degree of each node in the *minimal-ranks-tree*.

2- The initial distance of nodes of pairs from node zero.

3- Count of non-minimal descendants is zero for all pairs as all pairs are minimal.

Maintenance:

The augmented attributes has to be maintained after every promotion with other data structures. The changes that may occur in these data when a pair is promoted are:

1- Degree of uncovered node is decremented.

2- Degree of covered node is incremented.

3- Distance from node zero should be calculated for the promoted pair (using Equation 2).

4- Distance from node zero should be calculated for all pairs in the subtree of the promoted node (using Equation 3).

5- Count of non-minimal descendants for all ancestors of uncovered node should be decremented if promoted pair was non-minima before promotion.

6- Count of non-minimal descendants for all ancestors of covered node should be incremented.

Algorithm 8 shows the maintenance actions required for this data structure after promoting a pair.

```
MAINTAINSPANNINGTREEATTRIBUTES(pilot)
    degree(uncovered-node) ← degree(uncovered-node) – 1
    degree(covered-node) ← degree(covered-node) + 1
    length of path from pilot to v0 ← length of path from covered-node to v0 + 1
    For each p in subtree(node(pilot))
            Calculate length of path from p to v0
    If pilot (before promotion) is non-minimal
            Decrement count-of-non-minimal-descendants of all previous ancestors of
    pilot
    Increment count-of-non-minimal-descendants of all ancestors of pilot
```

Algorithm 8: Maintenance of the attributes of the spanning tree after a pair promotion

### 3.4.2. Data structure 2: ***Promotable-pairs***:

This data structure will be used to traverse pairs for promotion. The data structure should ensure that:

1- All possible promotions that may generate child spanning trees are included,

2- Unnecessary promotions that do not generate child spanning trees are eliminated (or at least reduced).

For these requirements to hold we shall use the following features of member pairs to build and maintain the data structure.

Properties of member pairs:

Property 3.1: The pair should not be maximal.

As maximal pairs (by Definition 8) are saturated and cannot be promoted.

Property 3.2: Promotion of the pair should not uncover a pair with better standings for pilot appointment.

The 3 nodes affected by the promotion are $v_p$, the node of the promoted pair, $v_u$, the uncovered node, and $v_c$ the covered node. From Property 2.3 - Property 2.5 of pair promotion, the depth of the covered and uncovered pairs will only change if the pair becomes non-leaf or leaf, respectively. The depth of the promoted pair after the promotion is:

$depth(pair(v_p))$ (after promotion) = $depth(pair(v_c))$ – 1 or 0 if leaf pair.

By Definition 11 and Definition 13, the promotion uncovers a pair with higher pilot standings if:

$$pair(v_u) \text{ is non minimal and } \begin{cases} depth(pair(v_u)) < depth(pair(v_c)) - 1 \\ or \\ depth(pair(v_u)) = depth(pair(v_c)) - 1 \text{ and} \\ order(v_u) > order(v_p) \end{cases}$$

As the promoted pair should be the pilot, any promotion that uncovers a non-minimal pair of better pilot standings than the promoted pair should be excluded from the data structure.

**Property 3.3:** The promoted pair should become the pilot. Specifically, the promoted pair should have, after promotion, higher standings for pilot appointment than the current pilot.

As per Definition 13, standings for pilot appointment are based on pair depth and node order. From Equation 2, the depth of the promoted pair can be calculated from the depth of the covered pair. We need to consider pairs that have lower depth than the current pilot or same depth and higher node order.

In order to ensure that only pairs with better pilot standings are considered, the pairs in this data structure shall be ordered by depth (ascending) and node order (descending). By this ordering we can access the head of the data structure till we reached the pilot (or the last element with standings better than pilot). All these elements will certainly produce a new child as the promoted pair will be higher in pilot standings than parent pilot.

**Property 3.4:** All pairs in the subtree rooted at the promoted pair should be minimal.

As these pairs have depth less than the promoted pair (by Lemma 3) and, therefore, better standings for the pilot appointment.

**Property 3.5:** There is an edge of higher order than current pair rank that is not present in the tree, to allow opportunity for promotion, and,

**Property 3.6:** The edge described in Property 3.5 retains the connectedness of the tree.

In Lemma 13 we show a necessary and sufficient condition for Property 3.6 to hold.

**Lemma 13:**

A promotion disconnects the spanning tree if (and only if) the covered node is in the subtree of the promoted pair.

**Proof:**

Let the node of the promoted pair be $v_p$, the covered node be $v_c$ and the uncovered node be $v_u$ in the spanning tree $T$.

A promotion will exchange the edge $v_p,v_u$ with $v_p,v_c$.

When the edge $v_p,v_u$ is removed the tree $T$ will be partitioned to: $\mathit{subtree}(v_p)$ and $T - \mathit{subtree}(v_p)$, where $v_p \in \mathit{subtree}(v_p)$.

The edge $v_p,v_c$ reconnects the tree $T$ if (and only if) $v_c \in T - \mathit{subtree}(v_p)$. ■

Figure 11 illustrates Lemma 13. (a) shows a graph $G(\{v_0,v_1,\ldots,v_6\},(e_1,e_2,\ldots,e_{10}\})$ with a spanning tree. (b) is an illegal promotion of $(v_2,e_2)$ to $(v_2,e_7)$ that disconnects the tree. Note that the covered node $v_6 \in \mathit{subtree}(v_2)$. (c) is a legal promotion of $(v_1,e_1)$ to $(v_1,e_6)$ that retains connectedness of the tree. Note that the covered node $v_0 \notin \mathit{subtree}(v_1)$.

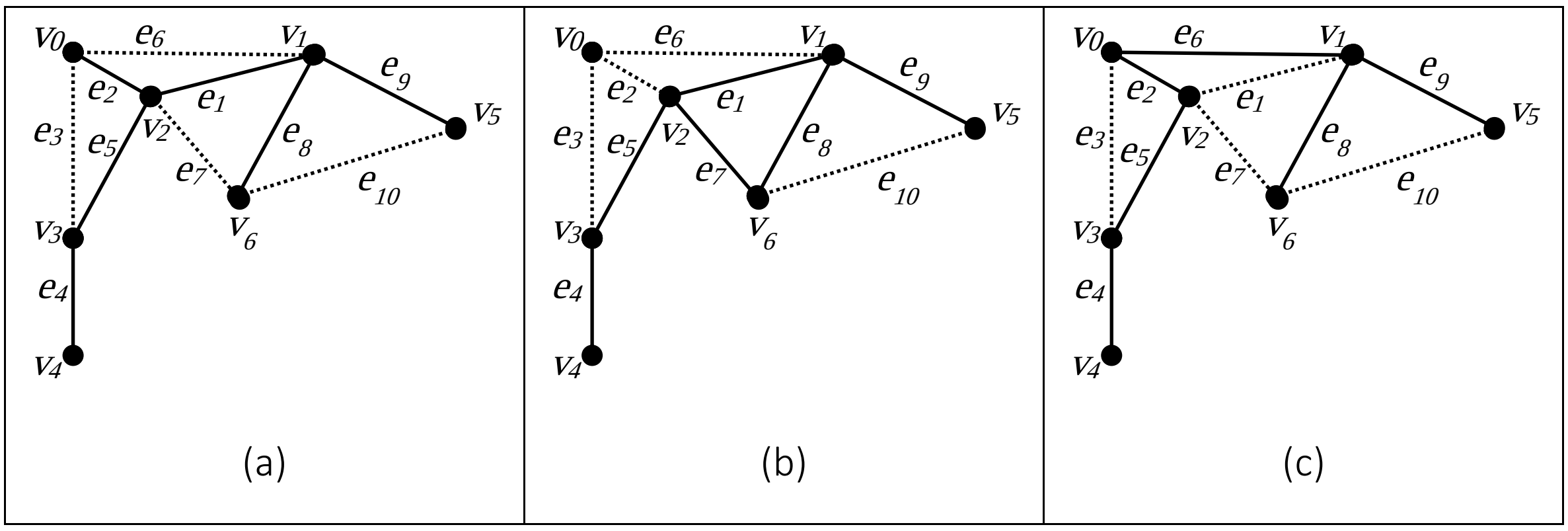

Figure 11: Promotion must retain the connectedness of the spanning tree

Initialization:

For initializing the data structure *promotable-pairs*, it is obvious that all pairs in *minimal-ranks-tree* comply with Property 3.2, Property 3.3 and Property 3.4 of *promotable-pairs*. Property 3.1 and Property 3.5 will hold for all, and only for all, pairs of nodes that have degree in *minimal-ranks-tree* less than their degree in graph. Only Property 3.6 has to be explicitly tested using Lemma 13.

Algorithm 9 shows the pseudo-code for initialization of this data structure.

1. INITIALIZE-PROMOTABLE-PAIRS()
2. *promotable-pairs* ← Φ
3. **For each** *p* **in** pairs of *minimal-ranks-tree*
4. **If** *degree*(*node*(*p*) in *G*) > *degree*(*node*(*p*) in *minimal-ranks-tree*)
5. **and** *pair*(*p*) has a promotion not in *subtree*(*p*)
6. *promotable-pairs* ← *promotable-pairs* ∪ *p*

Algorithm 9: Initialization of *promotable-pairs* structure

Maintenance:

We shall introduce the requirements for maintaining *promotable-pairs* data structure by going through the properties of members of *promotable-pairs* and examining how promotion may affect these properties. It is important to note that the effect of promotion may remove existing members from *promotable-pairs*, add non-members to *promotable-pairs* and relocate some members.

Maintaining Property 3.1:

1- The promoted pair may become maximal. It should be removed from the structure.

Maintaining Property 3.2:

2- The covered pair may be of better standings for pilot appointment than the promoted pair. The promoted pair should be removed from the structure.

3- The uncovered pair may become of higher standings for pilot appointment than another promotable pair that covers it. This promotable pair should be removed from the structure.

4- The covered pair may become of lower standings for pilot appointment than another promotable pair that covers it. This promotable pair should be added to the structure.

Maintaining Property 3.3:

5- The depth of pairs in the subtree of the promoted pair should be adjusted. This may also affect the standings for pilot appointment (Property 3.2).

Maintaining Property 3.4:

6- The ancestors of the promoted pair, after the promotion, should be removed from *promotable-pairs* as they have a non-minimal descendent.

7- If the promoted pair was non-minimal before the promotion then the ancestors of the promoted pair, before the promotion, should be examined as they may have no non-minimal descendent (and thus may be added to the structure).

Maintaining Property 3.5:

8- If the uncovered pair was not in *promotable-pairs* breaching Property 3.5. It may comply with Property 3.5 after the promotion and, therefore, should be added to the structure.

9- If the covered pair was in *promotable-pairs*. It may breach Property 3.5 after the promotion and, therefore, should be removed from the structure.

Maintaining Property 3.6:

10- Some permitted promotions that result in a connected tree may become illegitimate and will disconnect the tree. And vice-versa. In order to examine such situations we state the following statements.

Lemma 14:

A promotion $p_1$ that retains the connectedness of a spanning tree will disconnect the spanning tree after performing another promotion $p_2$ if (and only if):

Covered node by $p_1 \in subtree$(node of $p_2$) and

Covered node by $p_2 \in subtree$(node of $p_1$) and

Uncovered node by $p_2 \notin subtree$(node of $p_1$)

**Proof:**

For the promotion $p_1$, let the node of the promoted pair be $v_{p1}$, the covered node $v_{c1}$ and the uncovered node be $v_{u1}$. Likewise, for promotion $p_2$: $v_{p2}$, $v_{c2}$ and $v_{u2}$.

($\Rightarrow$)

Before promotion $p_2$:

Promotion $p_1$ reconnects the spanning tree $\Leftrightarrow v_{c1} \notin subtree(v_{p1})$ by Lemma 13

After promotion $p_2$:

Promotion $p_1$ disconnects the spanning tree $\Leftrightarrow v_{c1} \in subtree(v_{p1})$ by Lemma 13

The promotion $p_2$ takes $v_{c1}$ from $T - subtree(v_{p1})$ to $subtree(v_{p1})$ but promotion $p_2$ only moves $subtree(v_{p2})$

$\Rightarrow$ $v_{c1} \in subtree(v_{p2})$ and $v_{p2}$ (before promotion $p_2$) $\notin subtree(v_{p1})$ and $v_{p2}$ (after promotion $p_2$) $\in subtree(v_{p1})$

$\Rightarrow$ $v_{c1} \in subtree(v_{p2})$ and $v_{c2} \in subtree(v_{p1})$ and $v_{u2} \notin subtree(v_{p1})$

($\Leftarrow$)

If we take the other way round,

$v_{c1} \in subtree(v_{p2})$ and $v_{c2} \in subtree(v_{p1})$ and $v_{u2} \notin subtree(v_{p1})$

It is clear we have:

$v_{c1}$ (before promotion $p_2$) $\notin subtree(v_{p1})$ and $v_{c1}$ (after promotion $p_2$) $\in subtree(v_{p1})$

$\Rightarrow$ Promotion $p_1$ retains connectedness of the tree before promotion $p_2$ and disconnects the tree after promotion $p_2$. ∎

Similarly, we can state the following with the same argument for Lemma 14.

**Lemma 15:**

A promotion $p_1$ that disconnects a spanning tree will retain the connectedness of the spanning tree after another promotion $p_2$ if (and only if):

Covered node by $p_1 \in subtree$(node of $p_2$) and

Covered node by $p_2 \notin subtree$(node of $p_1$) and

Uncovered node by $p_2 \in subtree$(node of $p_1$).

Algorithm 10 provides an in depth look into the maintenance of *promotable-pairs* data structure.

```
MaintainPromotablePairs(pilot)
  If pilot (after promotion) is maximal
      Remove pilot from promotable-pairs
  If covered-node (before promotion) is leaf
      Relocate pair(covered-node) in promotable-pairs
  If uncovered-node (after promotion) is leaf
      Relocate pair(uncovered-node) in promotable-pairs
  Check non-promotable pairs that cover covered-node if became promotable
  Check promotable pairs that cover uncovered-node if became non-promotable
  For each p in subtree(node(pilot))
      Relocate p in promotable-pairs as per new depth(p)
  Find (before promotion) 1st-non-minimal-ancestor
  If count-of-minimal-descendants of 1st-non-minimal-ancestor = 0
      Add all ancestors till 1st-non-minimal-ancestor to promotable-pairs
  Find (after promotion) 1st-non-minimal-ancestor
  If count-of-minimal-descendants of 1st-non-minimal-ancestor = 1
      Remove 1st-non-minimal-ancestor from promotable-pairs
  If covered-node in promotable-pairs
      If not have available edge for promotion
          Remove pair(covered-node) from promotable-pairs
  If uncovered-node not in promotable-pairs
      If have available edge for promotion
          Add pair(uncovered-node) to promotable-pairs
  For each v in subtree(node(pilot))
      If v is cover-node of a promotion p
          Check status of p using Lemma 14 and Lemma 15
```

Algorithm 10: Maintenance of *promotable-pairs* structure after a pair promotion

### 3.4.3. Data structure 3: *Pilot-candidates*:

This data structure contains all non-minimal pairs that are considered for pilot appointment in the order of their standings. As all the pairs in the *minimal-ranks-tree* are minimal, by Definition 18, this structure holds pairs that were promoted at least once to generate one of the ancestors of current tree.

This data structure is used to dictate the promotions that will produce a pilot pair in the child tree. A promotion of any pair that will not make it a pilot pair should not be considered. Practically, at any stage we need to track only the 1st two members of this structure. The 1st member is the pilot of current tree while the second member is the next higher in standings for pilot appointment. We shall use the current pilot to force all promotions of higher (or equal) standings to be generated. The promotions having standings between pilot and *next-to-pilot* will occur if (and only if) it degrades the standings of the current pilot and becomes the new pilot. Other promotions, less than *next-to-pilot* are completely neglected as it cannot degrade 2 pilot candidates higher than it.

**Properties of member pairs:**

**Property 4.1**: The pair should not be minimal. As by Definition 11 and Definition 13 only non-minimal pairs are considered for pilot appointment.

**Property 4.2**: The pairs in the structure are ordered by their standings for pilot appointment (i.e. by depth (ascending) then node order (descending)).

**Property 4.3**: A promoted pair should prove to be the 1st pair in the structure after the promotion in order to be considered for promotion. Here we do not need to find the pair's actual position as any position other than the head of the structure is rejected.

**Initialization:**

Initializing *pilot-candidates* is trivial as all pairs in *minimal-ranks-tree* are minimal and do not comply with Property 4.1.

Initially, this structure is empty as all pairs are minimal.

1. INITIALIZE-PILOT-CANDIDATES()
2. *pilot-candidates* ← Φ

**Algorithm 11: Initialization of *pilot-candidates* structure**

**Maintenance:**

The structure is used during pair promotion by the following inspections:

1- The promoted pair should be the new pilot. It should be the head element of the structure.

2- The promoted pair should be relocated in the structure according to the new depth.

3- The second element in the structure is appointed as *next-to-pilot*.

We shall examine the effect of promotion on the properties used to assign these 2 members of this data structure:

Maintaining Property 4.1:

1. The promoted pair may become non-minimal. It should be added to the structure.

Maintaining Property 4.2:

2. The depth of the promoted pair may change. It should be relocated in the structure. Fortunately, pairs in the subtree of the promoted pair are all minimal (by Property 3.4) and do not belong to this structure (by Property 4.1).
3. If the covered node is leaf before the promotion it becomes non-leaf. This will degrade its standings for pilot appointment.
4. If the uncovered node becomes leaf after the promotion. This will upgrade its standings for pilot appointment.

Algorithm 12 provides an in depth look into the maintenance of *pilot-candidates* data structure.

```
MAINTAINPILOTCANDIDATES(pilot)
    If pilot (before promotion) is minimal
        Add pilot to pilot-candidates
    Else
        Relocate pilot in pilot-candidates
    If covered-node (before promotion) is leaf
        Relocate pair(covered-node) in pilot-candidates
    If uncovered-node (after promotion) is leaf
        Relocate pair(uncovered-node) in pilot-candidates
    next-to-pilot ← 2nd element in pilot-candidates
```

Algorithm 12: Maintenance of *pilot-candidates* structure after a pair promotion

A complete example of using MP algorithm to enumerate the spanning trees on a simple graph is shown in Figure 12. MP algorithm is executed to enumerate all the spanning trees of a graph $G(\{v_0,v_1,\ldots,v_4\},(e_1,e_2,\ldots,e_6\})$. The arrows indicate the promotion of pairs to generate child spanning trees. The three data structures are initialized at the root, and maintained on each step once the promotion is performed. The spanning trees nodes are augmented by depth (up) and count of non-minimal children (down). The degree of node is not displayed (as it is obvious). The *pilot-candidates* and *promotable-pairs* data structures are shown at each step as maintained by the algorithm.

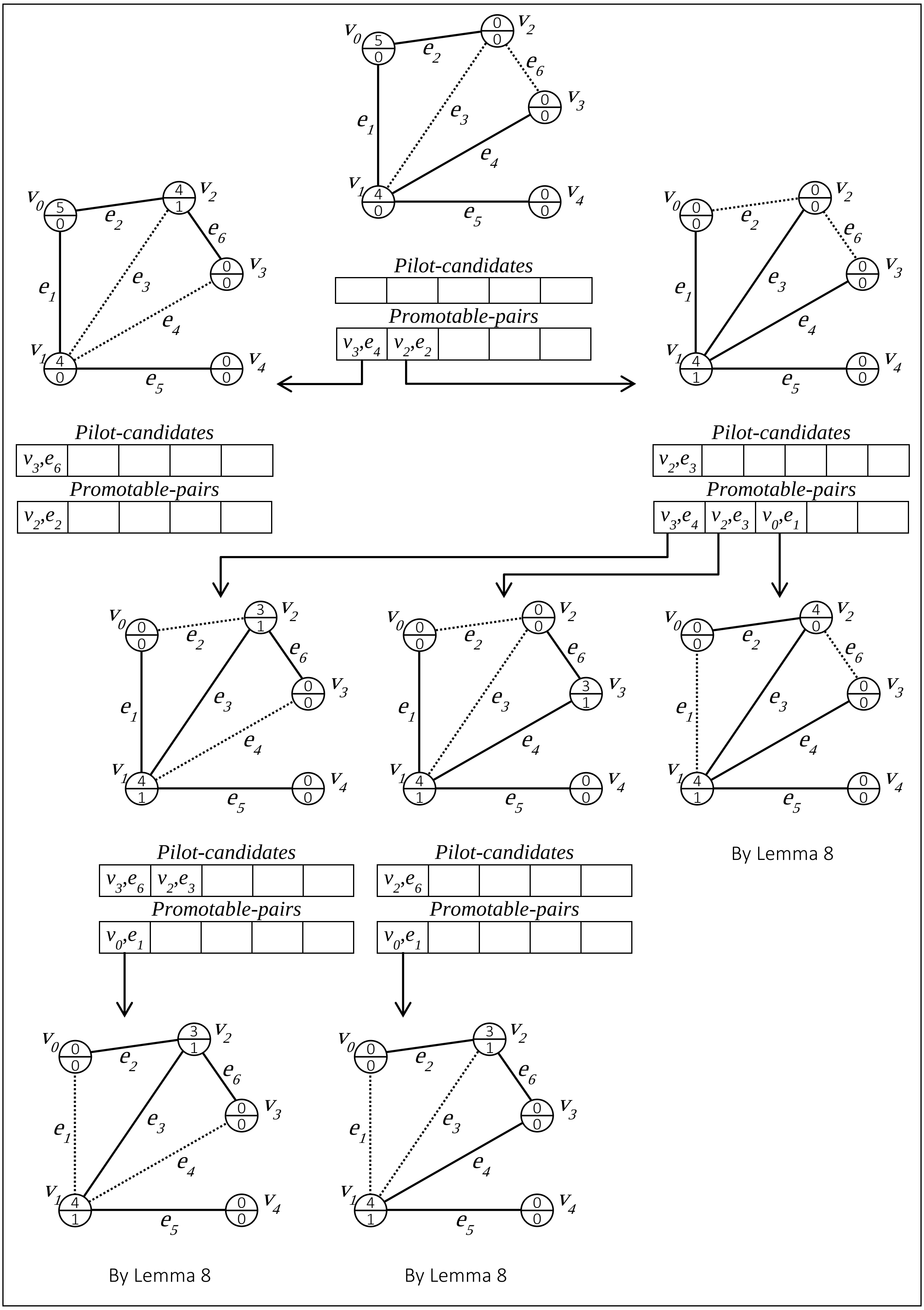

Figure 12: MP algorithm with data structures to enumerate spanning tree

### 3.4.4. Algorithm listing using the data structures:

We shall re-write Algorithm 4 to Algorithm 7 using the 3 data structures introduced for more precise description of the process and better understanding of the operations performed. In Algorithm 13, the 4 possible scenarios for generating spanning trees are explained using the data structures. Moreover, the subroutines for generating type 1 and type 2 children are merged for each type of parent.

```
ENUMERATECHILDRENOFTYPE1(T)
  MAINTAINSPANNINGTREEATTRIBUTES(pilot(T))
  MAINTAINPROMOTABLEPAIRS(pilot(T))
  MAINTAINPILOTCANDIDATES(pilot(T))
  For each p in promotable-pairs until next-to-pilot
  ▷ type-1-from-1
      If degree(node(p) in T)) > 1
          T_C ← Promote p in T
          If pilot(T_C) = p
              S(G) ← S(G) ∪ T_C
              ENUMERATECHILDRENOFTYPE1(T_C)
  ▷ type-2-from-1
      Else
          T_C ← Promote p on T
          S(G) ← S(G) ∪ T_C
          ENUMERATECHILDRENOFTYPE2(T_C)

ENUMERATECHILDRENOFTYPE2(T)
  MAINTAINSPANNINGTREEATTRIBUTES (pilot(T))
  MAINTAINPROMOTABLEPAIRS(pilot(T))
  MAINTAINPILOTCANDIDATES(pilot(T))
  v_T ← node(pilot(T)
  ▷ type-1-from-2
  If next-to-pilot is non-leaf
      For each v in adjacency-list(v_T)
          T_C ← Promote pair(v) on T
          If promotion of pair(v) covers v_T and pilot(T_C) = pair(v)
              S(G) ← S(G) ∪ T_C
              ENUMERATECHILDRENOFTYPE1(T_C)
  ▷ type-2-from-2a
```

**For each** *p* **in** *promotable-pairs* **until** *pilot*

$T_C$ ← Promote *v* on *T*

$S(G) \leftarrow S(G) \cup T_C$

ENUMERATECHILDRENOFTYPE2($T_C$)

▷ *type-2-from-2b*

**For each** *v* **in** *adjacency-list*($v_T$) **between** $v_T$ **and** node(*next-to-pilot*)

**If** *degree*(*v* in *T*) = 1 **and** promotion of *pair*(*v*) covers $v_T$ **and** *pilot*($T_C$) = *pair*(*v*)

$T_C$ ← Promote *pair*(*v*) on *T*

$S(G) \leftarrow S(G) \cup T_C$

ENUMERATECHILDRENOFTYPE2($T_C$)

Algorithm 13: Enumeration of children spanning trees employing data structures

Chapter 4

# Complexity Analysis

After providing a precise description of MP algorithm, in the previous chapter, we shall proceed with analyzing the performance of the algorithm. We start by selecting an implementation for required data structures that ultimately supports the operations executed within MP algorithm.

## 4.1 Implementation of data structures:

In order to select a suitable implementation for each data structure used, it is essential to have an inclusive list of all operations performed in the data structure. The implementation selected should support these operations with least computational and space complexity. We shall survey the operations on each data structure and the complexity of each operation with a suitable implementation.

### 4.1.1. Data structure 1: Augmented spanning tree:

This data structure holds the current spanning tree at each step of the enumeration process. The data is, obviously, initialized by the *minimal-ranks-tree* with the extra attributes: degree in tree, length of path to node zero and count of non-minimal descendants. This has no extra cost than generating the root spanning tree itself:

1- The degree of nodes in tree can be directly calculated while adding the $V - 1$ edges of the tree and incrementing the degree of the nodes at each end.

2- The initial distance of pairs from node zero can be obtained directly during the breadth-first-search.

3- Count of non-minimal descendants is zero for all pairs as all pairs are minimal.

After being initialized, no items are ever deleted from or added to this structure.

The maintenance tasks required for this data structure (Algorithm 8) involves direct access to pairs in the structure, access to ancestors and access to subtree of a pair. These types of access are required by several tasks in the algorithm and have to be supported by the spanning tree.

This data structure is used while generating children of the spanning tree (Algorithm 13) for direct access of promoted, covered and uncovered nodes.

The tree is stored as an array of member pairs to support direct access in constant time. The structure of the tree is added as a pointer to parent pair at each element in the structure. This pointer to parent is sufficient to preserve the tree structure but does not facilitate access of descendants or subtree of a pair. An adjacency list for the spanning tree should be implemented to support this requirement.

Table 1 specifies the operations performed on the spanning tree and the complexity of each operation under the described implementation.

| | |
|---|---|
| Initialization of attributes in *minimal-ranks-tree* | *No extra cost* |
| Accessing element by index | $O(1)$ |
| Accessing ancestors | $O(log\ V)$ |
| Accessing all elements of subtree | $O(V)$ |
| Check pair is minimal/maximal/leaf | $O(1)$ |
| Get count of non-minimal descendants | $O(1)$ |
| Check if element in subtree of element | $O(log\ V)$ |

Table 1: Complexity analysis of operations on the augmented-spanning-tree structure

### 4.1.2. Data structure 2: *Promotable-pairs:*

This data structure is initialized by adding all pairs in the *minimal-ranks-tree* that comply with the properties of members in the structure. The test of compliance has two parts: checking the existence of edge for promotion which is constant time action and checking that this available edge actually reconnects the tree (as per Lemma 13) which requires $O(log\ V)$ time for a maximum of $E$ possible promotion (Algorithm 9). This makes the initialization of the structure an $O(E\ log\ V)$ time cost.

While maintaining the structure, due to a promotion of a pair (Algorithm 10), the major amount of operations performed on the structure is addition and removal of pairs as their compliance with member properties change. Some pairs are also relocated as their order in the structure alters due to depth or rank changes. For simplicity, we can view relocation as a removal plus addition operations.

During children enumeration, the structure is traversed sequentially (Algorithm 13) till the lowest possible promotable pair to generate possible children from the current spanning tree using available promotions in the structure.

An appropriate implementation for *promotable-pairs* is the binary search tree. The binary tree should be ensured to be balanced. This will make all modifications in the structure during maintenance of cost $O(log\ V)$. The traversal of items in the structure will be of order $O(log\ V + k)$ where $k$ is the number of visited items. We shall see that these traversals can be charged to child trees as constant time operations.

Table 2 specifies the operations performed on *promotable-pairs* and the complexity of each operation under the binary-tree implementation.

| | |
|---|---|
| Test all pairs in *minimal-ranks-tree* to be added | $O(E\ log\ V)$ |
| Adding element to structure | $O(log\ V)$ |
| Removing element from structure | $O(log\ V)$ |
| Relocating element in structure | $O(log\ V)$ |
| Accessing an element | $O(log\ V)$ |
| Checking existence of an element | $O(log\ V)$ |
| Accessing a range of $k$ elements | $O(log\ V + k)$ |

Table 2: Complexity analysis of operations on *promotable-pairs* structure

### 4.1.3. Data structure 3: *Pilot-candidates:*

Initialization of this data structure (Algorithm 11) is obviously constant time.

The maintenance of this data structure is linked to the pair promotion in two ways. The promoted pair should always be inserted in the head of this data structure or relocated to the head if already in the data structure. The other link is the change in ordering of elements based on the standings for pilot appointment. This is changed for pairs in the subtree of the promoted pair. Fortunately, these pairs are all minimal pairs, by Property 3.4, and do not belong to the data structure. Additionally, covered or uncovered nodes may have depth change due to change of state from or to leaf node. Accordingly, the only maintenance changes that may occur in this data structure is the insertion, or relocation, of pilot pair to the head of the list and the relocation of covered and uncovered pairs.

The structure is used (Algorithm 13), during children enumeration, to read the first and second elements to be used as pilot and next-to-pilot.

Implementing *pilot-candidates* data structure as a list with special pointers from the spanning tree pairs is sufficient and will give constant time complexity to operations required to access and modify the head of the list. By using the other data structure, *promotable-pairs*, as sorting index, we can relocate the covered and uncovered nodes in $O(log\ V)$ time.

Table 3 specifies the operations performed on *pilot-candidates* and the complexity of each operation.

| | |
|---|---|
| Initialization | $O(1)$ |
| Adding or relocating pilot to head of structure | $O(1)$ |
| Relocating covered and uncovered pairs in structure | $O(log\ V)$ |
| Accessing 1st and 2nd element in structure | $O(1)$ |

Table 3: Complexity analysis of operations on *pilot-candidates* structure

## 4.2 Time complexity of MP algorithm:

We shall proceed with analyzing the time complexity of the algorithm by examining the cost of each task based on the implementation selected for the data structures.

### 4.2.1. Initialization:

The initialization phase of the algorithm consists of generating the root spanning tree (Algorithm 3) including the augmented attributes and initializing data structures used (Algorithm 9 and Algorithm 11). The operations involved in generating the root spanning tree are all linear in $V$ and $E$ except for sorting of adjacency lists of all nodes. We shall just go with any comparison sort that will provide $O(\delta\ log\ \delta)$ for each list where $\delta$ is the degree of the node. Summation over all nodes and as $\delta < V$, we will have $O(\sum\delta\ log\ V)$ = $O(E\ log\ V)$.

Initializing the data structures, *promotable-pairs* and *pilot-candidates*, is of cost $O(E\ log\ V)$ and $O(1)$ respectively as explained in the previous section. We do not need to consider the $O(V)$ cost as the graphs are always connected and $V - 1 \leq E$. This gives the whole initialization a complexity of $O(E\ log\ V)$.

Table 4 provides details of time complexity of each of the initialization steps.

| | |
|---|---|
| GENERATEMINIMUMRANKSSPANNINGTREE($G(V,E)$) | $O(E\ log\ V)$ |
| INITIALIZE-PROMOTABLE-PAIRS() | $O(E\ log\ V)$ |
| INITIALIZE-PILOT-CANDIDATES() | $O(1)$ |

Table 4: Complexity analysis of initialization phase of algorithm

From Table 4 and the accompanying argument we can deduce the result in Lemma 16.

Lemma 16:

Initialization of MP algorithm has total time complexity of $O(E\ log\ V)$. The initialization consists of generation of *minimal-ranks-tree* and initialization of data structures.

### 4.2.2. Data structures maintenance:

For the maintenance of data structures, we have to calculate the complexity of the subroutines in Algorithm 8, Algorithm 10 and Algorithm 12. Unlike initialization, these subroutines are executed for every generated spanning tree. This means that the cost of these subroutines is charged to every spanning tree enumerated.

We start by the maintenance of augmented spanning tree in Algorithm 8. Updating degree of covered and uncovered pairs and calculating path length of promoted pair are constant time operations. Accessing ancestors of the promoted pair to modify count-of-non-minimal-descendants is $O(log\ V)$ operation. Calculation of path length for pairs in the subtree of pilot pair is constant time operation (by adding or subtracting the same difference calculated for pilot pair) but should be performed for every pair in the subtree. When the spanning tree is type 1 the subtree is $O(V)$ which makes the operation of cost $O(V)$. When the spanning tree is of type 2 this operation is completely eliminated. This makes the whole maintenance of the augmented tree of cost $O(V)$ for type 1 spanning trees and $O(log\ V)$ for type 2 spanning trees.

For maintenance of *promotable-pairs* in Algorithm 10, the operations of testing a pair for being maximal, minimal, leaf or non-leaf are all constant time operations available from the adjacency list of $G$ and attributes of pairs in current tree $T$. Addition, removal and relocation of a pair in the structure are all $O(log\ V)$ operations. Finding the ancestor of a pair is tracing a path on the spanning tree and is also $O(log\ V)$.

This leaves only the two loops in steps 10-11 and 24-26 in Algorithm 10. The internals of both loops are $O(log\ V)$. The first loop involves depth calculation and relocation within the structure both of order $O(log\ V)$. The second loop involves, as per Lemma 14 and Lemma 15, 3 checks of membership in subtrees of $T$. These tests can be performed in $O(log\ V)$.

Now these loops are repeated for every pair in the subtree of pilot pair. When the spanning tree is type 1 the subtree is $O(V)$ which makes both loops of cost $O(V\ log\ V)$. When the spanning tree is type 2 the subtree is the pilot node only which makes the subtree of 1 node and the whole process of $O(log\ V)$.

Maintenance of *pilot-candidates* is much simpler as in Algorithm 12. We have operations of constant complexity for testing a pair for being minimal, leaf and non-leaf. Additionally we have operations for insertion and relocation of pairs in the data structure. There are suitable implementations as shown earlier that can perform all these operations in $O(log\ V)$ time.

Table 5 shows details of time complexity of each of the maintenance tasks. The extra time complexity required by type 1 spanning trees only is shown separately.

| | For type 1 | For type 2 |
|---|---|---|
| MAINTAINSPANNINGTREEATTRIBUTES(*pilot*) | $O(V)$ | $O(log\ V)$ |
| MAINTAINPROMOTABLEPAIRS(*pilot*) | $O(V\ log\ V)$ | $O(log\ V)$ |
| MAINTAINPILOTCANDIDATES(*pilot*) | $O(log\ V)$ | $O(log\ V)$ |

**Table 5: Complexity analysis of maintenance of data structures**

From Table 5 and the accompanying argument we can deduce the result in Lemma 17.

**Lemma 17:**

Maintenance of data structures for spanning trees enumeration has time complexity of $O(V\ log\ V)$ for type 1 spanning trees and $O(log\ V)$ for type 2 spanning trees.

### 4.2.3. Enumeration *Type-1-from-1*:

*Type-1-from-1* spanning trees are generated as in Algorithm 13 by checking each pair of promotable pairs using constant time test. The whole structure *promotable-pairs* has size $\leq V$ which makes the time complexity of this task to be $O(V)$.

### 4.2.4. Enumeration *Type-2-from-1*:

Enumerating a *type-2-from-1* spanning trees is definite for every iteration in the loop in Algorithm 13. Every promotable leaf in a type 1 tree will produce a type 2 child spanning tree. Thus, we may charge the time complexity of each iteration, which is constant, to the child spanning tree.

Lemma 18:

Enumeration of type 1 spanning trees children has expected time complexity of $O(V)$.

### 4.2.5. Enumeration *Type-1-from-2*:

Enumeration of *type-1-from-2* is by the loop 24-28 in Algorithm 13. This loop consists of a test of constant time complexity for the pair proposed for promotion. Unfortunately, this loop does not always generate a child tree as the control test may fail and consequently no child tree generated. Therefore, the cost of this loop has to be charged to the parent tree. The loop is repeated for a range of nodes in adjacency-list of pilot node which makes it of expected time complexity $O(E/V)$ and worst case complexity of $O(V)$. We should also note that this type of tree is generated only for parent trees that have only 1 non-minimal leaf. This should be considered for better expected complexity analysis.

### 4.2.6. Enumeration *Type-2-from-2a*:

Similar to *type-2-from-1*, enumerating a *type-2-from-2a* spanning trees is definite for every iteration in the loop 30-33 in Algorithm 13. Every promotable leaf in a type 2 tree that has node order higher than the current leaf will produce a type 2 child spanning tree. Thus, we may charge the time complexity of each iteration, which is constant, to the child spanning tree.

### 4.2.7. Enumeration *Type-2-from-2b*:

Enumeration of *type-2-from-2b* is by the loop 35-39 in Algorithm 13. This loop consists of a test of constant time complexity for of the pair proposed for promotion. Unfortunately, this loop does not always generate a child tree as the control test may fail and consequently no child tree generated. Therefore, the cost of this loop has to be charged to the parent tree. The loop is repeated for a range of nodes in adjacency-list of pilot node which makes it of expected time complexity $O(E/V)$ and worst case complexity of $O(V)$.

Lemma 19:

Enumeration of type 2 spanning trees children has expected time complexity of $O(E/V)$ and worst case complexity of $O(V)$.

From the results obtained in Lemma 16 to Lemma 19, we can conclude the time complexity of the algorithm in the following theorem.

### 4.2.8. Theorem 3: Time complexity of the algorithm:

The algorithm of spanning trees enumeration has expected time complexity of $O(E \log V + T_1 V \log V + T_2(\log V + E/V))$ and worst time complexity of $O(E \log V + T_1 V \log V + T_2 V)$ where $T_1$ and $T_2$ are the number of generated type 1 and type 2 spanning trees.

## 4.3 Space limit of MP algorithm:

In order to compute a limit for the space complexity of the algorithm, we should note that the space required for the algorithm consists of two factors:

1- The space required for data structures used for processing each spanning tree.
2- The depth of the computational spanning tree that traverses all spanning trees of $G$, as defined by Theorem 2.

For setting a limit for both factors we state and prove the following statements.

Lemma 20:

The data structures used for processing a spanning tree are $O(V)$ space limit.

Proof:

The data structures used for processing a spanning tree are:

1- The augmented spanning tree: This consists of $V - 1$ pairs plus node zero. The pairs are augmented with constant amount of data. The sum of entries in the adjacency lists of nodes in the tree is equal to $2(V - 1)$ as the number of edges in the tree is $V - 1$. Which makes the size of this structure $O(V)$.

2- *Promotable-pairs*: These are the pairs that can be promoted. This is clearly limited by the total number of pairs $V - 1$, i.e. $O(V)$.

3- *Pilot-candidates*: Similar to *promotable-pairs* this structure has size limited by $O(V)$.

$\Rightarrow$ The space required for processing any single spanning tree is limited by $O(V)$. ■

**Lemma 21:**

The modified elements during data structures maintenance is $O(V)$ for type 1 spanning trees and $O(log\ V)$ for type 2 spanning trees.

**Proof:**

The elements modified in data structure during maintenance are:

1- Pilot, covered and uncovered pairs: $O(1)$.

2- Ancestors of pilot pair before and after the promotion: $O(log\ V)$.

3- The subtree of pilot pair: $O(V)$ for type 1 and $O(1)$ for type 2 spanning trees.

$\Rightarrow$ The elements modified are $O(V)$ for type 1 spanning trees and $O(log\ V)$ for type 2 spanning trees. ■

**Lemma 22:**

The computational directed spanning tree inscribed by the parent/child relation over the set of all spanning trees of $G$ as nodes as defined by Theorem 2 has depth $\leq 2E - V$.

**Proof:**

The depth of this spanning tree is the maximum path length from the root spanning tree, *minimal-ranks-tree*, to any other spanning tree.

Now suppose this depth is from *minimal-ranks-tree* to another tree $T'$.

$T'$ is generated from *minimal-ranks-tree* by a sequence of pair promotions.

But the maximum number of promotions that can be applied to *minimal-ranks-tree* per pair $p$ is $degree(node(p)) - 1$.

The maximum number of promotions that can be applied sequentially to all pairs of *minimal-ranks-tree* is $\sum_{node(p) \in V}\left(degree\left(node(p)\right) - 1\right) = \sum_{v \in V}(degree(v) - 1) = 2E - V$

Therefore, the depth of the computational spanning tree is $\leq 2E - V$. ■

From the above results we can limit the space required by the algorithm to be:

### 4.3.1. Theorem 4: Maximum Space limit

The maximum space required by the algorithm is $O(EV)$.

**Proof:**

At any step of the algorithm we only need to save the information from the root spanning tree to the current tree in the data and stack buffers. Using Lemma 20 and Lemma 22, the computation of all spanning trees requires space limit of $O(EV)$. ■

In practice, the space required has much lower expected limit as changes for maintaining data structures for type 2 trees are $O(log\ V)$. We shall revisit this when we establish a limit for type 1 spanning trees.

The space required to output, or possibly store, the solution is similar to other algorithms that use edge exchange technique over a computational tree graph. We shall conclude the space analysis by stating the output space limit.

### 4.3.2. Theorem 5: Output space limit

The space required by MP algorithm to output the solution is $O(V + T)$.

**Proof:**

With reference to the computational tree graph, the algorithm needs to output the root spanning tree explicitly. This is a $(V - 1)$ edges.

Subsequently, for any other spanning tree, only a reference to its parent together with the pair of exchanged edges has to be output. This is constant data per spanning tree.

⇒ The space required to output the complete solution is $O(V + T)$. ■

## 4.4 Relative counts of type 1 and type 2 spanning trees:

Our main objective from developing this novel algorithm was to provide an enumeration scheme for spanning trees, through edge exchange, that swaps edges at the margin of the spanning trees. Therefore, and for the sake of ascertaining this feature, we need to prove that the solution provided, expectedly, generates spanning trees by exchanging leaf edges (type 2 as in Definition 10) with higher magnitude than internal edges (type 1 as in Definition 9).

Additionally, in the complexity analysis, we derived a time complexity limit for MP solution (Theorem 3) that is dependent on the type of tree. In order to make MP algorithm complexity comparable to other algorithms, and to make it comprehensible by all readers as well, we need to develop a complexity limit that is type independent. In order to achieve this, we shall establish an expected ratio between the counts of type 1 spanning trees and type 2 spanning trees. We shall start by describing some used terms.

### 4.4.1. Node deletion:

**Definition 20: Node deletion**

Node deletion from a graph is the removal of a node and all edges incident to it from the graph.

We shall denote the graph obtained by deletion of node $v$ from graph $G$ by $G - \{v\}$.

**Definition 21: Cut node**

A cut node is a node the deletion of which would disconnect the graph.

**Definition 22: Pendant node**

A pendant node is a node that has only one incident edge in the graph. By this a pendant node is a leaf node in the graph. Anyhow, we shall reserve the term leaf nodes for the leaves of the generated trees and use the term pendant nodes for graph leaves.

Figure 13 demonstrates a deletion of a node from a graph $G(V = \{v_0,v_1,\ldots,v_8\},E)$. The dashed edges are the edges removed together with node $\{v_2\}$ to get $G - \{v_2\}$. Besides, it shows samples of cut and pendant nodes. Nodes $v_1$ and $v_3$ are cut nodes while $v_4$ is a pendant node.

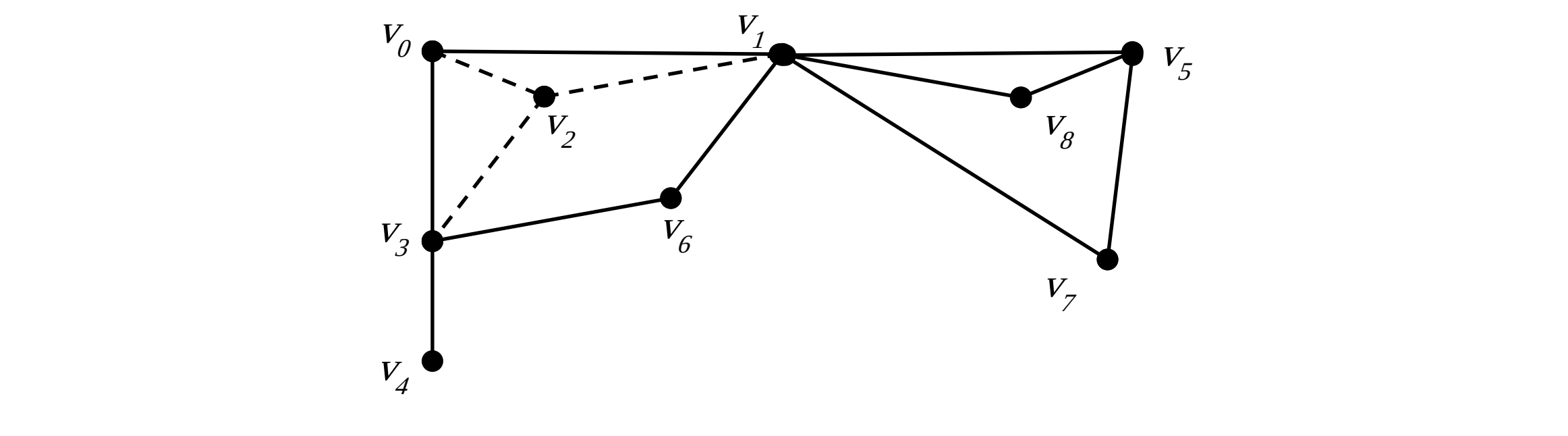

Figure 13: Node deletion, cut node, and pendant node

In our subsequent analysis, we shall use the notation $\tau(G)$ to denote the number of spanning trees of the graph $G$. We may extend the notation by a constraint $\tau(G, A)$ to denote the number of spanning trees of the graph $G$ that satisfy the condition $A$.

Lemma 23:

The count of spanning trees of $G$ where a node $v$ is internal node or minimal leaf is given by the formula:

$\tau(G, v$ is internal node or $v$ is minimal leaf$) = \tau(G) - \tau(G - \{v\}) * (degree(v) - 1)$.

Proof:

Initially, we shall examine the case when $v$ is a leaf of a spanning tree.

A spanning tree where node $v$ is leaf is a spanning tree of $G - \{v\} \cup 1$ edge incident to $v$. As shown in Figure 14, each spanning tree of $G - \{v\}$ can be extended to include node $v$ by any edge incident to $v$.

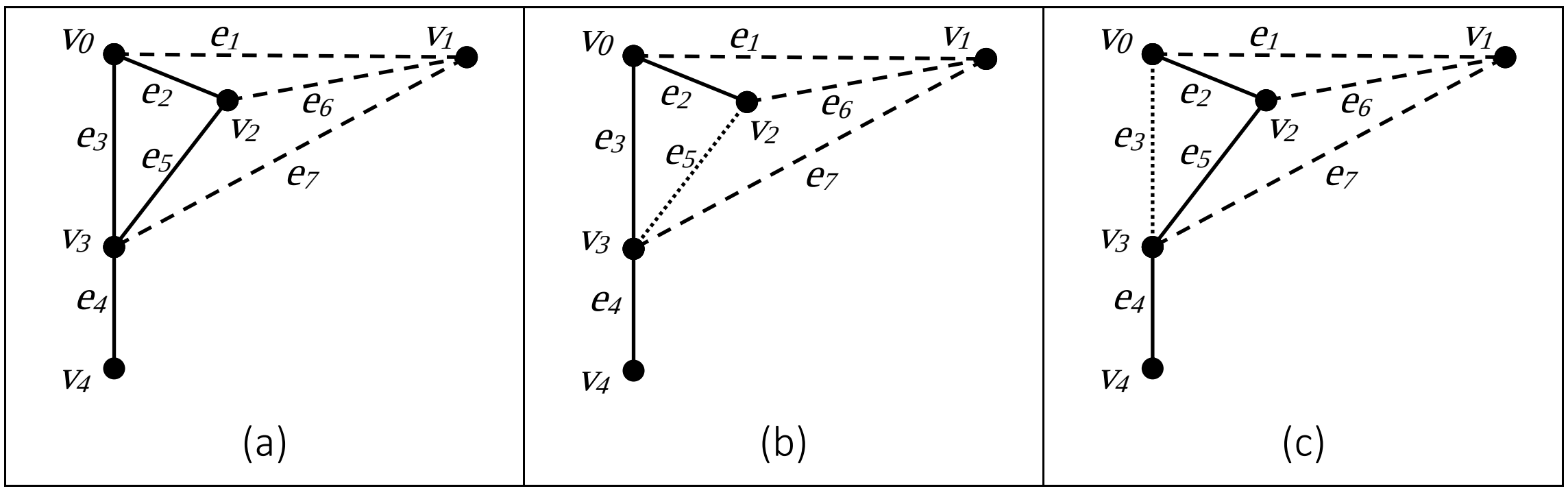

Figure 14: Spanning tree with specific leaf

Therefore, $\tau(G, v \text{ is leaf}) = \tau(G - \{v\}) * degree(v)$ **(Eq.4)**

Out of the edges incident to $v$, using the lowest order edge to connect $v$ will make it a minimal leaf while all other edges, if exist, will make $v$ a non-minimal leaf of the resultant spanning tree.

Therefore, $\tau(G, v \text{ is minimal leaf}) = \tau(G - \{v\})$ and, **(Eq.5)**

$\tau(G, v \text{ is non-minimal leaf}) = \tau(G, v \text{ is leaf}) - \tau(G, v \text{ is minimal leaf})$

From Equation 4 and 5:= $\tau(G - \{v\}) * (degree(v) - 1)$ **(Eq.6)**

Now, $\tau(G) = \tau(G, v \text{ is internal node}) + \tau(G, v \text{ is minimal leaf}) + \tau(G, v \text{ is non-minimal leaf})$

$\Rightarrow \tau(G, v \text{ is internal node or } v \text{ is minimal leaf}) = \tau(G, v \text{ is internal node}) + \tau(G, v \text{ is minimal leaf})$

$= \tau(G) - \tau(G, v \text{ is non-minimal leaf})$

$= \tau(G) - \tau(G - \{v\}) * (degree(v) - 1)$ ■

**Lemma 24:**

For a cut node or pendant node $v$, $\tau(G, v \text{ is internal node or } v \text{ is minimal leaf}) = \tau(G)$.

**Proof:**

By Lemma 23: $\tau(G, v \text{ is internal node or } v \text{ is minimal leaf}) = \tau(G) - \tau(G - \{v\}) * (degree(v) - 1)$

Now by Definition 21 for cut node, $G - \{v\}$ is disconnected $\Rightarrow \tau(G - \{v\}) = 0$.

And by Definition 22 for pendant node, $degree(v) = 1$.

$\Rightarrow$ For both cut nodes and pendant nodes: $\tau(G, v \text{ is internal node or } v \text{ is minimal leaf}) = \tau(G)$. ■

**Lemma 25:**

The expected count of spanning trees in a graph $G(V,E) = \frac{1}{V}\left(\frac{2E}{V-1}\right)^{V-1}$.

**Proof:**

For this argument we shall use the Erdős–Rényi random graph notation [51]. In this notation, a graph is denoted as $G(n,p)$ where $n$ is the number of nodes and $p$ is the probability that any two nodes are connected.

By Cayley formula [52] and its proof by Prüfer [53], we know that the total number of spanning trees for a complete $n$-nodes graph is $n^{n-2}$.

The probability that $n-1$ specific edges in the complete graph are all edges in $G(n,p)$ is $p^{n-1}$.

$\Rightarrow$ The expected number of spanning trees for the random graph is $G(n,p) = p^{n-1}n^{n-2}$.**(Eq.7)**

Now the probability $p$ = Actual edges / all possible edges = $\frac{2E}{V(V-1)}$ **(Eq.8)**

$\Rightarrow$ The expected number of spanning trees for the random graph $G(V,E)$ = $\frac{1}{V}\left(\frac{2E}{V-1}\right)^{V-1}$. ■

In the subsequent analysis we shall use Equation 7 derived above to derive an expectation of type 1 spanning trees. The final expectation derived by Lemma 25 will be used to verify the experimental results.

**Lemma 26:**

The expectation of spanning trees in a graph $G(V,E)$ where a random node $v$ is internal node or minimal leaf = 0.63 + 0.18 $V/E$.

**Proof:**

The required expectation = $\tau(G$, $v$ is internal node or $v$ is minimal leaf$)$ / $\tau(G)$

By Lemma 23: $= 1 - \tau(G - \{v\})*(degree(v) - 1) / \tau(G)$

Applying Equation 7: $= 1 - p^{n-2}(n-1)^{n-3}(degree(v) - 1) / p^{n-1}n^{n-2}$

Note that the same value for $p$, the expectation of existence of an edge between two nodes, should be used for both $G$ and $G - \{v\}$.

$$= 1 - (1 - {}^{1}/_{n})^{n-3}(degree(v) - 1) / pn$$

Now, let $p = \frac{2E}{V(V-1)}$ (by Equation 8), expectation of $degree(v) = \frac{2E}{V}$ and revert $n$ back to $V$.

$= 1 - (1 - \frac{1}{V})^{V-2}(1 - \frac{V}{2E})$

For sufficiently large $V$, $(1 - \frac{1}{V})^{V-2} \approx 1/e$, where $e$ is the base of the natural logarithm function [54].

$= 1 - \frac{1}{e} + \frac{V}{2eE}$

$\approx 0.63 + 0.18\ V/E$ ■

Note that from Lemma 26 and for random large graphs, the expectation of a node being internal node or minimal leaf in a spanning tree is bounded between 0.63 (for large dense graphs) and 0.82 (for sparse graphs).

**Lemma 27:**

For a random graph $G(V,E)$ the expectation of type 1 spanning trees is $(0.63 + 0.18\ V/E)^L$, where $L$ is the count of nodes that are not pendant or cut nodes.

**Proof:**

The expectation of type 1 spanning trees is the expectation that all nodes in the graph are internal nodes or minimal leaves. This can be calculated using joint probability equation, assuming that all the sub-probabilities are independent.

Expectation of type 1 spanning trees $= \prod_{v \in V}(v \text{ is internal node or minimal leaf})$

$= \prod_{over\ all\ V}(0.63 + 0.18\, V/E)$

But, by Lemma 24, cut nodes and pendant nodes should not be considered in the product.

$\Rightarrow$ Expectation of type 1 spanning trees = $(0.63 + 0.18\ V/E)^L$ ■

In order to give a better expectation of type 1 spanning trees, we need to find an expected count of non-cut and non-pendant nodes in a graph (denoted $L$ in Lemma 27). Many upper bounds for cut nodes were developed on bases of node degrees [55,56]. The probability that a random node be a non-cut node was also calculated for special ranges of $V$ and $E$ [57]. The expectation of a pendant node in a graph is much simpler to formulate. Anyhow, we shall derive a lower bound that is un-tight, but satisfactory, of the expectation of the value $L$ using a simple technique.

Lemma 28:

The expected number of cut nodes in a random large connected graph $G(V,E)$ is $O(\sqrt{V})$.

Proof:

Consider a random connected graph represented in Erdős–Rényi random graph notation as $G(n,p)$.

A cut node exists at a random node $v$ with pre-specified cut subsets of sizes $k$ and $n-k-1$ with probability $(1-p)^{k(n-1-k)}$ as every pair of nodes from different subsets should not have a common edge. Moreover, the cut set should have at least 1 edge to each of the partitions. This has probability $1-(1-p)^k$ and $1-(1-p)^{n-1-k}$. There are $\binom{n-1}{k}$ such possible subsets. Denote the expected number of cut nodes with partition sizes of $k$ and $n-k-1$ as $E_k$.

$$E_k = n\binom{n-1}{k}(1-p)^{k(n-k-1)}(1-(1-p)^k)(1-(1-p)^{n-1-k})$$

The expectation that $v$ is a cut node: $E_{Cut}$ =

$$E_{Cut} = \sum_{k=1}^{\lfloor n/2 \rfloor} E_k$$

$$= \sum_{k=1}^{\lfloor n/2 \rfloor} n\binom{n-1}{k}(1-p)^{k(n-k-1)}(1-(1-p)^k)(1-(1-p)^{n-1-k})$$

Consider the function:

$$f(a,b,x) = \frac{a^{x(b-x)}(1-a^x)(1-a^{b-x})}{\Gamma(x+1)\Gamma(b-x+1)}$$

$$E_k = n!\,f(1-p,n-1,k)$$

Where $\Gamma(x)$ is the Gamma function, for integers $= (x-1)!$,

We can show by positive value derivative $f'(a,b,x)$ of the summand that this function is increasing in the interval $(1, \frac{n}{2})$.

$f'(a,b,x) =$

$$-\frac{a^{2xb}\big((a^x-1)\psi(x+1)+(4a^x\ln a-4\ln a)x+\psi(b-x+1)+a^x(-\psi(b-x+1)-\ln a)\big)}{a^{2x^2}\Gamma(x+1)\Gamma(b-x+1)} -$$

$$\frac{a^{xb}\left(a^{x^2}(1-a^x)\psi(x+1)+a^{x^2}(2\ln a-2a^x\ln a)x+a^{x^2}\left(a^x(\psi(n-x+1)+\ln a)-\psi(b-x+1)\right)\right)}{a^{2x^2}\Gamma(x+1)\Gamma(b-x+1)}$$

$$=\frac{\left(1-a^{x(b-x)}\right)(1-a^x)\left(\psi(b-x+1)-\psi(x+1)\right)}{a^{-x(b-x)}\Gamma(x+1)\Gamma(b-x+1)}$$

$$+\frac{\left(a^x\left(1-a^{x(b-x)}\right)+2x\left(1-2a^{x(b-x)}\right)(1-a^x)\right)(-\ln a)}{a^{-x(b-x)}\Gamma(x+1)\Gamma(b-x+1)}$$

Where $\psi(x)$ is the Digamma function, for integers $= -\gamma + \sum_{t=1}^{x}\frac{1}{t}$,

$\gamma$ is Euler–Mascheroni constant $\approx 0.577$

Note that $0 < a < 1$ and $1 \le x \le b/2$

Let $r = \frac{pn}{\ln n}$, we will use a limit on $r$ by means of the asymptotic expectation of graph connectedness obtained by Erdős–Rényi [58]. They proved that as $n \to \infty$, $p = \frac{\ln n}{n}$ is a sharp threshold for graph connectedness. When $p = \frac{(1+\varepsilon)\ln n}{n}$ the graph is almost certainly connected and when $p = \frac{(1-\varepsilon)\ln n}{n}$ the graph is almost certainly disconnected (where ε > 0).

$\Rightarrow f'(1-p, n-1, x)$ is positive when $a^{x(b-x)} = (1-p)^{x(n-1-x)} < 1/2$

with $p = r\ln n/n$ this holds $\forall\, r > 1$ and $n \ge 4$

$\Rightarrow E_1 < E_2 < \cdots < E_{\lfloor n/2 \rfloor}$

Then:

$$E_{Cut} = \sum_{k=1}^{\lfloor n/2 \rfloor} E_k < \frac{n}{2}E_{\lfloor n/2 \rfloor} < \frac{n}{2}n!\,f(1-p, n-1, \frac{n}{2})$$

$$E_{Cut} < \frac{n\,n!\,(1-p)^{(n/2)^2}\left(1-(1-p)^{n/2}\right)^2}{2(n/2)!^2}$$

Substituting $p = \frac{r\ln n}{n}$ and using the identity $(1-\frac{x}{n})^n \le e^{-x}$

$$E_{Cut} < \frac{n!\,n^{1-rn/4}\left(1-n^{-r/2}\right)^2}{2(n/2)!^2}$$

Use Stirling's bounds [59] for $n!$: $\sqrt{2\pi}n^{n+1/2}e^{-n} \le n! \le e^{1-n}n^{n+1/2}$

$$\frac{n!}{(n/2)!^2} < \frac{\sqrt{2\pi}n^{n+1/2}e^{-n}}{\left(e^{1-n/2}\,(n/2)^{n/2+1/2}\right)^2} = \frac{\sqrt{2\pi}e^{-n}n^{n+1/2}}{e^{2-n}\,(n/2)^{n+1}} = \frac{2^{n+1}}{e^2}\sqrt{\frac{2\pi}{n}}$$

$$\Rightarrow E_{Cut} < \left(\frac{2}{n^{r/4}}\right)^n \frac{n}{e^2}\sqrt{\frac{2\pi}{n}}\left(1 - n^{-r/2}\right)^2 < \frac{\sqrt{2\pi n}}{e^2}\ \forall r > 1 \text{ and } n > 16$$

⇒ The expected number of cut vertices is $O\left(\sqrt{V}\right)$ ■

Lemma 29:

The expected number of pendant nodes in a random large connected graph $G(V,E)$ is $O(log\ V)$.

Proof:

A pendant node will occur only if a cut node with cut set 1 and $|V| - 2$ occurs (this is the only node to which the pendant node is connected).

Using the same terms and argument used in **Lemma** 28, the probability that a random node is a pendant node is:

$$E_1 = n(n-1)n^{-r(1-2/n)}\left(\frac{r\ln n}{n}\right)(1 - n^{-r(1-2/n)})$$

For large $n$:

$$E_1 < \frac{r\ln n}{n^{r-1}} \text{ where } r > 1$$

⇒ The expected number of pendant nodes in a random large connected graph is $O(log\ V)$. ■

### 4.4.2. Theorem 6: Expectation of type 1 spanning trees

For a random graph $G(V,E)$ the expectation of type 1 spanning trees is $(0.63 + 0.18\ V/E)^V$. Moreover, $O(1/V)$ can be used as a non-tight upper bound for expected number of type 1 spanning trees.

Proof:

By combining the results from **Lemma** 28 and Lemma 29, for random large graphs: $L \approx V$.

Then by **Lemma** 27, the expected value follows.

Now, for $(0.63 + 0.18\ V/E)^L < 1/V$ then:

$0.81^L < 1/V$

$\Rightarrow L > 5 \ln V$

$\Rightarrow$ With $L$ as low as $5 \ln V$ the expected count of type 1 spanning trees is $O(1/V)$. ■

This non-tight upper bound stated is sufficient for the subsequent analysis in this thesis as we shall see in the following sections.

Using Theorem 6, we can re-state the time complexity of the algorithm as follows.

### 4.4.3. Theorem 7: Time complexity of the algorithm (type-independent)

The algorithm of spanning trees enumeration has expected time complexity of $O(E \log V + (\log V + E/V)T)$ and worst time complexity of $O(E \log V + T\,V \log V)$ where $T$ is the number of spanning trees.

Proof:

Consider the expected time complexity in Theorem 3:

The expected complexity = $O(E \log V + T_1 V \log V + T_2(\log V + E/V))$

Take the term $T_1V$ and using Theorem 6 $T_1V < O(T)$.

$\Rightarrow$ The expected time complexity of the algorithm = $O(E \log V + (\log V + E/V)T)$.

For the worst time complexity, let $T_1 \gg T_2$ in Theorem 3, then:

The worst time complexity = $O(E \log V + T\ V \log V)$. ■

### 4.4.4. Theorem 8: Expected space limit

MP algorithm of spanning trees enumeration has expected space limit of $O(E \log V)$.

Proof:

Consider the limit on modified elements in data structures in Lemma 21 and the limit on the depth of computational graph in Lemma 22:

Let the number of type 1 spanning trees in the maximum depth be $t_1$.

The expected space limit = $t_1V + (2E - V - t_1)\log V$.

Using Theorem 6, then $t_1 < (2E - V)/V$:

The expected space limit < $2E - V + (2E - V)\log V$.

Note that $E \geq V - 1$ as graph is connected:

$\Rightarrow$ The expected space limit = $O(E \log V)$. ■

Chapter 5

# Experimental Results and Potential Applications

In order to demonstrate the power of MP algorithm, we show in this chapter experimental results of executing the algorithm on different graphs configurations, especially, the minimal partitioning feature. Besides, sample situations where MP algorithm outperforms other enumeration methods are examined.

## 5.1 Experimental Results:

To support our, so far, theoretical findings, we shall offer empirical results from implementation of the algorithm. The main objectives of presenting these results is to verify that:

1- The algorithm actually enumerates all the spanning trees of a given graph.

2- The algorithm enumerates the spanning trees uniquely. Spanning trees are not repeated during the enumeration process.

3- Type 2 spanning trees are enumerated with higher magnitude than type 1 spanning trees. In particular, we shall compare between the actual ratio of type 1 spanning trees and the expectation limit of type 1 spanning trees derived in Lemma 27 and Theorem 6.

We shall present two sets of experimental results:

1- Experimental results with random graphs of different node sizes and edge densities.

2- Experimental results for special known graphs in the literature.

While a complete set of experimental results is provided in Appendix A, we shall list here the results obtained from graphs of larger sizes where the values of different counts and ratios are analyzable and comparable.

### 5.1.1. Experimental results with random graphs:

For our experiments, we shall use a random graph generator that follows the Erdős–Rényi random graph generation method [51]. The graphs are generated by specifying the number of nodes and edges. The generated random graph has the same number of nodes specified while the number of edges is used to calculate the probability of including any edge independently of other edges and nodes. The final graph may have different number of edges than specified.

In Figure 15, sample graphs generated using this process are shown. Graph (a) was generated by $V$=10 and $E$=20. The generated graph has actual sizes $V$=10 and $E$=17. Graph (b) was generated by $V$=12 and $E$=20. The generated graph has actual sizes $V$=12 and $E$=23. Graph (c) was generated by $V$=13 and $E$=20. The generated graph has actual sizes $V$=13 and $E$=16.

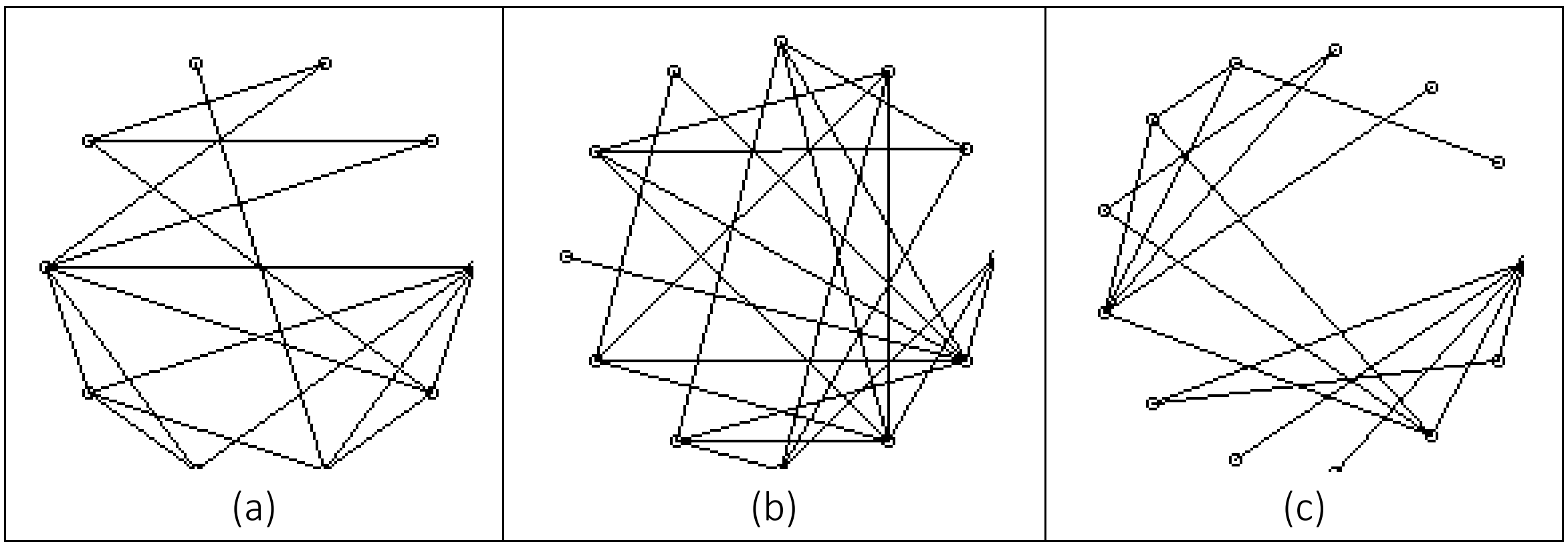

Figure 15: Samples of random graphs

In Table 6, the results of running the algorithm with randomly generated graphs, using the described process, are shown. For every graph of node size and edge size as listed, the number of spanning trees of each type 1 and type 2 are stated. Additionally the ratio of type 1 trees is calculated and compared to the expected ratio by Theorem 6.

| | $V$ | $E$ | Actual total spanning trees ($T_{Act}$) | Expected total spanning trees ($T_{Exp}$) | $\frac{T_{Exp}}{T_{Act}}$ | Type 1 spanning trees | Type 2 spanning trees | Actual type 1 % | Expected type 1 % |
|---|---|---|---|---|---|---|---|---|---|
| 1. | 10 | 32 | 3,148,933 | 4,649,831 | 1.5 | 1,471 | 3,147,462 | %0.0 | %2.4 |
| 2. | 10 | 32 | 3,188,554 | 4,649,831 | 1.5 | 1,774 | 3,186,780 | %0.1 | %2.4 |
| 3. | 11 | 29 | 1,768,527 | 3,916,402 | 2.2 | 2,205 | 1,766,322 | %0.1 | %2.0 |
| 4. | 11 | 30 | 2,644,640 | 5,496,925 | 2.1 | 871 | 2,643,769 | %0.0 | %2.0 |
| 5. | 11 | 36 | 22,330,668 | 34,035,511 | 1.5 | 15,707 | 22,314,961 | %0.1 | %1.6 |
| 6. | 11 | 37 | 33,182,983 | 44,763,549 | 1.3 | 6,984 | 33,175,999 | %0.0 | %1.6 |
| 7. | 11 | 39 | 50,763,759 | 75,779,780 | 1.5 | 12,358 | 50,751,401 | %0.0 | %1.5 |
| 8. | 11 | 40 | 69,780,818 | 97,612,893 | 1.4 | 12,614 | 69,768,204 | %0.0 | %1.5 |
| 9. | 11 | 41 | 96,431,271 | 124,952,756 | 1.3 | 15,855 | 96,415,416 | %0.0 | %1.5 |
| 10. | 11 | 43 | 158,424,609 | 201,183,254 | 1.3 | 9,179 | 158,415,430 | %0.0 | %1.4 |

| | $V$ | $E$ | Actual total spanning trees ($T_{Act}$) | Expected total spanning trees ($T_{Exp}$) | $\frac{T_{Exp}}{T_{Act}}$ | Type 1 spanning trees | Type 2 spanning trees | Actual type 1 % | Expected type 1 % |
|---|---|---|---|---|---|---|---|---|---|
| 11. | 12 | 33 | 9,415,136 | 30,233,088 | 3.2 | 6,967 | 9,408,169 | %0.1 | %1.4 |
| 12. | 12 | 34 | 13,261,628 | 41,985,300 | 3.2 | 5,380 | 13,256,248 | %0.0 | %1.3 |
| 13. | 12 | 36 | 29,655,725 | 78,732,978 | 2.7 | 16,235 | 29,639,490 | %0.1 | %1.2 |
| 14. | 12 | 37 | 54,052,984 | 106,426,097 | 2.0 | 12,887 | 54,040,097 | %0.0 | %1.2 |
| 15. | 12 | 39 | 90,370,804 | 189,906,507 | 2.1 | 26,363 | 90,344,441 | %0.0 | %1.1 |
| 16. | 12 | 40 | 136,717,386 | 250,893,306 | 1.8 | 55,554 | 136,661,832 | %0.0 | %1.1 |
| 17. | 12 | 41 | 206,082,150 | 329,193,759 | 1.6 | 58,662 | 206,023,488 | %0.0 | %1.1 |
| 18. | 12 | 46 | 689,866,136 | 1,167,253,858 | 1.7 | 47,514 | 689,818,622 | %0.0 | %1.0 |
| 19. | 13 | 31 | 3,445,488 | 27,834,407 | 8.1 | 8,814 | 3,436,674 | %0.3 | %1.1 |
| 20. | 13 | 32 | 9,833,573 | 40,741,910 | 4.1 | 10,361 | 9,823,212 | %0.1 | %1.1 |
| 21. | 13 | 32 | 13,835,055 | 40,741,910 | 2.9 | 11,435 | 13,823,620 | %0.1 | %1.1 |
| 22. | 13 | 40 | 218,147,166 | 592,872,818 | 2.7 | 98,323 | 218,048,843 | %0.0 | %0.8 |
| 23. | 13 | 41 | 297,670,444 | 797,348,027 | 2.7 | 54,495 | 297,615,949 | %0.0 | %0.8 |
| 24. | 13 | 46 | 1,649,875,393 | 3,172,017,855 | 1.9 | 118,020 | 1,649,757,373 | %0.0 | %0.7 |
| 25. | 13 | 46 | 1,924,233,332 | 3,172,017,855 | 1.6 | 217,036 | 1,924,016,296 | %0.0 | %0.7 |

Table 6: Experimental results with random graphs

In Table 6, the results are shown for different random node sizes and edge densities. Results where actual spanning trees were less than 1 million were dropped. The expectation of spanning trees total is derived by Lemma 25. The expected percentage of type 1 spanning trees is derived by Theorem 6. The results show how close our derived estimations for total spanning trees and expectation of type 1 spanning trees are. It suggests that the derived values in Lemma 25 and Theorem 6 are, as stated, upper bound expectations rather than average expectations. This is more desirable, as we need to find an upper limit for type 1 spanning trees rather than an average value. The data collected in Table 6 for actual and expected percentage of type 1 spanning trees is represented in the chart in Figure 16.

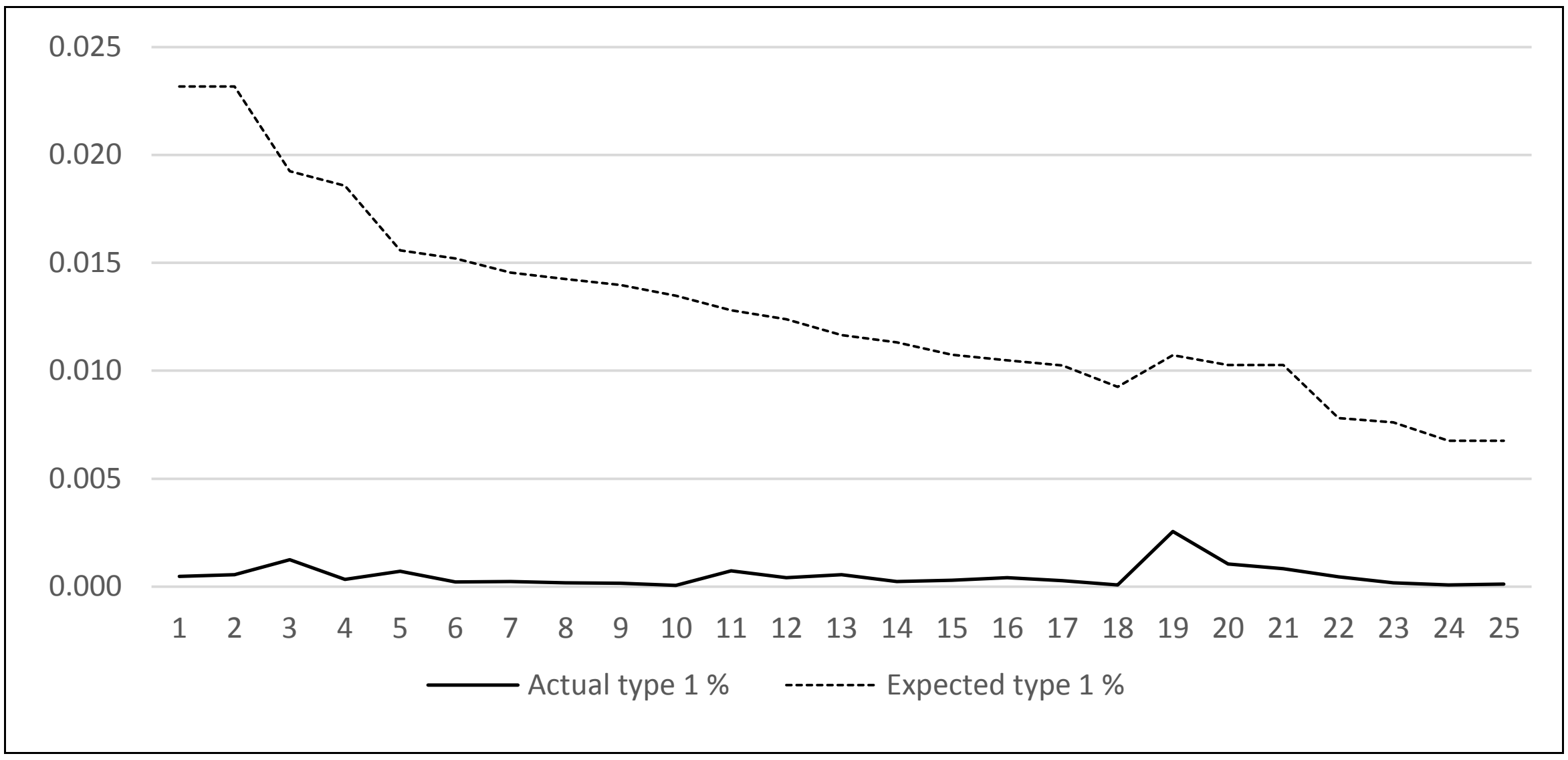

Figure 16: Actual and expected percentage of type 1 spanning trees in random graphs

The total counts, and percentages, of type 1 and type 2 spanning trees confirm the key objective of the algorithm: type 2 spanning trees are generated with much higher magnitude than type 1 spanning trees.

### 5.1.2. Experimental results with special graphs:

In Table 7, the results of running the algorithm for special graphs [60] that have distinct configurations are shown. The total number of spanning trees can be easily verified against the published number of spanning trees for these graphs. Additionally, the counts of each type of spanning trees is stated and the ratio of type 1 spanning trees is calculated and compared to the expected ratio of **Lemma** 27. Only graphs with spanning trees between 1 and 100 million were experimented.

| Graph Name | $V$ | $E$ | Total spanning trees | Type 1 spanning trees | Type 2 spanning trees | Actual type 1 % | Expected type 1 % |
|---|---|---|---|---|---|---|---|
| Antiprism-7 | 14 | 28 | 1,989,806 | 4,330 | 1,985,476 | %0.2 | %1.1 |
| Antiprism-8 | 16 | 32 | 15,586,704 | 18,335 | 15,568,369 | %0.1 | %0.6 |
| Barbell-6 | 12 | 31 | 1,679,616 | 47 | 1,679,569 | %0.0 | %1.5 |
| Cocktailparty-5 | 10 | 40 | 32,768,000 | 3,691 | 32,764,309 | %0.0 | %2.1 |
| Complete-9 | 9 | 36 | 4,782,969 | 1 | 4,782,968 | %0.0 | %3.0 |
| Crossedprism-10 | 20 | 30 | 3,932,160 | 8,920 | 3,923,240 | %0.2 | %0.4 |
| Crossedprism-12 | 24 | 36 | 75,497,472 | 72,004 | 75,425,468 | %0.1 | %0.1 |
| Crown-6 | 12 | 30 | 6,635,520 | 6,608 | 6,628,912 | %0.1 | %1.5 |
| Cubeconnectedcycle-3 | 24 | 36 | 32,400,000 | 46,453 | 32,353,547 | %0.1 | %0.1 |

| Graph Name | $V$ | $E$ | Total spanning trees | Type 1 spanning trees | Type 2 spanning trees | Actual type 1 % | Expected type 1 % |
|---|---|---|---|---|---|---|---|
| Cyclotomic-13 | 13 | 26 | 1,373,125 | 2,735 | 1,370,390 | %0.2 | %1.5 |
| Gear-11 | 23 | 33 | 1,956,242 | 2,036 | 1,954,206 | %0.1 | %0.2 |
| Gear-12 | 25 | 36 | 7,300,800 | 4,083 | 7,296,717 | %0.1 | %0.1 |
| Gear-13 | 27 | 39 | 27,246,962 | 8,178 | 27,238,784 | %0.0 | %0.1 |
| Hadamard-4 | 16 | 32 | 42,467,328 | 32,874 | 42,434,454 | %0.1 | %0.6 |
| Hanoi-3 | 27 | 39 | 20,503,125 | 24,035 | 20,479,090 | %0.1 | %0.1 |
| Helm-15 | 31 | 45 | 1,860,496 | 1,860,286 | 210 | %100.0 | %75.9 |
| Helm-16 | 33 | 48 | 4,870,845 | 4,870,605 | 240 | %100.0 | %75.9 |
| Helm-17 | 35 | 51 | 12,752,041 | 12,751,769 | 272 | %100.0 | %75.8 |
| Helm-18 | 37 | 54 | 33,385,280 | 33,384,974 | 306 | %100.0 | %75.8 |
| Helm-19 | 39 | 57 | 87,403,801 | 87,403,459 | 342 | %100.0 | %75.8 |
| Hypercube-4 | 16 | 32 | 42,467,328 | 32,874 | 42,434,454 | %0.1 | %0.6 |
| Ladder-12 | 24 | 34 | 2,107,560 | 8,119 | 2,099,441 | %0.4 | %0.1 |
| Ladder-13 | 26 | 37 | 7,865,521 | 19,601 | 7,845,920 | %0.2 | %0.1 |
| Ladder-14 | 28 | 40 | 29,354,524 | 47,321 | 29,307,203 | %0.2 | %0.0 |
| Moebiusladder-10 | 20 | 30 | 2,620,880 | 5,419 | 2,615,461 | %0.2 | %0.4 |
| Moebiusladder-11 | 22 | 33 | 10,759,353 | 13,433 | 10,745,920 | %0.1 | %0.2 |
| Moebiusladder-12 | 24 | 36 | 43,804,824 | 34,932 | 43,769,892 | %0.1 | %0.1 |
| Prism-10 | 20 | 30 | 2,620,860 | 4,722 | 2,616,138 | %0.2 | %0.4 |
| Prism-11 | 22 | 33 | 10,759,331 | 12,300 | 10,747,031 | %0.1 | %0.2 |
| Prism-12 | 24 | 36 | 43,804,800 | 30,270 | 43,774,530 | %0.1 | %0.1 |
| Sun-7 | 14 | 35 | 5,329,646 | 37,486 | 5,292,160 | %0.7 | %0.8 |
| Triangular-5 | 10 | 30 | 2,048,000 | 1,688 | 2,046,312 | %0.1 | %2.6 |
| Web-10 | 30 | 40 | 2,620,860 | 308,436 | 2,312,424 | %11.8 | %0.0 |
| Web-11 | 33 | 44 | 10,759,331 | 1,088,368 | 9,670,963 | %10.1 | %0.0 |
| Web-12 | 36 | 48 | 43,804,800 | 3,809,999 | 39,994,801 | %8.7 | %0.0 |
| Wheel-16 | 16 | 30 | 1,860,496 | 37 | 1,860,459 | %0.0 | %0.7 |
| Wheel-17 | 17 | 32 | 4,870,845 | 40 | 4,870,805 | %0.0 | %0.5 |
| Wheel-18 | 18 | 34 | 12,752,041 | 43 | 12,751,998 | %0.0 | %0.3 |
| Wheel-19 | 19 | 36 | 33,385,280 | 46 | 33,385,234 | %0.0 | %0.2 |
| Wheel-20 | 20 | 38 | 87,403,801 | 49 | 87,403,752 | %0.0 | %0.2 |

Table 7: Experimental results for special graphs

The data collected in Table 7 for actual and expected percentage of type 1 spanning trees is represented in the chart in Figure 17.

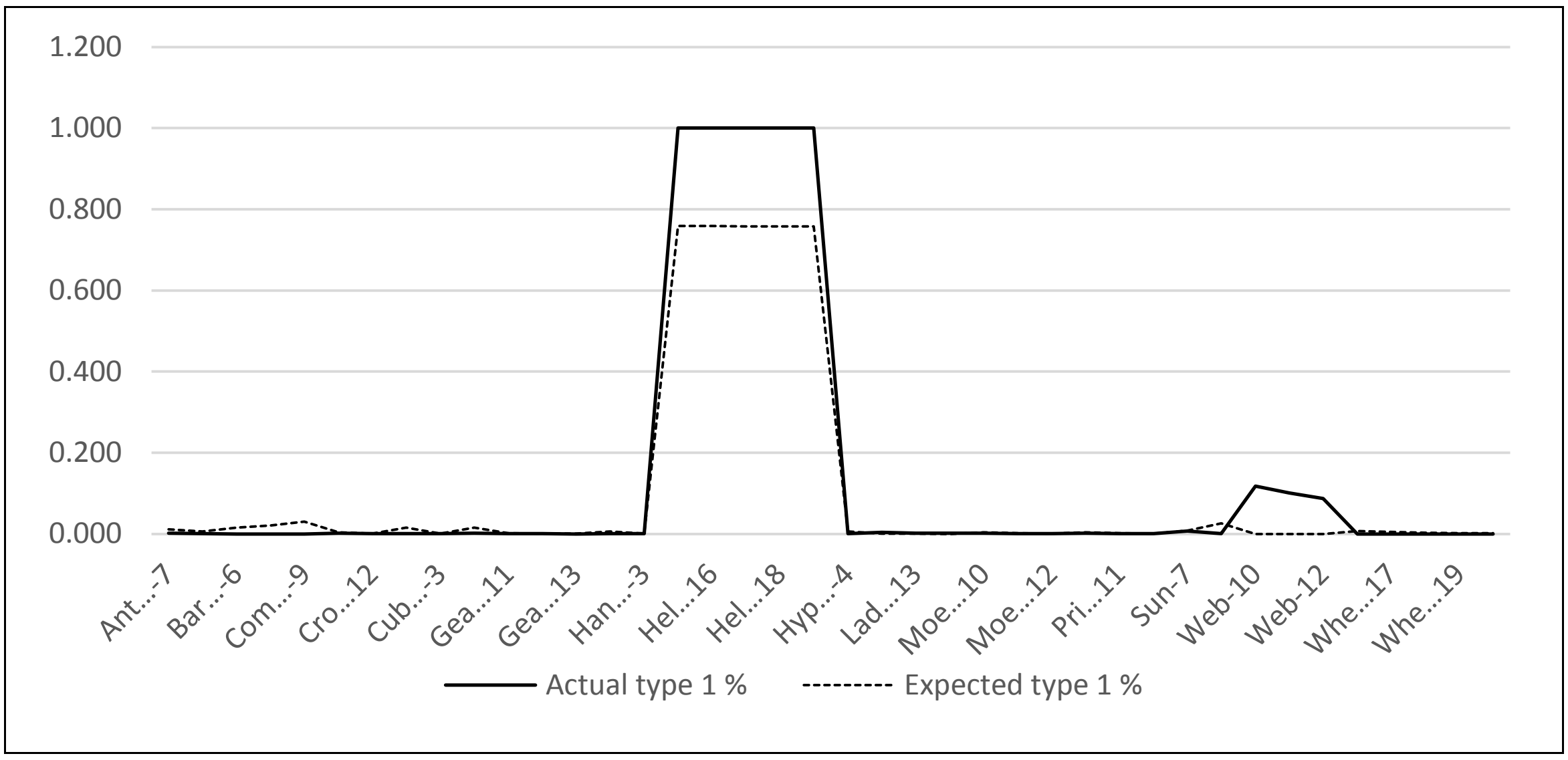

Figure 17: Actual and expected percentage of type 1 spanning trees in special graphs

From the results we obtained in Table 7, we certified that the algorithm actually generates the same number of spanning trees as expected which is confirmed by the number published [60]. This verifies that the implementation enumerates, uniquely, all the spanning trees of the given graphs. Additionally, the ratio of type 1 spanning trees remains within expectation for all the graphs except for some special graphs that have a structure consisting of high number of cut nodes and pendant nodes, namely, Web and Helm groups of graphs, shown in Figure 18. The calculated values of type 1 spanning tree ratio for these graphs is clearly deviating from the expected ratio as the expectation is drawn for random graphs.

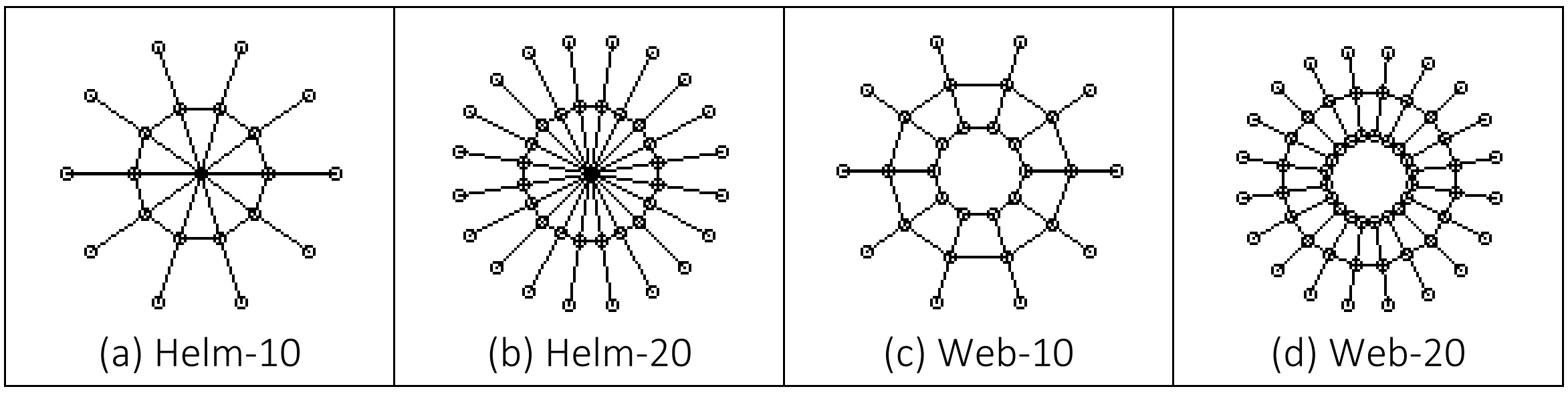

Figure 18: Layout of worst case special graphs

## 5.2 Potential applications of MP algorithm:

Real world problems, modelled using spanning trees, usually have multiple objectives that have to be optimized simultaneously. These optimizations, in many cases, may exhibit Pareto-optimal behaviors [61,62] that make it more difficult to find an optimum configuration. Consider, for instance, a communication network. The optimization objectives may include minimizing the hardware cost, maximizing the network capacity and minimizing the network delay. Furthermore, a single objective may have different, and not very consistent, interpretations when it is mapped to a spanning tree property. For example, minimizing network delay can be interpreted as minimizing average path length or minimizing longest path length.

In view of this situation, spanning trees have many properties that are desired to be optimized in different problems. For many of these properties, an optimum solution cannot be obtained in polynomial time and requires a sacrifice of correctness and adoption of an approximate solution. Alternatively, enumeration may be used to ensure the correctness of the solution in situations where an accurate solution is required. In this case, a costly, but unavoidable, approach would be to check all possible configurations of spanning trees for the given model. All spanning trees of the graph are generated and tested against the desired properties. Subsequently, a spanning tree that best fits the requirements is selected. This extensive enumeration process is, in general, a costly operation that is used as a last resort when approximation is not sufficient due to the nature of the problem [63].

Now, imagine we model our real-life problem using spanning trees and want to optimize our spanning tree for one or more measurements. In this case, a great advantage of the enumeration algorithm we adopt will be to facilitate the re-calculation of these measurements of interest while every new tree is generated. In the following sections, we discuss the benefits of using MP algorithm over other enumeration algorithms in some common applications of spanning trees.

### 5.2.1. Spanning tree partition applications:

As we are exploring the fields where MP algorithm surpasses other spanning trees enumeration algorithms, it is very appropriate to link beginnings with ends by revisiting our motivating application, construction of polyhedral nets (Section 1.4), and see how MP algorithm performs on this task.

We have demonstrated in Figure 1 that the computational cost is proportional to the size of the smaller partition of the spanning tree as we need to transform all the facets in that partition and realign them with the larger partition. In this section, we shall provide asymptotic comparison between MP algorithm and random partition size algorithms. Then, we shall present experimental results of using MP algorithm on standard polyhedra and report average partition sizes.

Lemma 30:

The expected size of partition with random partitioning edge exchange algorithm is $O(V)$.

Proof:

Let the partitioned parts of the tree be of sizes $s$ and $V - s$ where $s \in \{1,2,\ldots,V-1\}$ with equal expectation $\frac{1}{V-1}$ for all values.

The expected size of the smaller partition

$$= \frac{\sum_{1 \le s \le V-1} \min(s, V-s)}{V-1} = \frac{2 \times \sum_{1 \le s \le V/2} s}{V-1}$$

$$= \frac{2\,(V/2)(1+V/2)/2}{V-1} > \frac{V}{4} = O(V)$$

■

### 5.2.2. Theorem 9: Expected partition size

The expected size of partition by edge exchange with MP algorithm is $O(1)$.

Proof:

Let the partitioned parts of the tree be of sizes $s$ and $V - s$ where $s \in \{1,2,\ldots,V/2\}$ is the smaller part.

Let number of spanning trees generated by partition of size s be $T_s$ Where the total spanning trees $T = \sum_{1 \le s \le V/2} T_s$

The expected size of partition is:

$$= \frac{\sum_{1 \le s \le V/2} sT_s}{T} = \frac{T_1 \times 1 + \sum_{2 \le s \le V/2} sT_s}{T}$$

$$< \frac{T_1 + \frac{V}{2} \sum_{2 \le s \le V/2} T_s}{T}$$

From Theorem 6, $\sum_{2 \le s \le V/2} T_s < T/V$:

$$< \frac{T + \frac{V}{2} \times \frac{T}{V}}{T} = 1.5$$

$\Rightarrow$ The expected size of partition is $O(1)$. ∎

## Experimental results of partition size:

Here we shall examine generation of nets of polyhedra using MP algorithm to enumerate spanning trees. In this application, it is of special interest to minimize the average size of partition. The polyhedra used for experimenting are standard polyhedra [64,65]. The polyhedral facets are considered labelled, i.e. symmetry is ignored. The facets adjacency graphs of 4 of the experimented polyhedra are shown in Figure 19.

| | | | |
|---|---|---|---|
| 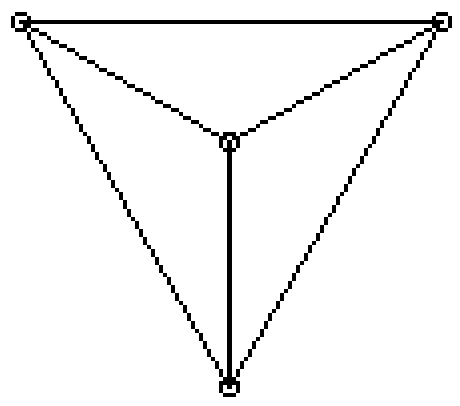 (a) Pyramid | 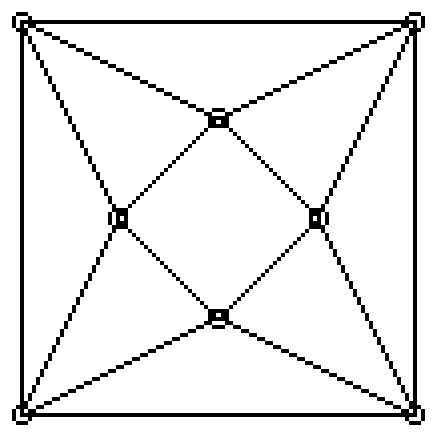 (b) Anti-prism 4-gon base | 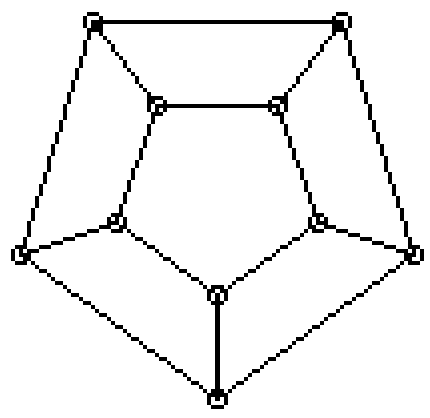 (c) Prism 5-gon base | 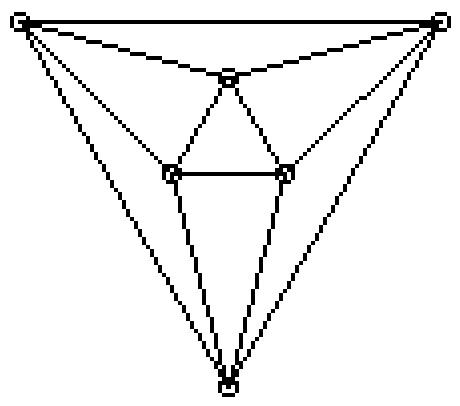 (d) Cube |

**Figure 19: Facet adjacency graph of sample polyhedra**

The results from the experiment is presented in Table 8. The average size of the partition is stated as 1+ϵ to emphasize the extra partition size due to type 1 spanning trees (Definition 9) as type 2 spanning trees (Definition 10) always have size 1 partition.

| Polyhedron | Nodes (facets) | Edges | Total spanning trees | Type 1 spanning trees | Type 2 spanning trees | Average partition size |
|---|---|---|---|---|---|---|
| Cube | 6 | 12 | 384 | 5 | 379 | 1+1.0E-02 |
| Octahedron | 8 | 12 | 384 | 13 | 371 | 1+3.4E-02 |
| Icosahedron | 20 | 30 | 5,184,000 | 8,808 | 5,175,192 | 1+2.1E-03 |
| Regular dodecahedron | 12 | 30 | 5,184,000 | 6,166 | 5,177,834 | 1+1.5E-03 |
| Rhombic dodecahedron | 12 | 24 | 331,776 | 1,066 | 330,710 | 1+4.0E-03 |
| Pyramid | 4 | 6 | 16 | 1 | 15 | 1+0.0E+00 |
| Pyramid 4-gon base | 5 | 8 | 45 | 2 | 43 | 1+2.3E-02 |
| Pyramid 5-gon base | 6 | 10 | 121 | 5 | 116 | 1+3.3E-02 |
| Pyramid 6-gon base | 7 | 12 | 320 | 10 | 310 | 1+3.4E-02 |
| Pyramid 7-gon base | 8 | 14 | 841 | 13 | 828 | 1+2.4E-02 |
| Pyramid 8-gon base | 9 | 16 | 2,205 | 16 | 2,189 | 1+1.4E-02 |
| Pyramid 9-gon base | 10 | 18 | 5,776 | 19 | 5,757 | 1+7.8E-03 |
| Pyramid 10-gon base | 11 | 20 | 15,125 | 22 | 15,103 | 1+3.8E-03 |
| Prism triangle base | 6 | 9 | 75 | 3 | 72 | 1+2.7E-02 |
| Prism 4-gon base | 8 | 12 | 384 | 13 | 371 | 1+3.4E-02 |
| Prism 5-gon base | 10 | 15 | 1,805 | 37 | 1,768 | 1+2.2E-02 |
| Prism 6-gon base | 12 | 18 | 8,100 | 105 | 7,995 | 1+1.5E-02 |
| Prism 7-gon base | 14 | 21 | 35,287 | 269 | 35,018 | 1+9.2E-03 |
| Prism 8-gon base | 16 | 24 | 150,528 | 694 | 149,834 | 1+5.8E-03 |
| Prism 9-gon base | 18 | 27 | 632,025 | 1,874 | 630,151 | 1+3.7E-03 |
| Prism 10-gon base | 20 | 30 | 2,620,860 | 4,722 | 2,616,138 | 1+2.3E-03 |
| Prism 11-gon base | 22 | 33 | 10,759,331 | 12,300 | 10,747,031 | 1+1.5E-03 |
| Prism 12-gon base | 24 | 36 | 43,804,800 | 30,270 | 43,774,530 | 1+9.2E-04 |
| Anti-prism triangle base | 6 | 12 | 384 | 5 | 379 | 1+1.0E-02 |
| Anti-prism 4-gon base | 8 | 16 | 3,528 | 35 | 3,493 | 1+1.0E-02 |
| Anti-prism 5-gon base | 10 | 20 | 30,250 | 189 | 30,061 | 1+7.3E-03 |
| Anti-prism 6-gon base | 12 | 24 | 248,832 | 917 | 247,915 | 1+4.7E-03 |
| Anti-prism 7-gon base | 14 | 28 | 1,989,806 | 4,330 | 1,985,476 | 1+3.0E-03 |
| Anti-prism 8-gon base | 16 | 32 | 15,586,704 | 18,335 | 15,568,369 | 1+1.8E-03 |

Table 8: Experimental results of partition size in enumerating nets of polyhedra

The data in Table 8 evidently shows a size of partition of approximately 1 facet for all polyhedra examined. For more clarification, the data is presented as a chart in Figure 20 and compared against minimum, average and maximum partition size with random partition edge exchange as in Lemma 30. The lines representing the lower bound and the actual for MP algorithm are almost overlapping.

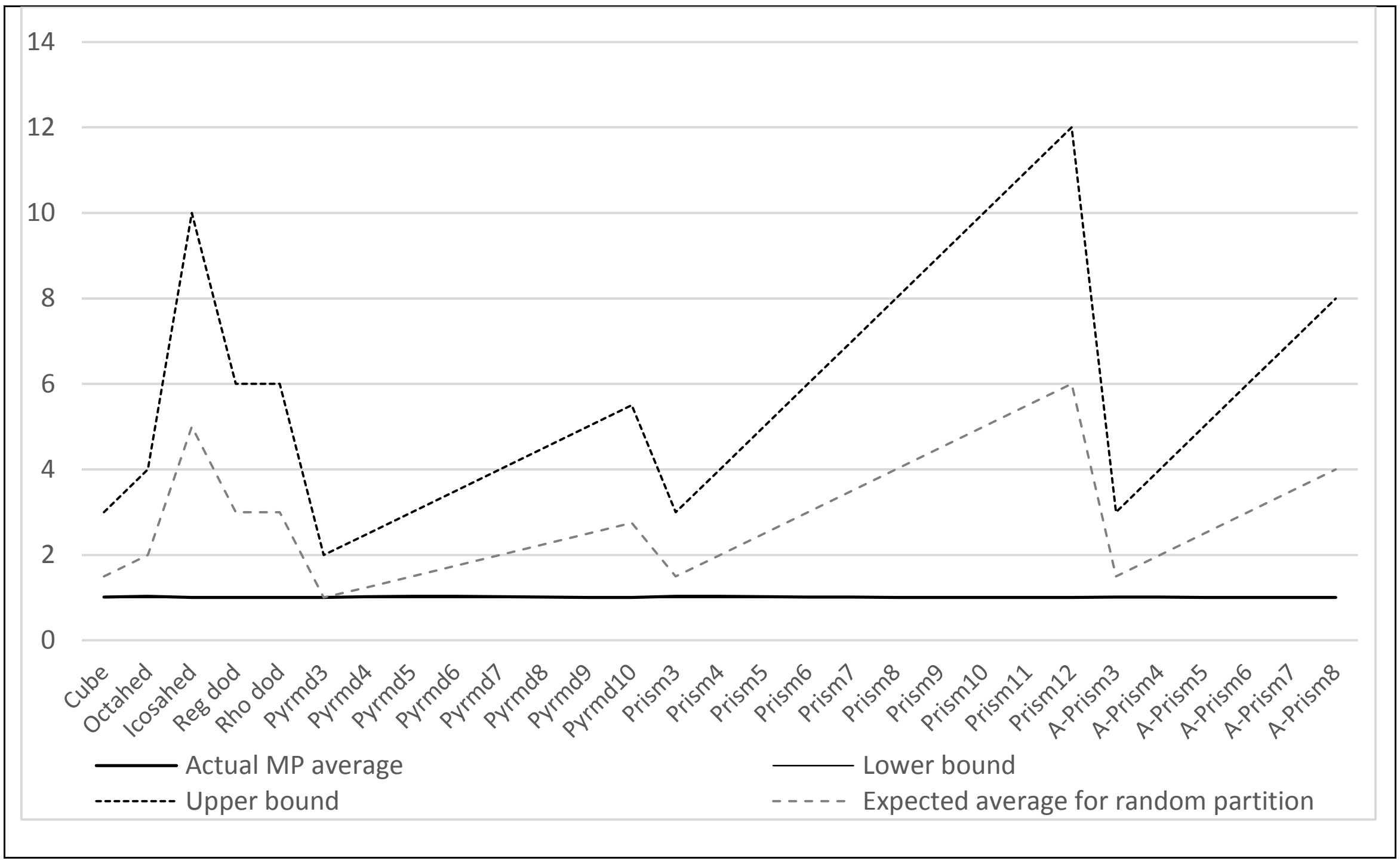

Figure 20: Actual (MP algorithm) and expected (random) partition sizes

### 5.2.3. Spanning tree path lengths application:

Path length between all pairwise nodes is a sample of spanning trees property that occurs very frequently in applications. It is computed as part of many optimization problems such as optimum communication spanning tree [66,67], minimum routing path spanning tree [11] and minimum diameter spanning tree [68]. The applications of these vary from motion planning, network design [69] and computational biology [70].

The naïve approach of calculating the path lengths of each spanning tree from scratch is $O(V^2)$ as we need to calculate the distance between every pair of nodes. Therefore, it is desirable, while enumerating the spanning trees, to keep the disconnected paths as low as possible in order to decrease the number of paths lengths re-calculated. We will show that MP algorithm has better performance in this aspect compared to other edge exchange techniques used for enumerating spanning trees. Here the emphasis is that MP algorithm favors marginal edges over internal edges.

Lemma 31:

The expected number of disconnected paths in the spanning tree in edge exchange with random partition size is $O(V^2)$.

Proof:

Let the partitions of the tree be of sizes $s$ and $V - s$ where $s \in \{1,2,\ldots,V-1\}$ with equal expectation $\frac{1}{V-1}$ for all values.

Any path across the partition will be disconnected.

The number of disconnected paths is $s(V-s)$.

The expected number of disconnected paths is

$$= \sum \frac{s(V-s)}{V-1} = \frac{V\sum s - \sum s^2}{V-1}$$

$$= \frac{V\left(\frac{V(V-1)}{2}\right) - \left(\frac{V(V-1)(2V-1)}{6}\right)}{V-1}$$

$$= (V^2 - V)/6 \qquad \text{(Eq.9)}$$

$= O(V^2)$. ■

### 5.2.4. Theorem 10: Expected disconnected paths

The expected number of disconnected paths in the spanning tree using MP algorithm is $O(V)$.

Proof:

Let the partitions of the tree be of sizes $s$ and $V - s$ where $s \in \{1,2,\ldots,V-1\}$.

Let expectation of $s$ = 1 be $E_1$ and of $s \in \{2,\ldots,V-1\}$ be $E_2$. Where $0 < E_1, E_2 < 1$ and $E_1 + E_2 = 1$

Any path across the partition will be disconnected.

The number of disconnected paths is $s(V-s)$.

The expected number of disconnected paths is

$$= E_1(V-1) + E_2 \sum_{s=2}^{V-1} \frac{s(V-s)}{V-2}$$

$$= (E_1 - E_2)(V-1) + E_2 \sum_{s=1}^{V-1} \frac{s(V - s)}{V - 2}$$

$$= (E_1 - E_2)(V-1) + E_2(V^2 - V)/6$$

Substituting $E_1 = 1 - E_2$: $= E_2/6 * V^2 + (1 - 11 * E_2/6)V + 2 * E_2 - 1$

$$= V-1 + E_2(V^2 - 13V + 6)/6 \quad \text{(Eq.10)}$$

From Theorem 6, $E_2 = O(1/V)$

$\Rightarrow$ The expected number of disconnected paths is $= V - 1 + O(V) = O(V)$. ■

### 5.2.5. Theorem 11: Half path calculation

Any disconnected path in MP algorithm has to be calculated for only 1 side of the edge exchange.

Proof:

Note that the edge exchange under promotion, by Definition 16, will replace an existing edge, $(v_1,v_2)$, in the spanning tree $T$ with another edge which is incident to either $v_1$ or $v_2$. Let it be $(v_1,v_3)$.

If a path between $u$ and $w$ in the spanning tree $T$ passes through $(v_1,v_2)$: $u,\ldots,v_1,v_2,\ldots,w$ then the path after edge promotion should pass by $(v_1,v_3)$ as $(v_1,v_3)$ is the only edge that connects the nodes from the 2 partitions after removing $(v_1,v_2)$.

$\Rightarrow$ The new path after promotion $= u,\ldots,v_1,v_3,\ldots,w$.

The part $u,\ldots,v_1$ remains same and only $v_3,\ldots,w$ has to be computed. ■

The result from Theorem 11 has to be contrasted with the general edge exchange case where a disconnected path has to be entirely computed.

### 5.2.6. Spanning tree based motion planning for multi-robots:

Spanning trees based routing [71] and the spanning tree coverage (STC) problems, in motion planning, are typical applications that depend on the paths between pairwise nodes in the spanning tree. Motion planning is attained by substituting the navigation graph of an environment of interest with a spanning tree of the graph [72]. The problem comes with many flavors where single agent or multi-agents are deployed to cover the environment for many practical reasons such as guarding or exploring the environment [73], sweeping or coating facilities [74], and hazardous waste cleaning [75].

It is a well-established fact that while performance and efficiency of the motion plan depends greatly on the selected spanning tree [76], many aspects of performance and efficiency can only be tested on the actual spanning trees and no algorithms to forecast suitable spanning trees are available [77]. We shall compare the use of MP algorithm to other edge exchange algorithms for calculating routing paths, i.e. paths used by robots to reach targets. Paths calculation is naturally the basic step in spanning tree evaluation. We shall start by defining the problem, then expected costs using both approaches will be calculated.

Problem statement: Spanning tree multi-robots routing

Let $G(V,E)$ be a navigation graph with $r$ robots and $t$ targets all stationed at random nodes of $G$. The initial spanning tree $T_0$ of $G$ is given and all $rt$ paths from robots to targets are calculated. Our objective is to find the expected cost for re-calculating paths in a spanning tree generated from $T_0$ by edge exchange.

Lemma 32:

The expected number of recalculated paths in the spanning tree routing problem with random partition size edge exchange is $O(rt)$.

Proof:

From Equation 9: Expected number of disconnected paths = $\frac{V^2 - V}{6}$

The total number of paths in the graph = $\frac{V^2 - V}{2}$

$\Rightarrow$ 1/3 of the paths will be disconnected (need recalculation).

In this problem we need to maintain all paths from robots to targets. This is $rt$ paths.

With random placement of robots and targets $rt/3 = O(rt)$ paths need to be re-computed after the edge exchange. ■

Lemma 33:

The expected number of recalculated paths in the spanning tree routing problem with MP algorithm is $O(rt/V)$.

Proof:

Consider the different cases of edge exchange:

*Case 1-1*: The leaf edge promoted connects a robot.

Then only $t$ paths from the robot to all targets should be re-computed. Other paths are not disconnected.

The expectation of having a robot at a random node = $r/V$

$\Rightarrow$ $rt/V$ of the paths will expectedly be disconnected (need recalculation).

*Case 1-2*: The leaf edge promoted connects a target.

Then only $r$ paths from all robots to the target should be re-computed. Other paths are not disconnected.

The expectation of having a robot at a random node = $t/V$

$\Rightarrow$ $rt/V$ of the paths will expectedly be disconnected (need recalculation).

*Case 1-3*: The leaf edge promoted does not connect a robot or a target.

$\Rightarrow$ No paths will be disconnected.

*Case 2*: An internal edge is promoted.

Let $E_I$ denote the expectation that an internal edge is promoted.

Assume that all sizes of partitions 2 to $V - 2$ have same expectation = $E_I/(V - 3)$

From Lemma 32, 1/3 of paths will expectedly be disconnected.

$\Rightarrow$ $E_I rt/3$ of the paths will expectedly be disconnected.

But, by Theorem 11, only 1 side of the path need to be recalculated.

Note that we assumed all partitions occur with equal expectation.

$\Rightarrow$ $E_I rt/6$ of the paths will expectedly be recalculated.

Adding expectation from Case 1-1, 1-2, 1-3 and 2:

Expected number of paths that should be re-computed = $2rt/V + E_I rt/6$.

By Theorem 6, $E_I = O(1/V)$.

$\Rightarrow O(rt/V)$ paths will expectedly be recalculated. ■

By comparing the results from Lemma 32 and Lemma 33, we realize that the expected cost per spanning tree in routing problem is $O(V)$ times higher in other algorithms compared to MP algorithm.

**Experimental results of disconnected paths:**

In order to experimentally evaluate the number of required path calculations during the process of enumerating all spanning trees for motion planning, we shall simulate the environment of motion planning using the same configurations we used in a previous paper on multi-robots motion planning [78]. The two different environments together with three robot/target configurations for each environment are shown in Figure 21.

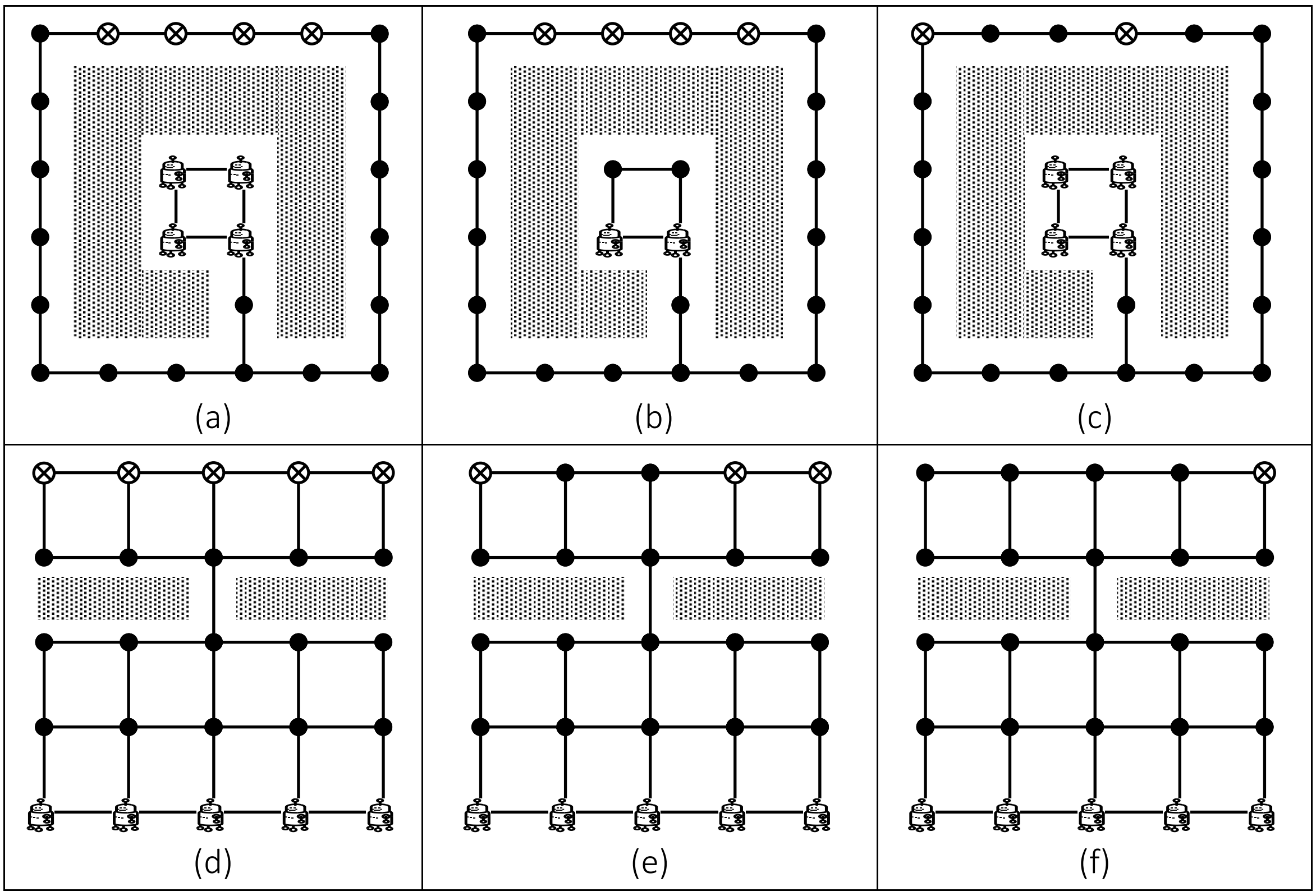

**Figure 21: Multi-robots motion planning simulations**

Figure 21 (a, b, c), shows a sparse simulated environment with (a) 4 robots located at (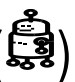) and 4 targets located at (⊗), (b) 2 robots and 4 targets, and (c) 4 robots and 2 targets. Figure 21 (d, e, f), shows a dense environment with 5 robots and (d) 5 targets, (e) 3 targets, and (f) 1 target.

In Table 9, we show the results of experimenting MP algorithm on the environments in Figure 21. The expected disconnected paths for MP algorithm are computed using Lemma 33 as $2rt/V$. The expected disconnected paths for random partition sizes are computed using Lemma 32 as $rt/3$.

| Case | Robots | Targets | Nodes | Spanning Trees | Disconnected paths | Disconnected paths per tree | | |
|---|---|---|---|---|---|---|---|---|
| | | | | | | Actual MP | Expected MP | Expected Random |
| (a) | 4 | 4 | 25 | 80 | 124 | 1.55 | 1.28 | 5.33 |
| (b) | 2 | 4 | 25 | 80 | 36 | 0.45 | 0.64 | 2.67 |
| (c) | 4 | 2 | 25 | 80 | 62 | 0.78 | 0.64 | 2.67 |
| (d) | 5 | 5 | 25 | 6,333,745 | 16,043,546 | 2.53 | 2.00 | 8.33 |
| (e) | 5 | 3 | 25 | 6,333,745 | 7,228,762 | 1.14 | 1.20 | 5.00 |
| (f) | 5 | 1 | 25 | 6,333,745 | 22,243 | 0.004 | 0.40 | 1.67 |

Table 9: Experimental results of disconnected paths in enumerating motion paths

The results in Table 9 verify the limit on disconnected paths computed by Lemma 33. These results are represented as chart in Figure 22. It shows how the actual and expected disconnected paths for MP algorithm overlap and how they compare to expected disconnected paths in random partition size algorithms.

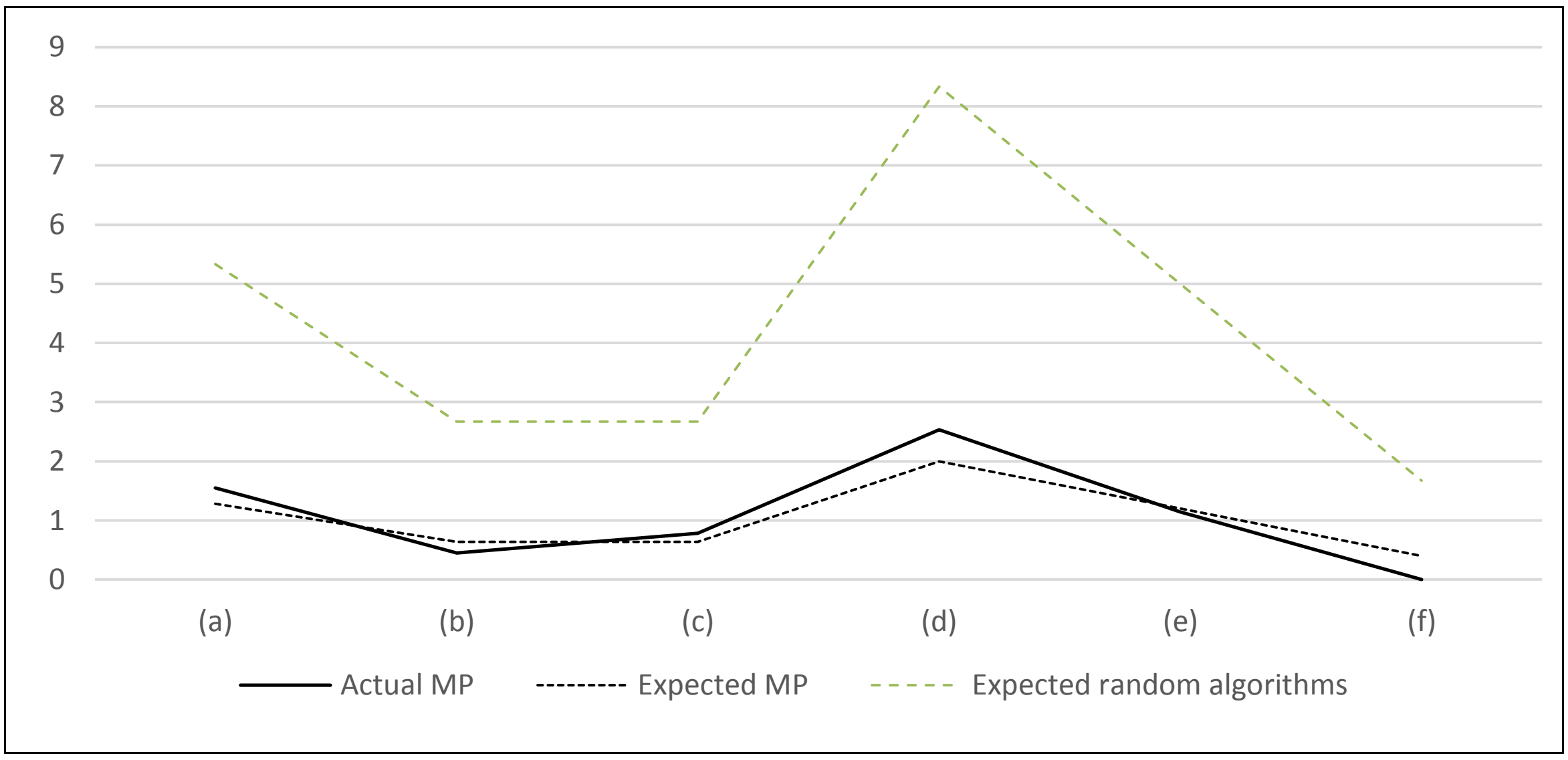

Figure 22: Actual, expected (MP algorithm) and expected (random) disconnected paths

### 5.2.7. Calculation of current in electrical network:

An important application from electrical engineering that requires enumeration of spanning trees is the calculation of electric current in a network. Given an electrical network, applying a power source at 2 nodes will cause different amounts of electric current to flow through edges. The current flow in/out a node and the current flow in/out the whole network should comply with Kirchhoff's first and second laws [79]. The current flow through any edge in the network is given by [80]:

$$C_G(e) = \frac{\sum_{T \in S(G)} C_T(e)}{|S(G)|}$$

Where $C_F(e)$ denotes electric current in edge $e$ under the graph $F$ and $S(G)$ is the set of all spanning trees of $G$.

This equation requires calculation of path between the 2 nodes of electrical source in all spanning trees of the network. The resistance of the path is calculated and the electric current in the edges of the path computed using Ohm's law [79]. By Kirchhoff's first law, the current in all edges in the path is equal while the current in all other edges is zero. We shall examine the effect of using MP algorithm on the solution of this problem.

Problem definition: Calculation of current in an electrical network

Given an electrical network $G(V,E)$ with power source applied at 2 nodes $u$ and $v$, we would like to compute the path between $u$ and $v$ in all spanning trees of $G$ (in order to calculate the electric current in each edge).

Lemma 34:

The expectation that we need to re-compute the path between $u$ and $v$ in a generated spanning tree using edge exchange with random partition size is 1/3.

Proof:

By Equation 9, the expectation that the path between $u$ and $v$ is disconnected is 1/3. ■

Lemma 35:

The expectation that we need to re-compute the path between $u$ and $v$ in the generated spanning tree using MP algorithm is $O(1/V)$.

Proof:

Consider the different cases of edge exchange:

*Case 1-1*: The leaf edge promoted connects a power source node.

Then the path between $u$ and $v$ should be re-computed.

The expectation of having a power source node at a random node = $2/V$

$\Rightarrow$ The path between $u$ and $v$ will be recomputed with expectation of $2/V$.

*Case 1-2*: The leaf edge promoted does not connect a power source node.

$\Rightarrow$ The path will remain connected.

*Case 2*: An internal edge is promoted.

Let $E_I$ denote the expectation that an internal edge is promoted.

Assume that all sizes of partitions 2 to $V - 2$ have same expectation = $E_I/(V - 3)$

From Lemma 32, 1/3 of paths will expectedly be disconnected.

$\Rightarrow$ The path between $u$ and $v$ will be recomputed with expectation of $E_I/3$.

But, by Theorem 11, only 1 side of the path need to be recalculated.

Note that we assumed all partitions occur with equal expectation.

$\Rightarrow$ $E_I/6$ of the paths will expectedly be recalculated.

Adding expectation from Case 1-1, 1-2 and 2:

Expected number of paths that should be re-computed = $2/V + E_I/6$.

By Theorem 6, $E_I = O(1/V)$.

$\Rightarrow$ The expectation that the path between $u$ and $v$ will be re-computed is $O(1/V)$. ■

By comparing the results from Lemma 34 and Lemma 35, we conclude that the expected cost of calculating the current in electrical network using MP algorithm is optimized by $O(V)$ compared to other edge exchange algorithms.

We shall skip experimental results here as the problem, in abstract level, is similar to the previous motion planning problem with 1 robot and 1 target.

Chapter 6

# Conclusion and Future Work

In this thesis, our main aim was to develop a new enumeration algorithm of spanning trees with a distinctive feature: spanning trees partitioning during the enumeration process is minimized. In fulfilment of this objective, MP spanning trees enumeration algorithm in undirected graphs was described, proved, analyzed and practically examined. The promising areas of application in some applied fields were studied.

We started by providing theoretical description of the algorithm. This part delivered the high level explanation of the solution together with the proof of correctness. The edge promotion technique, which is basically single side edge exchange, was proven to prescribe a directed spanning tree over the graph of all spanning trees of any given graph. Further description of the algorithm was then presented with detailed process description and data structures.

Further, the complexity of the algorithm was analyzed. Expected and worst case time and space complexity were computed for the algorithm. Additionally, the minimal partitioning feature of the algorithm was confirmed. An upper bound on spanning trees generated by a partition greater than 1 node was reported.

Subsequently, an experimental verification of the algorithm was performed. MP algorithm was utilized to enumerate spanning trees for a set of random and a set of nonrandom graphs. The results were consistent with all the theoretical findings.

In order to confirm the applicability of the proposed novel algorithm, three applications were examined to demonstrate the expected advantage of adoption of MP algorithm in the solution. These applications are:

1- Construction of nets of a polyhedron where we need to enumerate all the spanning trees over the adjacency graph of the polyhedron. MP algorithm is very suitable here as it decreases the number of unaligned facets in successive spanning trees. An improvement of $O(V)$ over random edge exchange was proven and experimental results on several standard polyhedra were conducted.

2- Optimization of spanning tree motion planning where all spanning trees of the navigation graph has to be generated and examined for certain properties. We have shown that the number of paths that need to be calculated in each spanning tree is decreased by $O(V)$. Here also we conducted experimental results on sample multi-robots/multi-targets motion environments.

3- Calculation of electric current in edges of an electrical network. We need to enumerate all the spanning trees and calculate the current in each spanning tree. We proved that the number of times we need to compute the electrical path in new spanning trees is $O(V)$ less than other edge exchange algorithms.

## 6.1 Future work:

Extensions to this research can be suggested in three main directions:

1- Calculation of better asymptotic bounds for some of the limits used,

2- Enhancement of MP algorithm to achieve better performance, and

3- Examination of ranges of potential application more thoroughly and deeply.

At first, some bounds used in calculating expected cost of the algorithm are loose bounds. These bounds can be tightened to report better expected cost. The expected modifications in data structures, the expected accessed nodes in adjacency list for enumeration of *type-2-from-2b*, and the expected depth of the tree graph are few example of such bounds.

In the side of enhancing MP algorithm, several possible enhancements can be proposed. One enhancement comes from the important observation of the negligible count of type 1 spanning trees (generated by internal edge exchange). Asymptotically, the count of type 2 spanning trees (generated by leaf exchanges) is, exponentially, more than the count of type 1 spanning trees (Theorem 6). This leads to an essential idea: Data structures can be maintained for type 2 spanning trees only, instead of both type 1 and type 2 spanning trees. The cost of permanently maintaining the data for type 1 spanning trees is, in many cases, higher than reconstructing the data structures from scratch when such a tree is encountered. To evaluate the outcome of this approach, we have to consider the count of type 1 spanning trees that are generated from type 2 spanning trees rather than the total count of all type 1 spanning trees. This naturally gives more favorable appraisal to this approach. Appendix B shows the count of type 1 trees generated from type 2 for some special and random graphs. The collected data verifies the result that this type of trees is extremely rarely encountered. MP algorithm can be improved and, positively, the expected cost of generating spanning trees can be reduced by this modification.

In order to make these modifications theoretically admissible, a tight expectation for type 1 spanning trees and, equally important, an expectation for type 1 spanning trees generated from type 2 spanning trees has to be computed.

Another area of improvement is using more efficient data structures for the operations in MP algorithm. In the current implementation, the most costly operations are performed in maintenance of data after promotion. Implementing data structures using different techniques, augmenting the spanning trees with more attributes, or introducing additional data structures can be considered in order to reduce the complexity of the algorithm.

On the other hand, application space of MP algorithm is an important field for further research. The significance of preserving the structure of the spanning trees in constructing nets of polyhedra was the motivation of this research. Additionally, we explored, in the thesis, some applications that depend on pairwise paths in the spanning trees. The expected outcome of employing MP algorithm in these application is appealing. Many other spanning trees application can probably benefit from the important feature of preserving the spanning trees structure. Sequence assembly and its applications in bioinformatics, node and edge coloring and its applications in scheduling and optimization, network flow and routing algorithms are all candidate applications that can benefit from minimal partitioning features. Further study of MP algorithm on these applications may reveal strengths and weaknesses of the algorithm.

Approximation algorithms where spanning trees are used to solve hard problems is another area of application. By using MP algorithm, slightly different spanning trees are generated from every given spanning tree. The effect of this gradual change can be utilized in refining and improving the result of many approximation algorithms that depend on spanning trees. Steiner tree [81] and travelling salesman [3] are samples of problems that may be examined. The presumed solution obtained for either problem using a spanning tree, $T$, can be re-computed for a set of spanning trees closely related to $T$, under MP algorithm.

Finally, the entire concept of minimal partitioning during spanning trees generation is a novel area that proves to be very interesting in various applications. A wide scope of research in many tracks that may not be currently foreseen may be demanded through further exploration.

# Appendix A: Experimental Result

| Graph Name [1] | Nodes | Edges | Total spanning trees [2] | Type 1 count [3] | Type 2 count [4] | Type 2 % |
|---|---|---|---|---|---|---|
| Andrasfai-3 | 8 | 12 | 392 | 12 | 380 | 97% |
| Andrasfai-4 | 11 | 22 | 130,691 | 544 | 130,147 | 100% |
| Antiprism-3 | 6 | 12 | 384 | 5 | 379 | 99% |
| Antiprism-4 | 8 | 16 | 3,528 | 35 | 3,493 | 99% |
| Antiprism-5 | 10 | 20 | 30,250 | 189 | 30,061 | 99% |
| Antiprism-6 | 12 | 24 | 248,832 | 917 | 247,915 | 100% |
| Antiprism-7 | 14 | 28 | 1,989,806 | 4,330 | 1,985,476 | 100% |
| Barbell-3 | 6 | 7 | 9 | 1 | 8 | 89% |
| Barbell-4 | 8 | 13 | 256 | 4 | 252 | 98% |
| Barbell-5 | 10 | 21 | 15,625 | 12 | 15,613 | 100% |
| Barbell-6 | 12 | 31 | 1,679,616 | 47 | 1,679,569 | 100% |
| Book-3 | 8 | 10 | 54 | 4 | 50 | 93% |
| Book-4 | 10 | 13 | 189 | 5 | 184 | 97% |
| Book-5 | 12 | 16 | 648 | 6 | 642 | 99% |
| Book-6 | 14 | 19 | 2,187 | 7 | 2,180 | 100% |
| Book-7 | 16 | 22 | 7,290 | 8 | 7,282 | 100% |
| Book-8 | 18 | 25 | 24,057 | 9 | 24,048 | 100% |
| Book-9 | 20 | 28 | 78,732 | 10 | 78,722 | 100% |
| Cocktailparty-3 | 6 | 12 | 384 | 5 | 379 | 99% |
| Cocktailparty-4 | 8 | 24 | 82,944 | 93 | 82,851 | 100% |
| Complete-3 | 3 | 3 | 3 | 1 | 2 | 67% |
| Complete-4 | 4 | 6 | 16 | 1 | 15 | 94% |
| Complete-5 | 5 | 10 | 125 | 1 | 124 | 99% |
| Complete-6 | 6 | 15 | 1,296 | 1 | 1,295 | 100% |
| Complete-7 | 7 | 21 | 16,807 | 1 | 16,806 | 100% |
| Complete-8 | 8 | 28 | 262,144 | 1 | 262,143 | 100% |
| Complete-9 | 9 | 36 | 4,782,969 | 1 | 4,782,968 | 100% |
| Crossedprism-10 | 20 | 30 | 3,932,160 | 8,920 | 3,923,240 | 100% |
| Crossedprism-4 | 8 | 12 | 384 | 13 | 371 | 97% |
| Crossedprism-6 | 12 | 18 | 9,216 | 142 | 9,074 | 98% |
| Crossedprism-8 | 16 | 24 | 196,608 | 1,065 | 195,543 | 99% |
| Crown-3 | 6 | 6 | 6 | 1 | 5 | 83% |
| Crown-4 | 8 | 12 | 384 | 13 | 371 | 97% |
| Crown-5 | 10 | 20 | 40,500 | 258 | 40,242 | 99% |
| Crown-6 | 12 | 30 | 6,635,520 | 6,608 | 6,628,912 | 100% |

| Graph Name [1] | Nodes | Edges | Total spanning trees [2] | Type 1 count [3] | Type 2 count [4] | Type 2 % |
|---|---|---|---|---|---|---|
| Cycle-3 | 3 | 3 | 3 | 1 | 2 | 67% |
| Cycle-4 | 4 | 4 | 4 | 1 | 3 | 75% |
| Cycle-5 | 5 | 5 | 5 | 1 | 4 | 80% |
| Cycle-6 | 6 | 6 | 6 | 1 | 5 | 83% |
| Cycle-7 | 7 | 7 | 7 | 1 | 6 | 86% |
| Cycle-8 | 8 | 8 | 8 | 1 | 7 | 88% |
| Cycle-9 | 9 | 9 | 9 | 1 | 8 | 89% |
| Cycle-10 | 10 | 10 | 10 | 1 | 9 | 90% |
| Cycle-11 | 11 | 11 | 11 | 1 | 10 | 91% |
| Cycle-12 | 12 | 12 | 12 | 1 | 11 | 92% |
| Cycle-13 | 13 | 13 | 13 | 1 | 12 | 92% |
| Cycle-14 | 14 | 14 | 14 | 1 | 13 | 93% |
| Cycle-15 | 15 | 15 | 15 | 1 | 14 | 93% |
| Cycle-16 | 16 | 16 | 16 | 1 | 15 | 94% |
| Cycle-17 | 17 | 17 | 17 | 1 | 16 | 94% |
| Cycle-18 | 18 | 18 | 18 | 1 | 17 | 94% |
| Cycle-19 | 19 | 19 | 19 | 1 | 18 | 95% |
| Cycle-20 | 20 | 20 | 20 | 1 | 19 | 95% |
| Cyclotomic-13 | 13 | 26 | 1,373,125 | 2,735 | 1,370,390 | 100% |
| Cyclotomic-7 | 7 | 7 | 7 | 1 | 6 | 86% |
| Foldedcube-3 | 4 | 6 | 16 | 1 | 15 | 94% |
| Foldedcube-4 | 8 | 16 | 4,096 | 46 | 4,050 | 99% |
| Gear-3 | 7 | 9 | 50 | 4 | 46 | 92% |
| Gear-4 | 9 | 12 | 192 | 10 | 182 | 95% |
| Gear-5 | 11 | 15 | 722 | 26 | 696 | 96% |
| Gear-6 | 13 | 18 | 2,700 | 57 | 2,643 | 98% |
| Gear-7 | 15 | 21 | 10,082 | 120 | 9,962 | 99% |
| Gear-8 | 17 | 24 | 37,632 | 247 | 37,385 | 99% |
| Gear-10 | 21 | 30 | 524,172 | 1,013 | 523,159 | 100% |
| Gear-11 | 23 | 33 | 1,956,242 | 2,036 | 1,954,206 | 100% |
| Haar-3 | 4 | 4 | 4 | 1 | 3 | 75% |
| Haar-5 | 6 | 6 | 6 | 1 | 5 | 83% |
| Haar-7 | 6 | 9 | 81 | 5 | 76 | 94% |
| Haar-9 | 8 | 8 | 8 | 1 | 7 | 88% |
| Haar-11 | 8 | 12 | 384 | 13 | 371 | 97% |
| Haar-15 | 8 | 16 | 4,096 | 46 | 4,050 | 99% |
| Haar-17 | 10 | 10 | 10 | 1 | 9 | 90% |
| Haar-19 | 10 | 15 | 1,815 | 43 | 1,772 | 98% |

| Graph Name [1] | Nodes | Edges | Total spanning trees [2] | Type 1 count [3] | Type 2 count [4] | Type 2 % |
|---|---|---|---|---|---|---|
| Halvedcube-3 | 4 | 6 | 16 | 1 | 15 | 94% |
| Halvedcube-4 | 8 | 24 | 82,944 | 93 | 82,851 | 100% |
| Helm-3 | 7 | 9 | 16 | 10 | 6 | 38% |
| Helm-4 | 9 | 12 | 45 | 33 | 12 | 27% |
| Helm-5 | 11 | 15 | 121 | 101 | 20 | 17% |
| Helm-6 | 13 | 18 | 320 | 290 | 30 | 9% |
| Helm-7 | 15 | 21 | 841 | 799 | 42 | 5% |
| Helm-8 | 17 | 24 | 2,205 | 2,149 | 56 | 3% |
| Helm-9 | 19 | 27 | 5,776 | 5,704 | 72 | 1% |
| Helm-10 | 21 | 30 | 15,125 | 15,035 | 90 | 1% |
| Helm-11 | 23 | 33 | 39,601 | 39,491 | 110 | 0% |
| Hypercube-3 | 8 | 12 | 384 | 13 | 371 | 97% |
| Ladder-3 | 6 | 7 | 15 | 3 | 12 | 80% |
| Ladder-4 | 8 | 10 | 56 | 7 | 49 | 88% |
| Ladder-5 | 10 | 13 | 209 | 17 | 192 | 92% |
| Ladder-6 | 12 | 16 | 780 | 41 | 739 | 95% |
| Ladder-7 | 14 | 19 | 2,911 | 99 | 2,812 | 97% |
| Ladder-8 | 16 | 22 | 10,864 | 239 | 10,625 | 98% |
| Ladder-9 | 18 | 25 | 40,545 | 577 | 39,968 | 99% |
| Moebiusladder-3 | 6 | 9 | 81 | 5 | 76 | 94% |
| Moebiusladder-4 | 8 | 12 | 392 | 12 | 380 | 97% |
| Moebiusladder-5 | 10 | 15 | 1,815 | 43 | 1,772 | 98% |
| Moebiusladder-6 | 12 | 18 | 8,112 | 112 | 8,000 | 99% |
| Moebiusladder-7 | 14 | 21 | 35,301 | 295 | 35,006 | 99% |
| Mycielski-3 | 5 | 5 | 5 | 1 | 4 | 80% |
| Mycielski-4 | 11 | 20 | 38,642 | 243 | 38,399 | 99% |
| Nonlinesubgraph-3 | 5 | 9 | 75 | 3 | 72 | 96% |
| Nonlinesubgraph-4 | 6 | 7 | 8 | 4 | 4 | 50% |
| Nonlinesubgraph-5 | 6 | 9 | 40 | 8 | 32 | 80% |
| Nonlinesubgraph-6 | 6 | 11 | 192 | 5 | 187 | 97% |
| Nonlinesubgraph-7 | 6 | 8 | 32 | 2 | 30 | 94% |
| Nonlinesubgraph-9 | 6 | 10 | 121 | 5 | 116 | 96% |
| Odd-3 | 10 | 15 | 2,000 | 41 | 1,959 | 98% |
| Paley-5 | 5 | 5 | 5 | 1 | 4 | 80% |
| Paley-9 | 9 | 18 | 11,664 | 89 | 11,575 | 99% |
| Pan-3 | 4 | 4 | 3 | 2 | 1 | 33% |
| Pan-4 | 5 | 5 | 4 | 2 | 2 | 50% |
| Pan-5 | 6 | 6 | 5 | 2 | 3 | 60% |

| Graph Name [1] | Nodes | Edges | Total spanning trees [2] | Type 1 count [3] | Type 2 count [4] | Type 2 % |
|---|---|---|---|---|---|---|
| Pan-6 | 7 | 7 | 6 | 2 | 4 | 67% |
| Pan-7 | 8 | 8 | 7 | 2 | 5 | 71% |
| Pan-8 | 9 | 9 | 8 | 2 | 6 | 75% |
| Pan-9 | 10 | 10 | 9 | 2 | 7 | 78% |
| Pan-10 | 11 | 11 | 10 | 2 | 8 | 80% |
| Pan-11 | 12 | 12 | 11 | 2 | 9 | 82% |
| Pan-12 | 13 | 13 | 12 | 2 | 10 | 83% |
| Pan-13 | 14 | 14 | 13 | 2 | 11 | 85% |
| Pan-14 | 15 | 15 | 14 | 2 | 12 | 86% |
| Pan-15 | 16 | 16 | 15 | 2 | 13 | 87% |
| Pan-16 | 17 | 17 | 16 | 2 | 14 | 88% |
| Pan-17 | 18 | 18 | 17 | 2 | 15 | 88% |
| Pan-18 | 19 | 19 | 18 | 2 | 16 | 89% |
| Pan-19 | 20 | 20 | 19 | 2 | 17 | 89% |
| Pan-20 | 21 | 21 | 20 | 2 | 18 | 90% |
| Permutationstar-3 | 6 | 6 | 6 | 1 | 5 | 83% |
| Prism-3 | 6 | 9 | 75 | 3 | 72 | 96% |
| Prism-4 | 8 | 12 | 384 | 13 | 371 | 97% |
| Prism-5 | 10 | 15 | 1,805 | 37 | 1,768 | 98% |
| Prism-6 | 12 | 18 | 8,100 | 105 | 7,995 | 99% |
| Prism-7 | 14 | 21 | 35,287 | 269 | 35,018 | 99% |
| Sun-4 | 8 | 14 | 600 | 27 | 573 | 96% |
| Sun-5 | 10 | 20 | 9,610 | 242 | 9,368 | 97% |
| Sunlet-3 | 6 | 6 | 3 | 3 | - | 0% |
| Sunlet-4 | 8 | 8 | 4 | 4 | - | 0% |
| Sunlet-5 | 10 | 10 | 5 | 5 | - | 0% |
| Sunlet-6 | 12 | 12 | 6 | 6 | - | 0% |
| Sunlet-7 | 14 | 14 | 7 | 7 | - | 0% |
| Sunlet-8 | 16 | 16 | 8 | 8 | - | 0% |
| Sunlet-9 | 18 | 18 | 9 | 9 | - | 0% |
| Sunlet-10 | 20 | 20 | 10 | 10 | - | 0% |
| Sunlet-11 | 22 | 22 | 11 | 11 | - | 0% |
| Sunlet-12 | 24 | 24 | 12 | 12 | - | 0% |
| Sunlet-13 | 26 | 26 | 13 | 13 | - | 0% |
| Sunlet-14 | 28 | 28 | 14 | 14 | - | 0% |
| Sunlet-15 | 30 | 30 | 15 | 15 | - | 0% |
| Sunlet-16 | 32 | 32 | 16 | 16 | - | 0% |
| Sunlet-17 | 34 | 34 | 17 | 17 | - | 0% |

| Graph Name [1] | Nodes | Edges | Total spanning trees [2] | Type 1 count [3] | Type 2 count [4] | Type 2 % |
|---|---|---|---|---|---|---|
| Sunlet-18 | 36 | 36 | 18 | 18 | - | 0% |
| Sunlet-19 | 38 | 38 | 19 | 19 | - | 0% |
| Sunlet-20 | 40 | 40 | 20 | 20 | - | 0% |
| Triangular-3 | 3 | 3 | 3 | 1 | 2 | 67% |
| Triangular-4 | 6 | 12 | 384 | 5 | 379 | 99% |
| Web-3 | 9 | 12 | 75 | 27 | 48 | 64% |
| Web-4 | 12 | 16 | 384 | 115 | 269 | 70% |
| Web-5 | 15 | 20 | 1,805 | 457 | 1,348 | 75% |
| Web-6 | 18 | 24 | 8,100 | 1,763 | 6,337 | 78% |
| Web-7 | 21 | 28 | 35,287 | 6,567 | 28,720 | 81% |
| Wheel-4 | 4 | 6 | 16 | 1 | 15 | 94% |
| Wheel-5 | 5 | 8 | 45 | 2 | 43 | 96% |
| Wheel-6 | 6 | 10 | 121 | 5 | 116 | 96% |
| Wheel-7 | 7 | 12 | 320 | 10 | 310 | 97% |
| Wheel-8 | 8 | 14 | 841 | 13 | 828 | 98% |
| Wheel-9 | 9 | 16 | 2,205 | 16 | 2,189 | 99% |
| Wheel-10 | 10 | 18 | 5,776 | 19 | 5,757 | 100% |
| Wheel-11 | 11 | 20 | 15,125 | 22 | 15,103 | 100% |
| Wheel-12 | 12 | 22 | 39,601 | 25 | 39,576 | 100% |

Notes:

1. Graphs generated by *Wolfram Mathematica*. Sample drawings of each class of graphs are provided in Appendix C.
2. Total spanning trees generated from implementation and verified against total calculated by *Wolfram Mathematica*.
3. Count of spanning trees generated by internal edge exchange in addition to the root spanning tree.
4. Count of spanning trees generated by leaf edge exchange.

## Appendix B: Experiments on Type 1 Generated from Type 2

| Graph Name [1] | Nodes | Edges | Total spanning trees | Type 1 count | Type 1 from 2 count [2] | Type 1 from 2 % |
|---|---|---|---|---|---|---|
| ANDRASFAI-3 | 8 | 12 | 392 | 12 | 8 | 2.04% |
| ANDRASFAI-4 | 11 | 22 | 130,691 | 544 | 461 | 0.35% |
| ANTIPRISM-3 | 6 | 12 | 384 | 5 | 2 | 0.52% |
| ANTIPRISM-4 | 8 | 16 | 3,528 | 35 | 26 | 0.74% |
| ANTIPRISM-5 | 10 | 20 | 30,250 | 189 | 140 | 0.46% |
| ANTIPRISM-6 | 12 | 24 | 248,832 | 917 | 650 | 0.26% |
| ANTIPRISM-7 | 14 | 28 | 1,989,806 | 4,330 | 2,650 | 0.13% |
| ANTIPRISM-8 | 16 | 32 | 15,586,704 | 18,335 | 11,593 | 0.07% |
| BARBELL-3 | 6 | 7 | 9 | 1 | - | 0.00% |
| BARBELL-4 | 8 | 13 | 256 | 4 | 1 | 0.39% |
| BARBELL-5 | 10 | 21 | 15,625 | 12 | 8 | 0.05% |
| BARBELL-6 | 12 | 31 | 1,679,616 | 47 | 42 | 0.00% |
| BOOK-3 | 8 | 10 | 54 | 4 | 2 | 3.70% |
| BOOK-4 | 10 | 13 | 189 | 5 | 3 | 1.59% |
| BOOK-5 | 12 | 16 | 648 | 6 | 4 | 0.62% |
| BOOK-6 | 14 | 19 | 2,187 | 7 | 5 | 0.23% |
| BOOK-7 | 16 | 22 | 7,290 | 8 | 6 | 0.08% |
| BOOK-8 | 18 | 25 | 24,057 | 9 | 7 | 0.03% |
| BOOK-9 | 20 | 28 | 78,732 | 10 | 8 | 0.01% |
| CENTIPEDE-10 | 20 | 19 | 1 | 1 | - | 0.00% |
| CENTIPEDE-3 | 6 | 5 | 1 | 1 | - | 0.00% |
| CENTIPEDE-4 | 8 | 7 | 1 | 1 | - | 0.00% |
| CENTIPEDE-5 | 10 | 9 | 1 | 1 | - | 0.00% |
| CENTIPEDE-6 | 12 | 11 | 1 | 1 | - | 0.00% |
| CENTIPEDE-7 | 14 | 13 | 1 | 1 | - | 0.00% |
| CENTIPEDE-8 | 16 | 15 | 1 | 1 | - | 0.00% |
| CENTIPEDE-9 | 18 | 17 | 1 | 1 | - | 0.00% |
| COCKTAILPARTY-3 | 6 | 12 | 384 | 5 | 2 | 0.52% |
| COCKTAILPARTY-4 | 8 | 24 | 82,944 | 93 | 88 | 0.11% |
| COCKTAILPARTY-5 | 10 | 40 | 32,768,000 | 3,691 | 3,684 | 0.01% |
| COMPLETE-3 | 3 | 3 | 3 | 1 | - | 0.00% |
| COMPLETE-4 | 4 | 6 | 16 | 1 | - | 0.00% |
| COMPLETE-5 | 5 | 10 | 125 | 1 | - | 0.00% |
| COMPLETE-6 | 6 | 15 | 1,296 | 1 | - | 0.00% |
| COMPLETE-7 | 7 | 21 | 16,807 | 1 | - | 0.00% |

| Graph Name [1] | Nodes | Edges | Total spanning trees | Type 1 count | Type 1 from 2 count [2] | Type 1 from 2 % |
|---|---|---|---|---|---|---|
| COMPLETE-8 | 8 | 28 | 262,144 | 1 | - | 0.00% |
| COMPLETE-9 | 9 | 36 | 4,782,969 | 1 | - | 0.00% |
| CROSSEDPRISM-10 | 20 | 30 | 3,932,160 | 8,920 | 4,313 | 0.11% |
| CROSSEDPRISM-12 | 24 | 36 | 75,497,472 | 72,004 | 34,781 | 0.05% |
| CROSSEDPRISM-4 | 8 | 12 | 384 | 13 | 9 | 2.34% |
| CROSSEDPRISM-6 | 12 | 18 | 9,216 | 142 | 110 | 1.19% |
| CROSSEDPRISM-8 | 16 | 24 | 196,608 | 1,065 | 785 | 0.40% |
| CROWN-3 | 6 | 6 | 6 | 1 | - | 0.00% |
| CROWN-4 | 8 | 12 | 384 | 13 | 9 | 2.34% |
| CROWN-5 | 10 | 20 | 40,500 | 258 | 205 | 0.51% |
| CROWN-6 | 12 | 30 | 6,635,520 | 6,608 | 5,542 | 0.08% |
| CUBECONNECTEDCYCLE-3 | 24 | 36 | 32,400,000 | 46,453 | 20,176 | 0.06% |
| CYCLE-10 | 10 | 10 | 10 | 1 | - | 0.00% |
| CYCLE-11 | 11 | 11 | 11 | 1 | - | 0.00% |
| CYCLE-12 | 12 | 12 | 12 | 1 | - | 0.00% |
| CYCLE-13 | 13 | 13 | 13 | 1 | - | 0.00% |
| CYCLE-14 | 14 | 14 | 14 | 1 | - | 0.00% |
| CYCLE-15 | 15 | 15 | 15 | 1 | - | 0.00% |
| CYCLE-16 | 16 | 16 | 16 | 1 | - | 0.00% |
| CYCLE-17 | 17 | 17 | 17 | 1 | - | 0.00% |
| CYCLE-18 | 18 | 18 | 18 | 1 | - | 0.00% |
| CYCLE-19 | 19 | 19 | 19 | 1 | - | 0.00% |
| CYCLE-20 | 20 | 20 | 20 | 1 | - | 0.00% |
| CYCLE-3 | 3 | 3 | 3 | 1 | - | 0.00% |
| CYCLE-4 | 4 | 4 | 4 | 1 | - | 0.00% |
| CYCLE-5 | 5 | 5 | 5 | 1 | - | 0.00% |
| CYCLE-6 | 6 | 6 | 6 | 1 | - | 0.00% |
| CYCLE-7 | 7 | 7 | 7 | 1 | - | 0.00% |
| CYCLE-8 | 8 | 8 | 8 | 1 | - | 0.00% |
| CYCLE-9 | 9 | 9 | 9 | 1 | - | 0.00% |
| CYCLOTOMIC-13 | 13 | 26 | 1,373,125 | 2,735 | 2,146 | 0.16% |
| CYCLOTOMIC-7 | 7 | 7 | 7 | 1 | - | 0.00% |
| FOLDEDCUBE-3 | 4 | 6 | 16 | 1 | - | 0.00% |
| FOLDEDCUBE-4 | 8 | 16 | 4,096 | 46 | 45 | 1.10% |
| GEAR-10 | 21 | 30 | 524,172 | 1,013 | 16 | 0.00% |
| GEAR-11 | 23 | 33 | 1,956,242 | 2,036 | 18 | 0.00% |
| GEAR-12 | 25 | 36 | 7,300,800 | 4,083 | 20 | 0.00% |

| Graph Name [1] | Nodes | Edges | Total spanning trees | Type 1 count | Type 1 from 2 count [2] | Type 1 from 2 % |
|---|---|---|---|---|---|---|
| GEAR-13 | 27 | 39 | 27,246,962 | 8,178 | 22 | 0.00% |
| GEAR-3 | 7 | 9 | 50 | 4 | 2 | 4.00% |
| GEAR-4 | 9 | 12 | 192 | 10 | 2 | 1.04% |
| GEAR-5 | 11 | 15 | 722 | 26 | 6 | 0.83% |
| GEAR-6 | 13 | 18 | 2,700 | 57 | 8 | 0.30% |
| GEAR-7 | 15 | 21 | 10,082 | 120 | 10 | 0.10% |
| GEAR-8 | 17 | 24 | 37,632 | 247 | 12 | 0.03% |
| GEAR-9 | 19 | 27 | 140,450 | 502 | 14 | 0.01% |
| HAAR-11 | 8 | 12 | 384 | 13 | 9 | 2.34% |
| HAAR-15 | 8 | 16 | 4,096 | 46 | 45 | 1.10% |
| HAAR-17 | 10 | 10 | 10 | 1 | - | 0.00% |
| HAAR-19 | 10 | 15 | 1,815 | 43 | 33 | 1.82% |
| HAAR-3 | 4 | 4 | 4 | 1 | - | 0.00% |
| HAAR-5 | 6 | 6 | 6 | 1 | - | 0.00% |
| HAAR-7 | 6 | 9 | 81 | 5 | 4 | 4.94% |
| HAAR-9 | 8 | 8 | 8 | 1 | - | 0.00% |
| HADAMARD-4 | 16 | 32 | 42,467,328 | 32,874 | 19,625 | 0.05% |
| HALVEDCUBE-3 | 4 | 6 | 16 | 1 | - | 0.00% |
| HALVEDCUBE-4 | 8 | 24 | 82,944 | 93 | 88 | 0.11% |
| HANOI-3 | 27 | 39 | 20,503,125 | 24,035 | 21,451 | 0.10% |
| HELM-10 | 21 | 30 | 15,125 | 15,035 | 41 | 0.27% |
| HELM-11 | 23 | 33 | 39,601 | 39,491 | 50 | 0.13% |
| HELM-12 | 25 | 36 | 103,680 | 103,548 | 60 | 0.06% |
| HELM-13 | 27 | 39 | 271,441 | 271,285 | 71 | 0.03% |
| HELM-14 | 29 | 42 | 710,645 | 710,463 | 83 | 0.01% |
| HELM-15 | 31 | 45 | 1,860,496 | 1,860,286 | 96 | 0.01% |
| HELM-16 | 33 | 48 | 4,870,845 | 4,870,605 | 110 | 0.00% |
| HELM-17 | 35 | 51 | 12,752,041 | 12,751,769 | 125 | 0.00% |
| HELM-18 | 37 | 54 | 33,385,280 | 33,384,974 | 141 | 0.00% |
| HELM-19 | 39 | 57 | 87,403,801 | 87,403,459 | 158 | 0.00% |
| HELM-3 | 7 | 9 | 16 | 10 | 2 | 12.50% |
| HELM-4 | 9 | 12 | 45 | 33 | 8 | 17.78% |
| HELM-5 | 11 | 15 | 121 | 101 | 15 | 12.40% |
| HELM-6 | 13 | 18 | 320 | 290 | 15 | 4.69% |
| HELM-7 | 15 | 21 | 841 | 799 | 20 | 2.38% |
| HELM-8 | 17 | 24 | 2,205 | 2,149 | 26 | 1.18% |
| HELM-9 | 19 | 27 | 5,776 | 5,704 | 33 | 0.57% |

| Graph Name [1] | Nodes | Edges | Total spanning trees | Type 1 count | Type 1 from 2 count [2] | Type 1 from 2 % |
|---|---|---|---|---|---|---|
| HYPERCUBE-3 | 8 | 12 | 384 | 13 | 9 | 2.34% |
| HYPERCUBE-4 | 16 | 32 | 42,467,328 | 32,874 | 19,625 | 0.05% |
| LADDER-10 | 20 | 28 | 151,316 | 1,393 | - | 0.00% |
| LADDER-11 | 22 | 31 | 564,719 | 3,363 | - | 0.00% |
| LADDER-12 | 24 | 34 | 2,107,560 | 8,119 | - | 0.00% |
| LADDER-13 | 26 | 37 | 7,865,521 | 19,601 | - | 0.00% |
| LADDER-14 | 28 | 40 | 29,354,524 | 47,321 | - | 0.00% |
| LADDER-3 | 6 | 7 | 15 | 3 | - | 0.00% |
| LADDER-4 | 8 | 10 | 56 | 7 | - | 0.00% |
| LADDER-5 | 10 | 13 | 209 | 17 | - | 0.00% |
| LADDER-6 | 12 | 16 | 780 | 41 | - | 0.00% |
| LADDER-7 | 14 | 19 | 2,911 | 99 | - | 0.00% |
| LADDER-8 | 16 | 22 | 10,864 | 239 | - | 0.00% |
| LADDER-9 | 18 | 25 | 40,545 | 577 | - | 0.00% |
| MOEBIUSLADDER-10 | 20 | 30 | 2,620,880 | 5,419 | 2,715 | 0.10% |
| MOEBIUSLADDER-11 | 22 | 33 | 10,759,353 | 13,433 | 7,308 | 0.07% |
| MOEBIUSLADDER-12 | 24 | 36 | 43,804,824 | 34,932 | 15,637 | 0.04% |
| MOEBIUSLADDER-3 | 6 | 9 | 81 | 5 | 4 | 4.94% |
| MOEBIUSLADDER-4 | 8 | 12 | 392 | 12 | 8 | 2.04% |
| MOEBIUSLADDER-5 | 10 | 15 | 1,815 | 43 | 33 | 1.82% |
| MOEBIUSLADDER-6 | 12 | 18 | 8,112 | 112 | 70 | 0.86% |
| MOEBIUSLADDER-7 | 14 | 21 | 35,301 | 295 | 196 | 0.56% |
| MOEBIUSLADDER-8 | 16 | 24 | 150,544 | 803 | 435 | 0.29% |
| MOEBIUSLADDER-9 | 18 | 27 | 632,043 | 2,071 | 1,255 | 0.20% |
| MYCIELSKI-3 | 5 | 5 | 5 | 1 | - | 0.00% |
| MYCIELSKI-4 | 11 | 20 | 38,642 | 243 | 202 | 0.52% |
| NONLINESUBGRAPH-3 | 5 | 9 | 75 | 3 | - | 0.00% |
| NONLINESUBGRAPH-4 | 6 | 7 | 8 | 4 | 1 | 12.50% |
| NONLINESUBGRAPH-5 | 6 | 9 | 40 | 8 | 4 | 10.00% |
| NONLINESUBGRAPH-6 | 6 | 11 | 192 | 5 | 2 | 1.04% |
| NONLINESUBGRAPH-7 | 6 | 8 | 32 | 2 | - | 0.00% |
| NONLINESUBGRAPH-9 | 6 | 10 | 121 | 5 | 2 | 1.65% |
| ODD-3 | 10 | 15 | 2,000 | 41 | 38 | 1.90% |
| PALEY-5 | 5 | 5 | 5 | 1 | - | 0.00% |
| PALEY-9 | 9 | 18 | 11,664 | 89 | 60 | 0.51% |
| PAN-10 | 11 | 11 | 10 | 2 | 1 | 10.00% |
| PAN-11 | 12 | 12 | 11 | 2 | 1 | 9.09% |

| Graph Name [1] | Nodes | Edges | Total spanning trees | Type 1 count | Type 1 from 2 count [2] | Type 1 from 2 % |
|---|---|---|---|---|---|---|
| PAN-12 | 13 | 13 | 12 | 2 | 1 | 8.33% |
| PAN-13 | 14 | 14 | 13 | 2 | 1 | 7.69% |
| PAN-14 | 15 | 15 | 14 | 2 | 1 | 7.14% |
| PAN-15 | 16 | 16 | 15 | 2 | 1 | 6.67% |
| PAN-16 | 17 | 17 | 16 | 2 | 1 | 6.25% |
| PAN-17 | 18 | 18 | 17 | 2 | 1 | 5.88% |
| PAN-18 | 19 | 19 | 18 | 2 | 1 | 5.56% |
| PAN-19 | 20 | 20 | 19 | 2 | 1 | 5.26% |
| PAN-20 | 21 | 21 | 20 | 2 | 1 | 5.00% |
| PAN-3 | 4 | 4 | 3 | 2 | - | 0.00% |
| PAN-4 | 5 | 5 | 4 | 2 | - | 0.00% |
| PAN-5 | 6 | 6 | 5 | 2 | 1 | 20.00% |
| PAN-6 | 7 | 7 | 6 | 2 | 1 | 16.67% |
| PAN-7 | 8 | 8 | 7 | 2 | 1 | 14.29% |
| PAN-8 | 9 | 9 | 8 | 2 | 1 | 12.50% |
| PAN-9 | 10 | 10 | 9 | 2 | 1 | 11.11% |
| PATH-10 | 10 | 9 | 1 | 1 | - | 0.00% |
| PATH-11 | 11 | 10 | 1 | 1 | - | 0.00% |
| PATH-12 | 12 | 11 | 1 | 1 | - | 0.00% |
| PATH-13 | 13 | 12 | 1 | 1 | - | 0.00% |
| PATH-14 | 14 | 13 | 1 | 1 | - | 0.00% |
| PATH-15 | 15 | 14 | 1 | 1 | - | 0.00% |
| PATH-16 | 16 | 15 | 1 | 1 | - | 0.00% |
| PATH-17 | 17 | 16 | 1 | 1 | - | 0.00% |
| PATH-18 | 18 | 17 | 1 | 1 | - | 0.00% |
| PATH-19 | 19 | 18 | 1 | 1 | - | 0.00% |
| PATH-20 | 20 | 19 | 1 | 1 | - | 0.00% |
| PATH-3 | 3 | 2 | 1 | 1 | - | 0.00% |
| PATH-4 | 4 | 3 | 1 | 1 | - | 0.00% |
| PATH-5 | 5 | 4 | 1 | 1 | - | 0.00% |
| PATH-6 | 6 | 5 | 1 | 1 | - | 0.00% |
| PATH-7 | 7 | 6 | 1 | 1 | - | 0.00% |
| PATH-8 | 8 | 7 | 1 | 1 | - | 0.00% |
| PATH-9 | 9 | 8 | 1 | 1 | - | 0.00% |
| PERMUTATIONSTAR-3 | 6 | 6 | 6 | 1 | - | 0.00% |
| PRISM-10 | 20 | 30 | 2,620,860 | 4,722 | 2,202 | 0.08% |
| PRISM-11 | 22 | 33 | 10,759,331 | 12,300 | 5,374 | 0.05% |

| Graph Name [1] | Nodes | Edges | Total spanning trees | Type 1 count | Type 1 from 2 count [2] | Type 1 from 2 % |
|---|---|---|---|---|---|---|
| PRISM-12 | 24 | 36 | 43,804,800 | 30,270 | 12,921 | 0.03% |
| PRISM-3 | 6 | 9 | 75 | 3 | - | 0.00% |
| PRISM-4 | 8 | 12 | 384 | 13 | 9 | 2.34% |
| PRISM-5 | 10 | 15 | 1,805 | 37 | 23 | 1.27% |
| PRISM-6 | 12 | 18 | 8,100 | 105 | 61 | 0.75% |
| PRISM-7 | 14 | 21 | 35,287 | 269 | 138 | 0.39% |
| PRISM-8 | 16 | 24 | 150,528 | 694 | 344 | 0.23% |
| PRISM-9 | 18 | 27 | 632,025 | 1,874 | 894 | 0.14% |
| SIERPINSKI-3 | 15 | 27 | 524,880 | 2,823 | 1,350 | 0.26% |
| SQUARE-3 | 9 | 24 | 129,600 | 361 | 301 | 0.23% |
| STAR-10 | 10 | 9 | 1 | 1 | - | 0.00% |
| STAR-11 | 11 | 10 | 1 | 1 | - | 0.00% |
| STAR-12 | 12 | 11 | 1 | 1 | - | 0.00% |
| STAR-13 | 13 | 12 | 1 | 1 | - | 0.00% |
| STAR-14 | 14 | 13 | 1 | 1 | - | 0.00% |
| STAR-15 | 15 | 14 | 1 | 1 | - | 0.00% |
| STAR-16 | 16 | 15 | 1 | 1 | - | 0.00% |
| STAR-17 | 17 | 16 | 1 | 1 | - | 0.00% |
| STAR-18 | 18 | 17 | 1 | 1 | - | 0.00% |
| STAR-19 | 19 | 18 | 1 | 1 | - | 0.00% |
| STAR-20 | 20 | 19 | 1 | 1 | - | 0.00% |
| STAR-3 | 3 | 2 | 1 | 1 | - | 0.00% |
| STAR-4 | 4 | 3 | 1 | 1 | - | 0.00% |
| STAR-5 | 5 | 4 | 1 | 1 | - | 0.00% |
| STAR-6 | 6 | 5 | 1 | 1 | - | 0.00% |
| STAR-7 | 7 | 6 | 1 | 1 | - | 0.00% |
| STAR-8 | 8 | 7 | 1 | 1 | - | 0.00% |
| STAR-9 | 9 | 8 | 1 | 1 | - | 0.00% |
| SUN-4 | 8 | 14 | 600 | 27 | 10 | 1.67% |
| SUN-5 | 10 | 20 | 9,610 | 242 | 76 | 0.79% |
| SUN-6 | 12 | 27 | 202,800 | 2,729 | 662 | 0.33% |
| SUN-7 | 14 | 35 | 5,329,646 | 37,486 | 7,225 | 0.14% |
| SUNLET-10 | 20 | 20 | 10 | 10 | - | 0.00% |
| SUNLET-11 | 22 | 22 | 11 | 11 | - | 0.00% |
| SUNLET-12 | 24 | 24 | 12 | 12 | - | 0.00% |
| SUNLET-13 | 26 | 26 | 13 | 13 | - | 0.00% |
| SUNLET-14 | 28 | 28 | 14 | 14 | - | 0.00% |

| Graph Name [1] | Nodes | Edges | Total spanning trees | Type 1 count | Type 1 from 2 count [2] | Type 1 from 2 % |
|---|---|---|---|---|---|---|
| SUNLET-15 | 30 | 30 | 15 | 15 | - | 0.00% |
| SUNLET-16 | 32 | 32 | 16 | 16 | - | 0.00% |
| SUNLET-17 | 34 | 34 | 17 | 17 | - | 0.00% |
| SUNLET-18 | 36 | 36 | 18 | 18 | - | 0.00% |
| SUNLET-19 | 38 | 38 | 19 | 19 | - | 0.00% |
| SUNLET-20 | 40 | 40 | 20 | 20 | - | 0.00% |
| SUNLET-3 | 6 | 6 | 3 | 3 | - | 0.00% |
| SUNLET-4 | 8 | 8 | 4 | 4 | - | 0.00% |
| SUNLET-5 | 10 | 10 | 5 | 5 | - | 0.00% |
| SUNLET-6 | 12 | 12 | 6 | 6 | - | 0.00% |
| SUNLET-7 | 14 | 14 | 7 | 7 | - | 0.00% |
| SUNLET-8 | 16 | 16 | 8 | 8 | - | 0.00% |
| SUNLET-9 | 18 | 18 | 9 | 9 | - | 0.00% |
| TRIANGULAR-3 | 3 | 3 | 3 | 1 | - | 0.00% |
| TRIANGULAR-4 | 6 | 12 | 384 | 5 | 2 | 0.52% |
| TRIANGULAR-5 | 10 | 30 | 2,048,000 | 1,688 | 1,451 | 0.07% |
| WEB-10 | 30 | 40 | 2,620,860 | 308,436 | 10,655 | 0.41% |
| WEB-11 | 33 | 44 | 10,759,331 | 1,088,368 | 72,111 | 0.67% |
| WEB-12 | 36 | 48 | 43,804,800 | 3,809,999 | 111,627 | 0.25% |
| WEB-3 | 9 | 12 | 75 | 27 | - | 0.00% |
| WEB-4 | 12 | 16 | 384 | 115 | 3 | 0.78% |
| WEB-5 | 15 | 20 | 1,805 | 457 | 52 | 2.88% |
| WEB-6 | 18 | 24 | 8,100 | 1,763 | 83 | 1.02% |
| WEB-7 | 21 | 28 | 35,287 | 6,567 | 636 | 1.80% |
| WEB-8 | 24 | 32 | 150,528 | 24,000 | 989 | 0.66% |
| WEB-9 | 27 | 36 | 632,025 | 86,549 | 6,878 | 1.09% |
| WHEEL-10 | 10 | 18 | 5,776 | 19 | 10 | 0.17% |
| WHEEL-11 | 11 | 20 | 15,125 | 22 | 13 | 0.09% |
| WHEEL-12 | 12 | 22 | 39,601 | 25 | 16 | 0.04% |
| WHEEL-13 | 13 | 24 | 103,680 | 28 | 19 | 0.02% |
| WHEEL-14 | 14 | 26 | 271,441 | 31 | 22 | 0.01% |
| WHEEL-15 | 15 | 28 | 710,645 | 34 | 25 | 0.00% |
| WHEEL-16 | 16 | 30 | 1,860,496 | 37 | 28 | 0.00% |
| WHEEL-17 | 17 | 32 | 4,870,845 | 40 | 31 | 0.00% |
| WHEEL-18 | 18 | 34 | 12,752,041 | 43 | 34 | 0.00% |
| WHEEL-19 | 19 | 36 | 33,385,280 | 46 | 37 | 0.00% |
| WHEEL-20 | 20 | 38 | 87,403,801 | 49 | 40 | 0.00% |

| Graph Name [1] | Nodes | Edges | Total spanning trees | Type 1 count | Type 1 from 2 count [2] | Type 1 from 2 % |
|---|---|---|---|---|---|---|
| WHEEL-4 | 4 | 6 | 16 | 1 | - | 0.00% |
| WHEEL-5 | 5 | 8 | 45 | 2 | - | 0.00% |
| WHEEL-6 | 6 | 10 | 121 | 5 | 2 | 1.65% |
| WHEEL-7 | 7 | 12 | 320 | 10 | 1 | 0.31% |
| WHEEL-8 | 8 | 14 | 841 | 13 | 4 | 0.48% |
| WHEEL-9 | 9 | 16 | 2,205 | 16 | 7 | 0.32% |

Notes:

1. Graphs generated by *Wolfram Mathematica*. Sample drawings of each class of graphs are provided in Appendix C.
2. Count of type 1 spanning trees (i.e. generated by internal edge exchange) from a type 2 parent spanning tree (i.e. generated by leaf edge exchange).

# Appendix C: Drawings of Graphs Used in Experiments

Andrasfai 3, 4, 5:

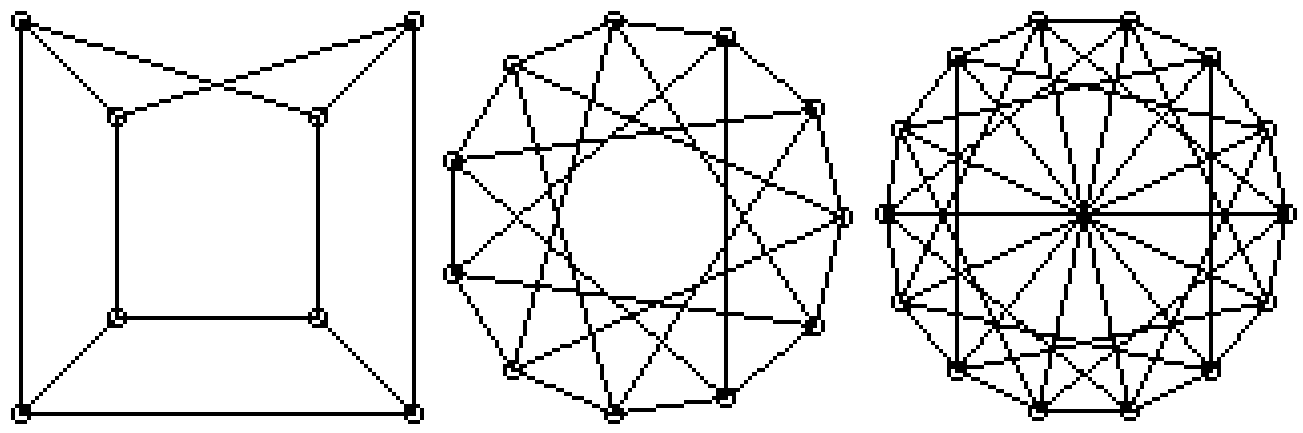

Antiprism 3, 4, 5, 6, 7:

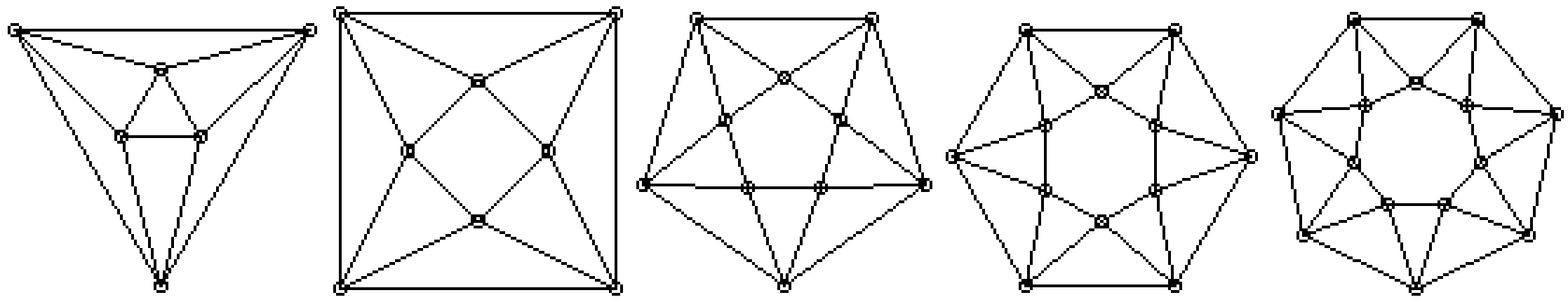

Barbell 3, 4, 5, 6, 7:

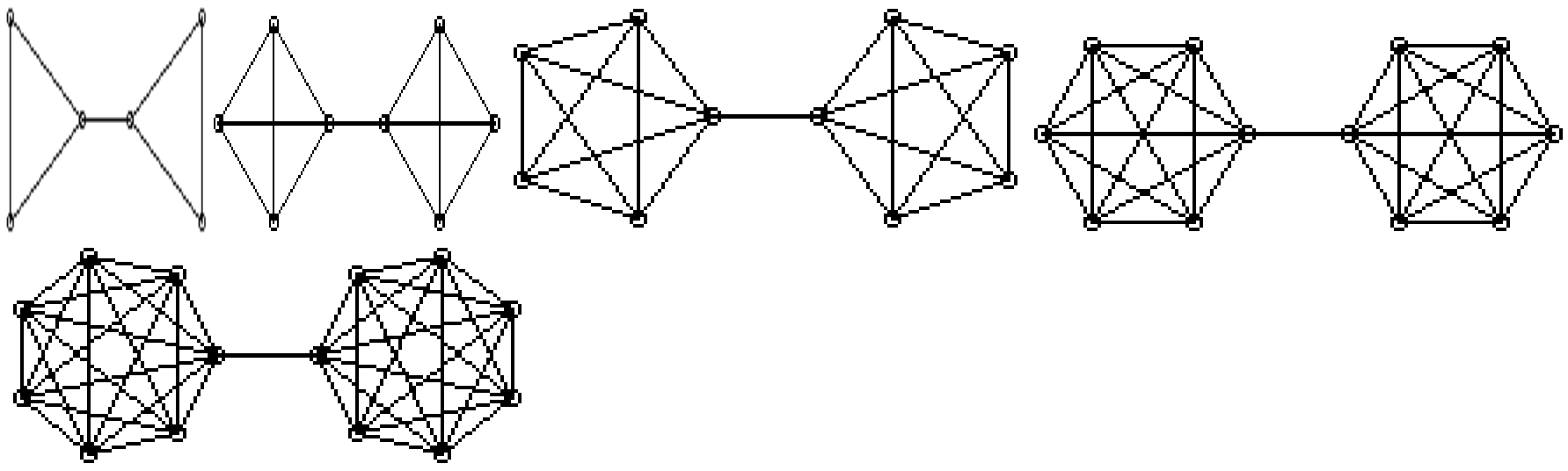

Book 3, 4, 5, 6, 7:

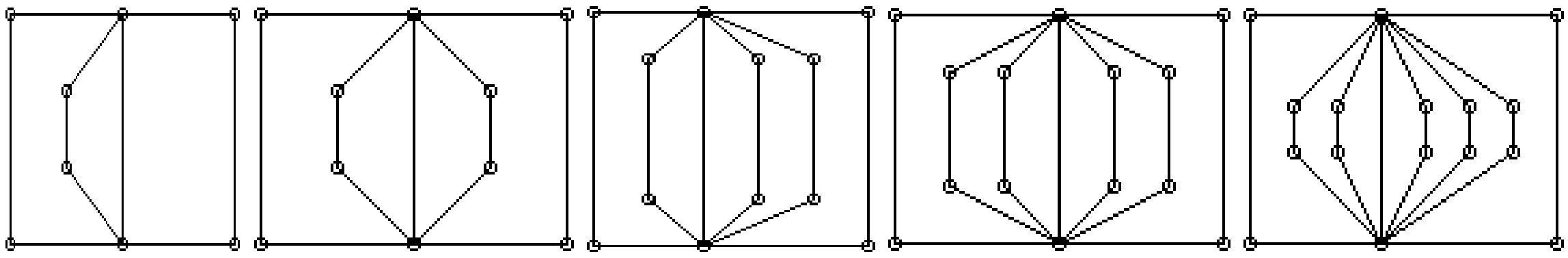

Cocktailparty 3, 4, 5:

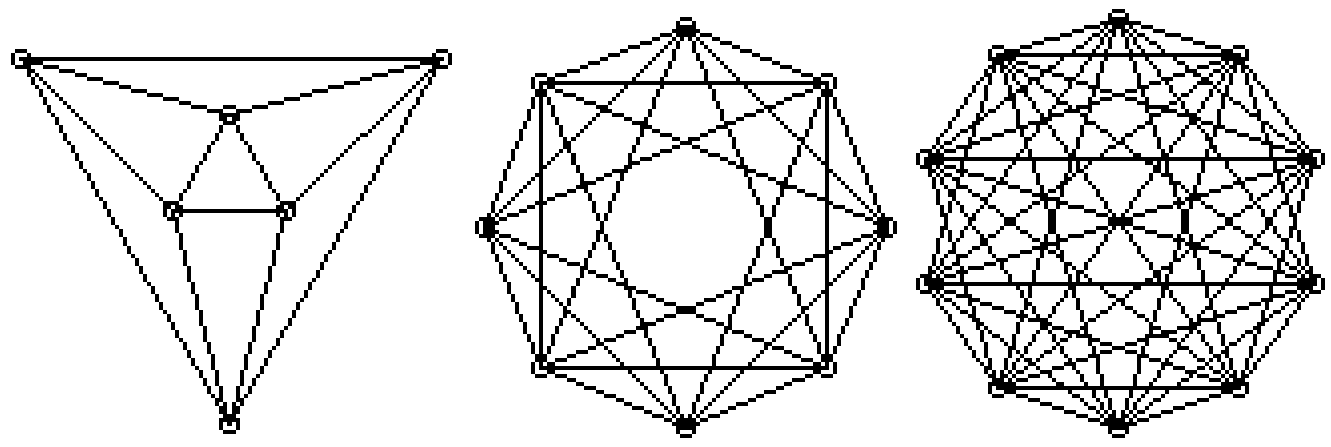

Complete 3, 4, 5, 6, 7:

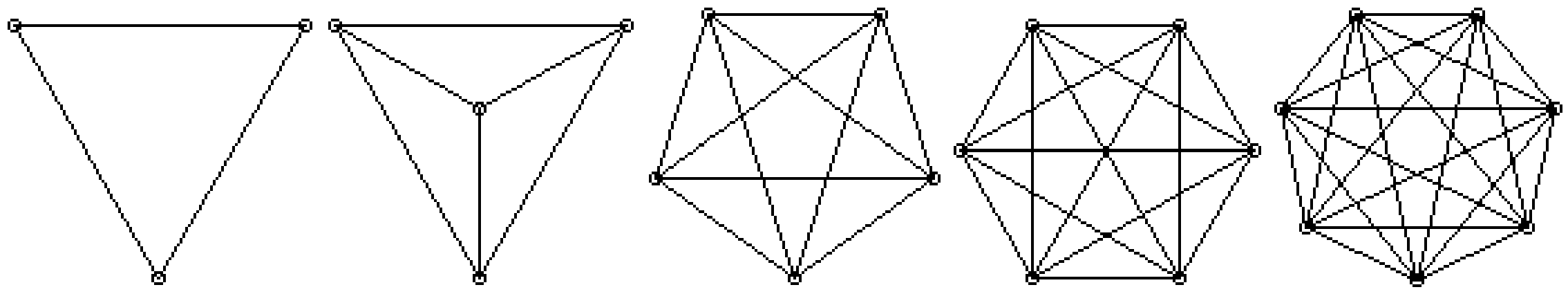

Crossedprism 4, 6, 8, 10, 12:

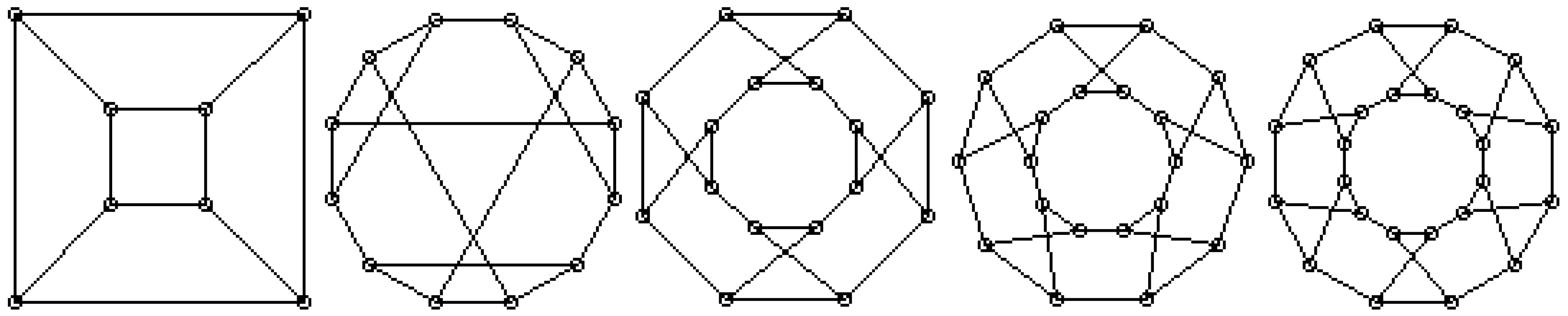

Crown 3, 4, 5:

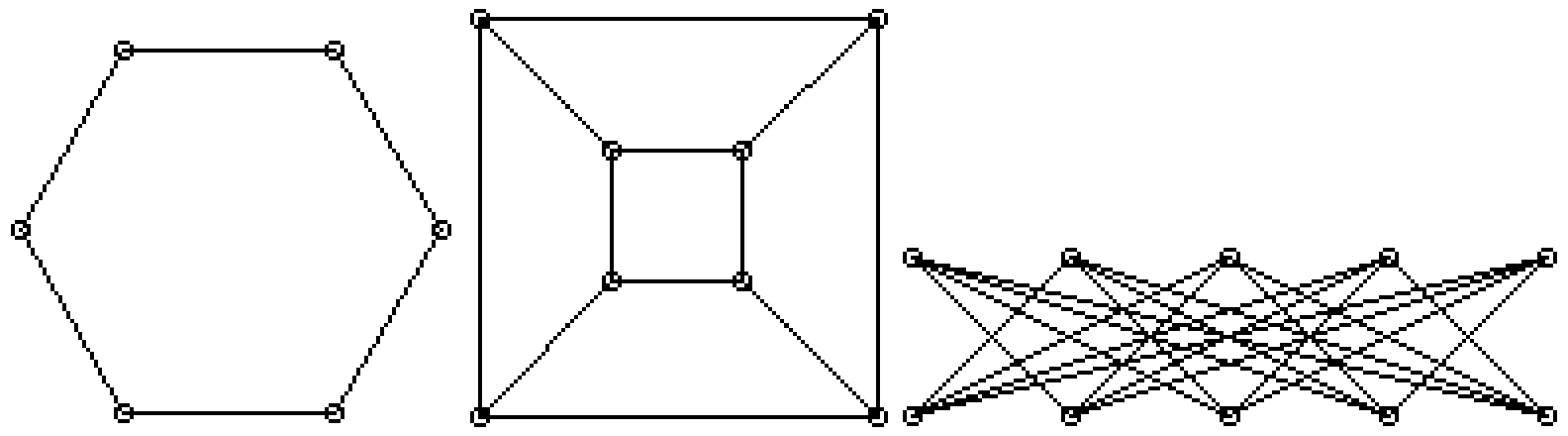

Cycle 3, 4, 5, 6, 7:

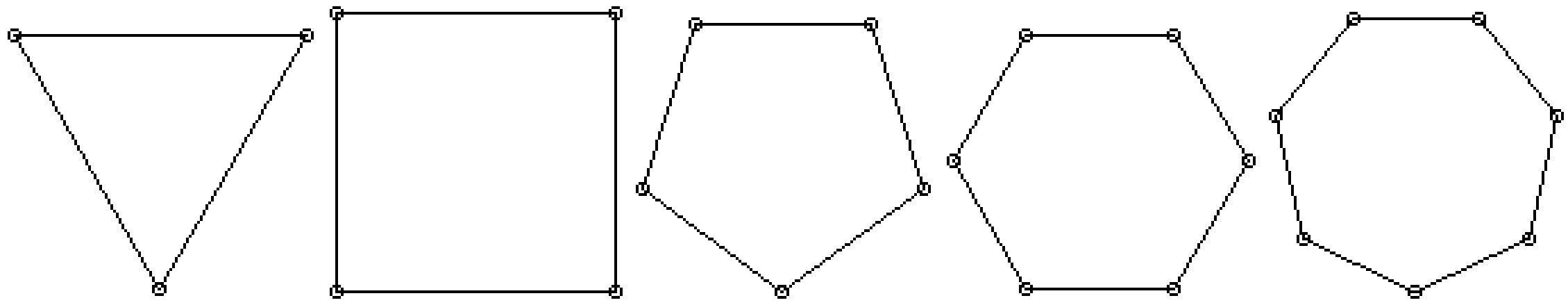

Cyclotomic 7, 13:

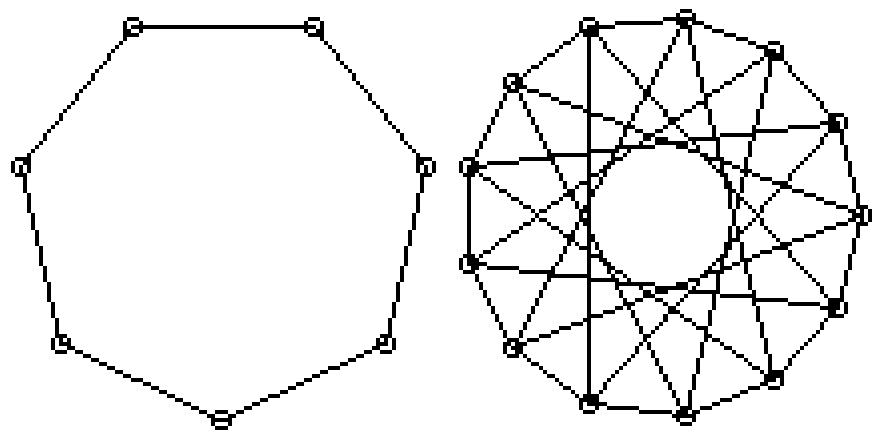

Foldedcube 3, 4:

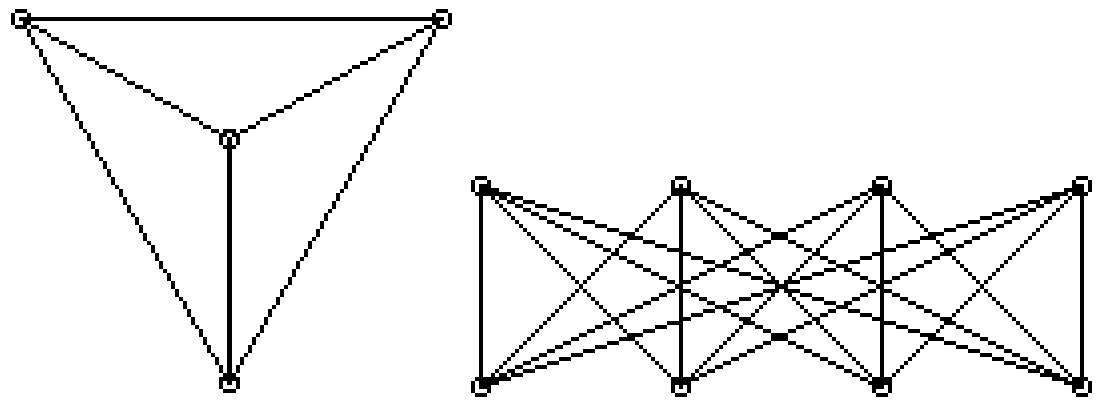

Gear 3, 4, 5, 6, 7:

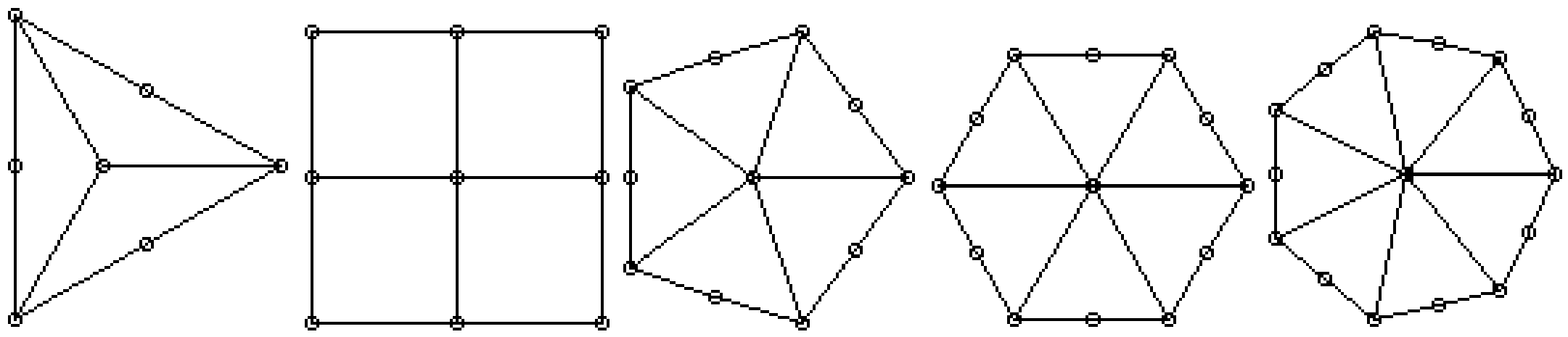

Haar 5, 7, 9, 11, 13:

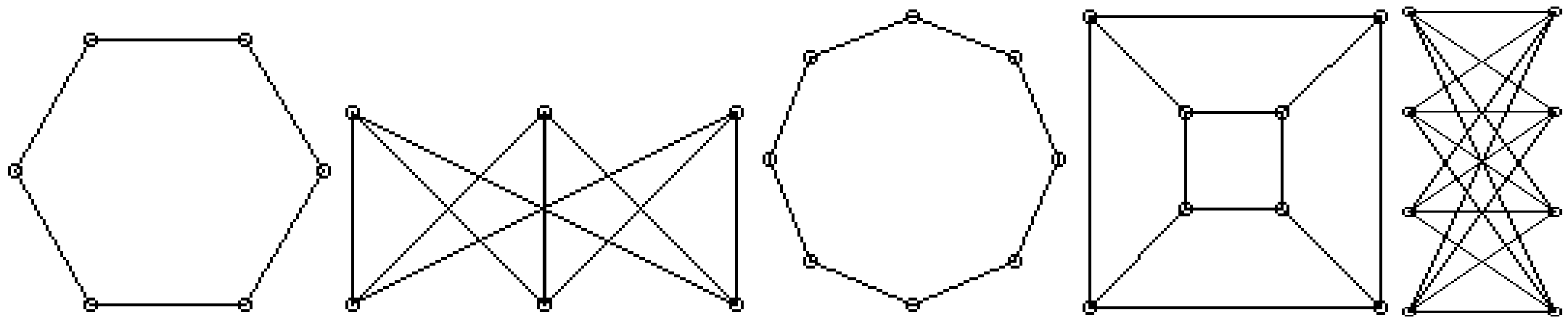

Halvedcube 3, 4:

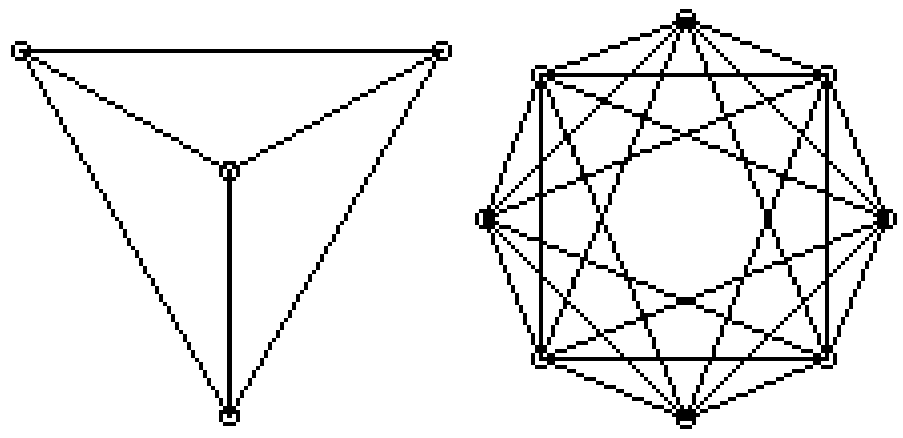

Helm 3, 4, 5, 6, 7:

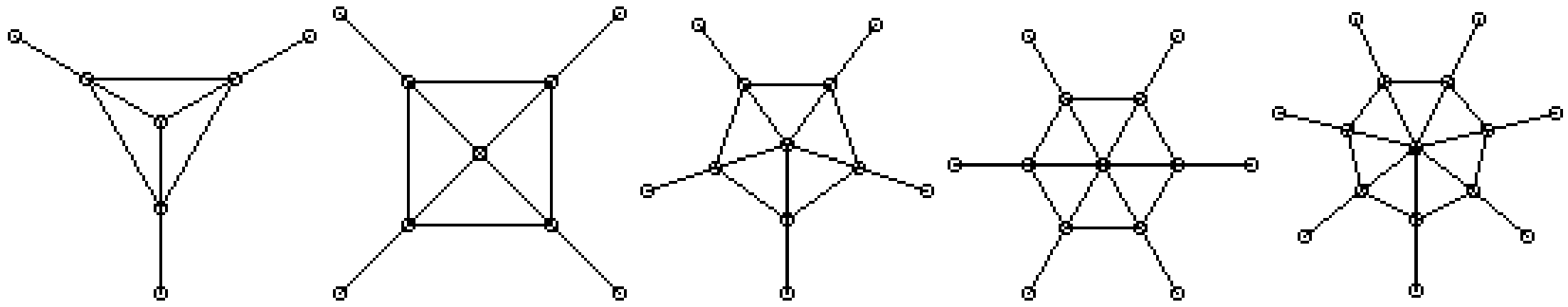

Hypercube 3, 4:

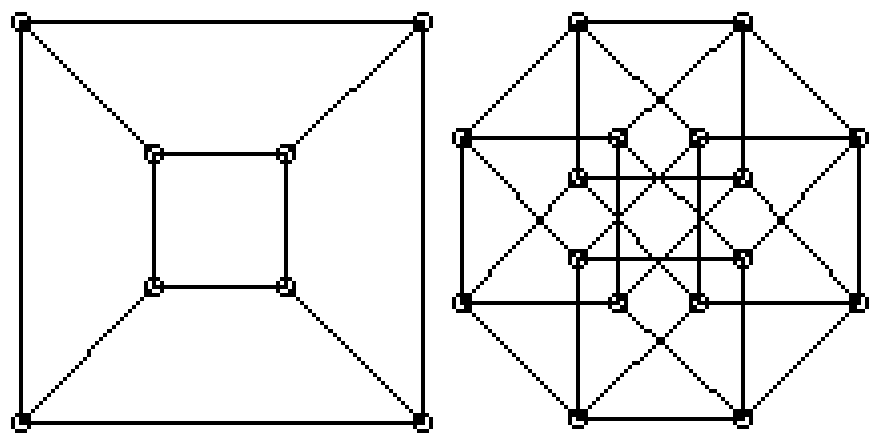

Ladder 3, 4, 5, 6, 7:

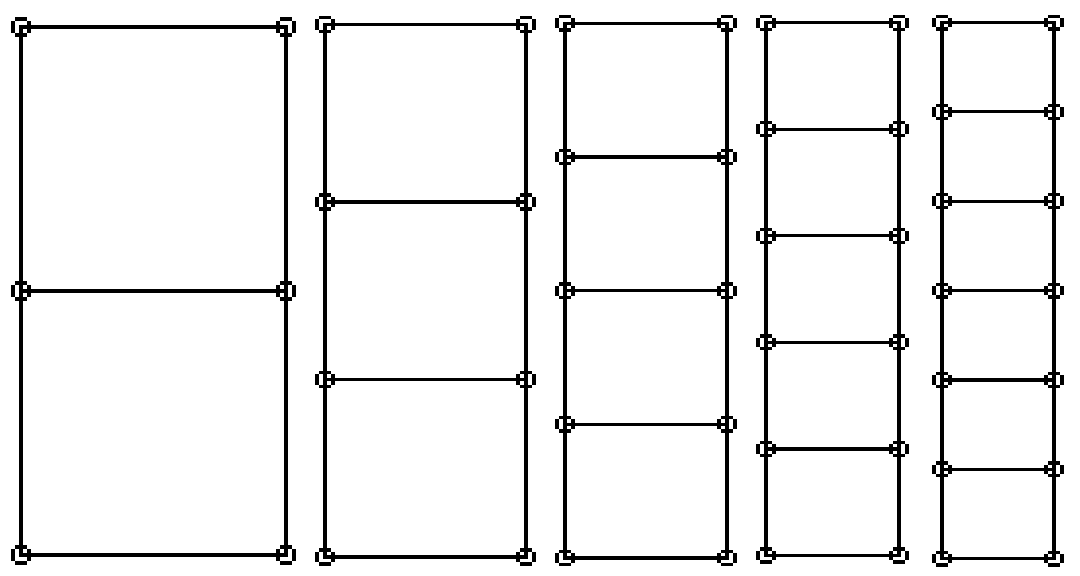

Moebiusladder 3, 4, 5, 6, 7:

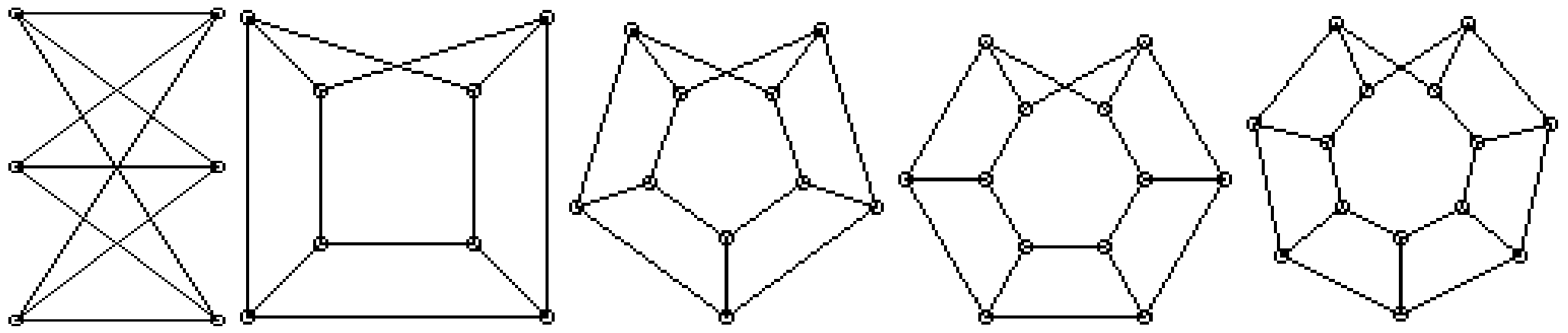

Mycielski 3, 4:

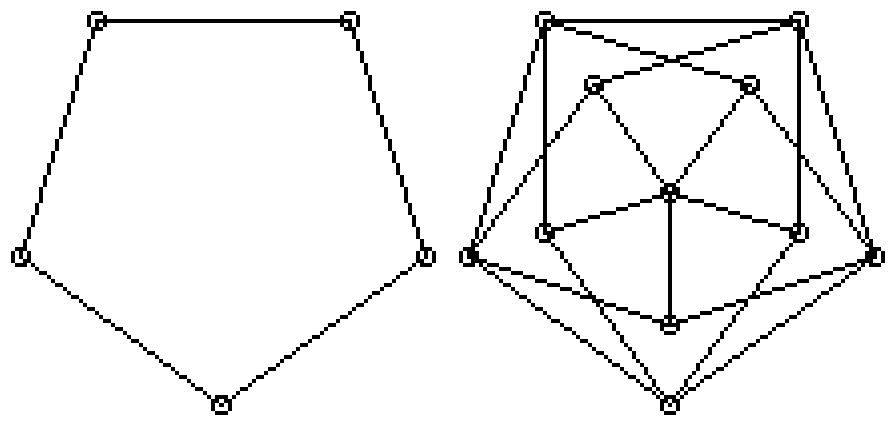

Nonlinesubgraph 3, 4, 5, 6, 7:

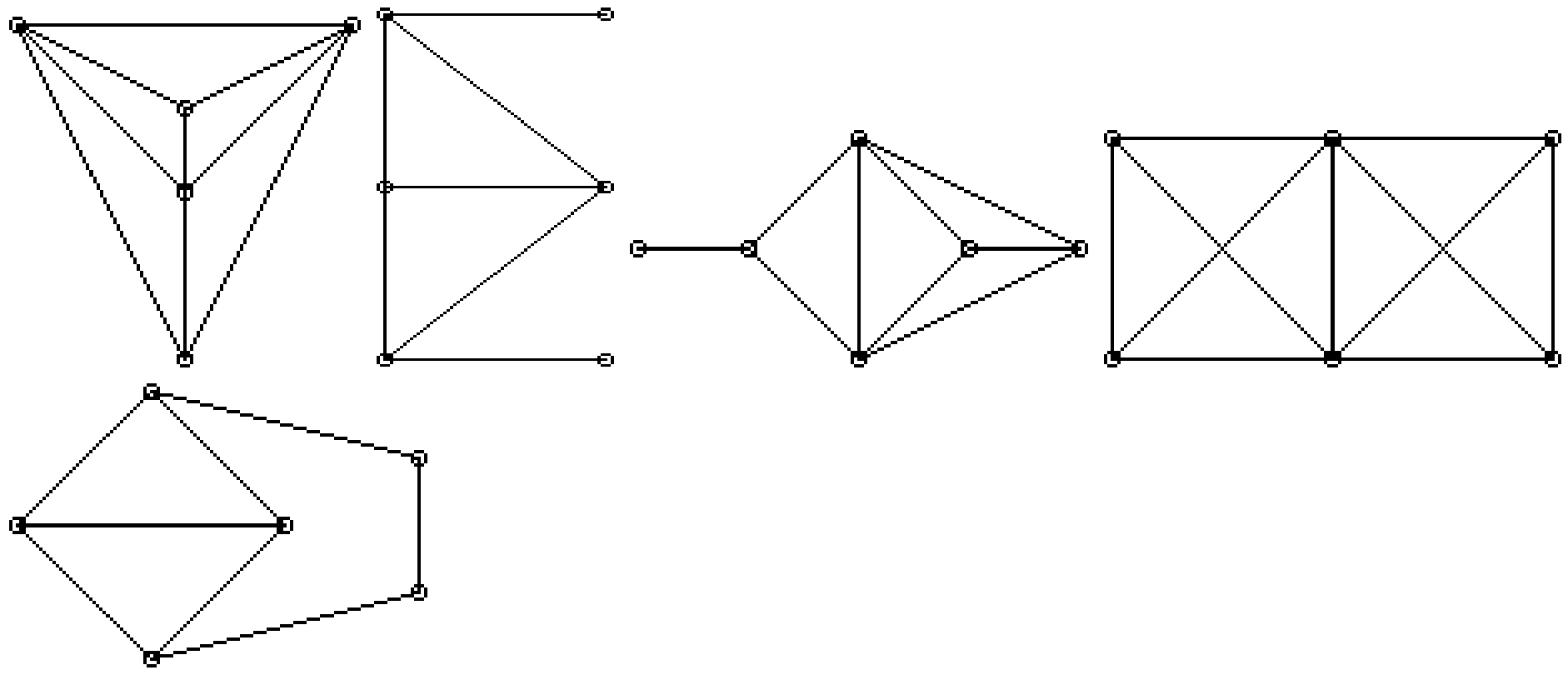

Odd 3:

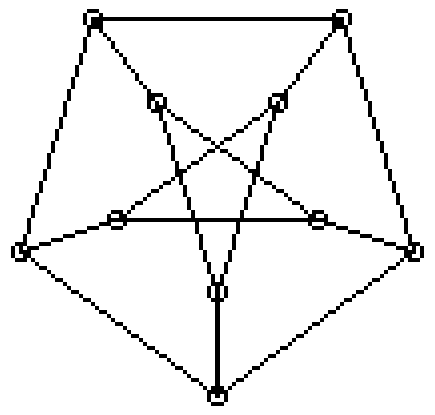

Paley 5, 9, 13:

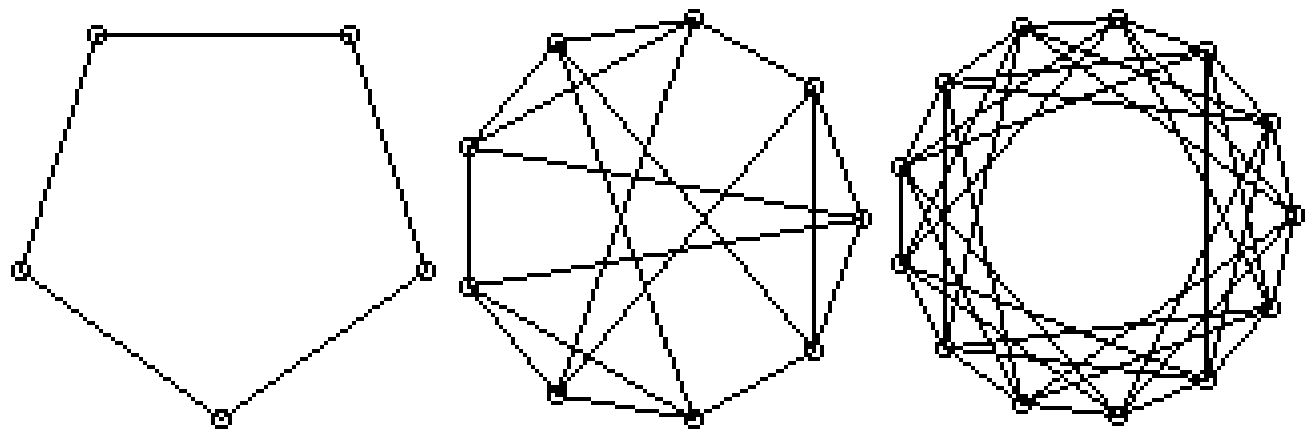

Pan 3, 4, 5, 6, 7:

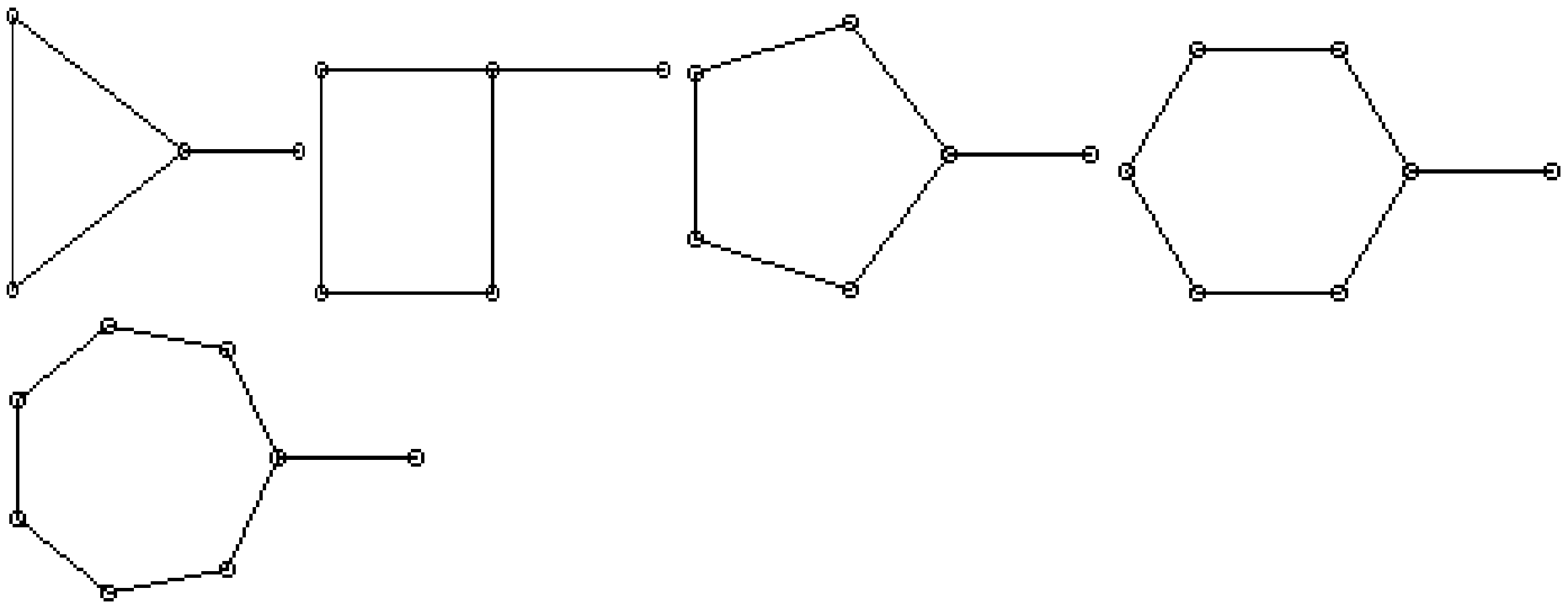

Permutationstar 3, 4:

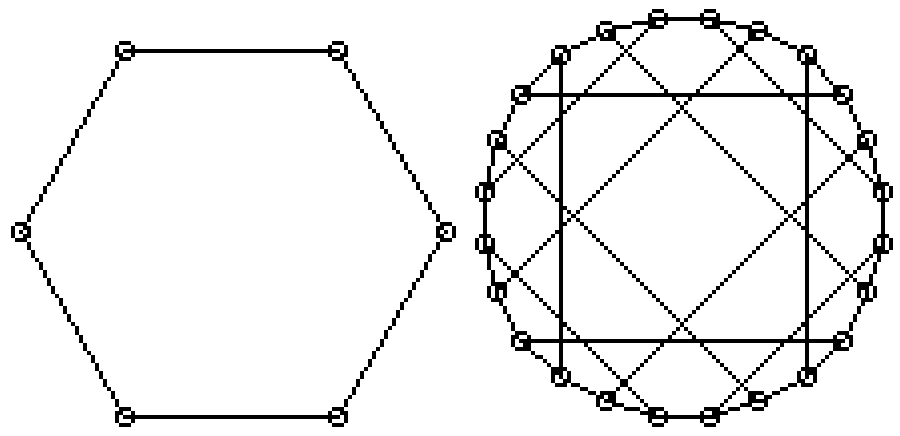

Prism 3, 4, 5, 6, 7:

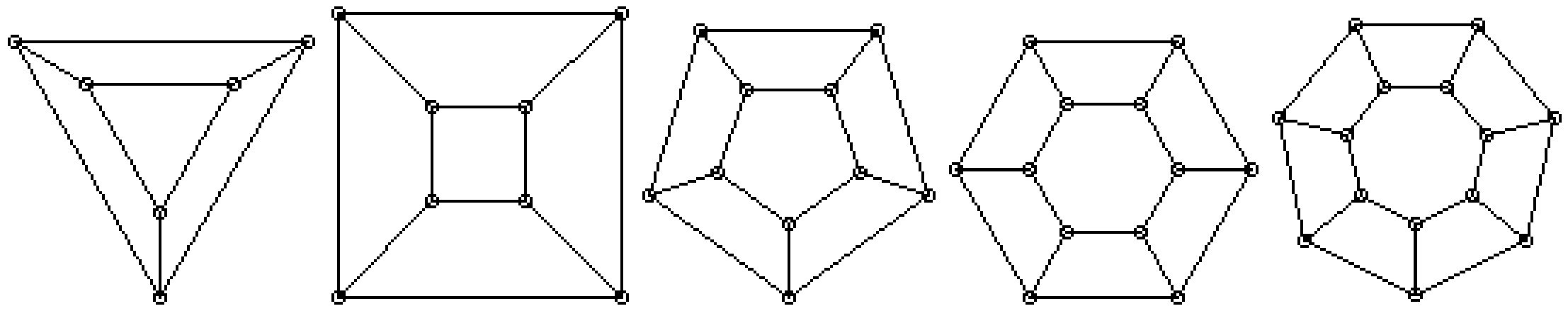

Sun 4, 5, 6, 7, 8:

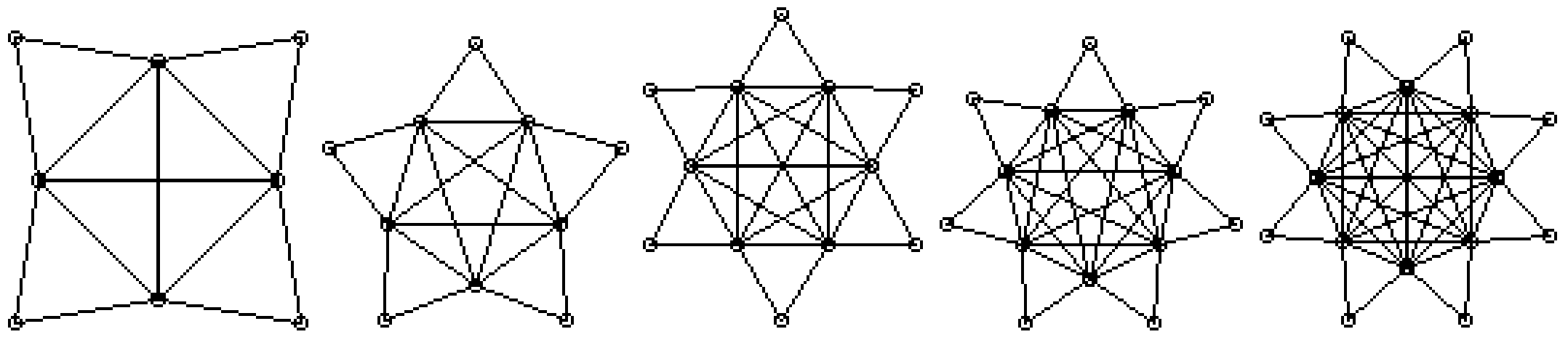

Sunlet 3, 4, 5, 6, 7:

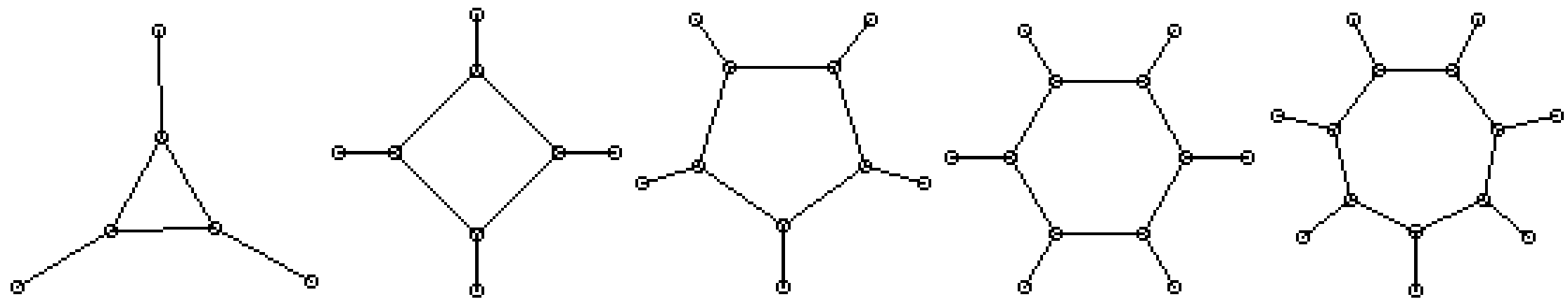

Triangular 3, 4, 5:

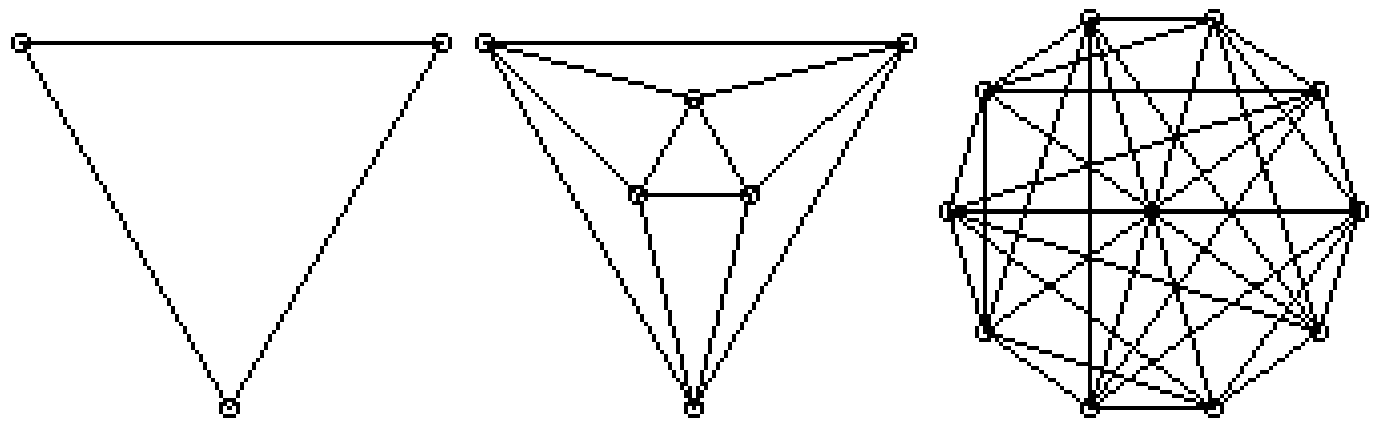

Web 3, 4, 5, 6, 7:

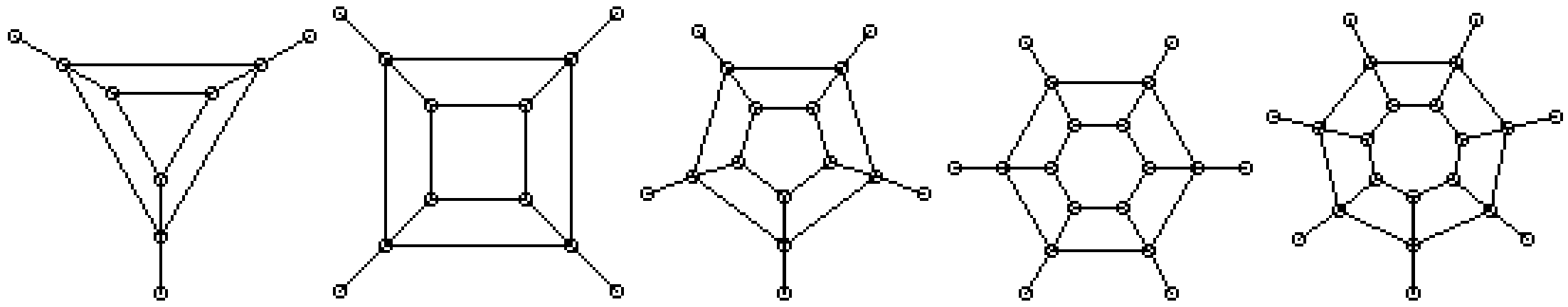

Wheel 4, 5, 6, 7, 8:

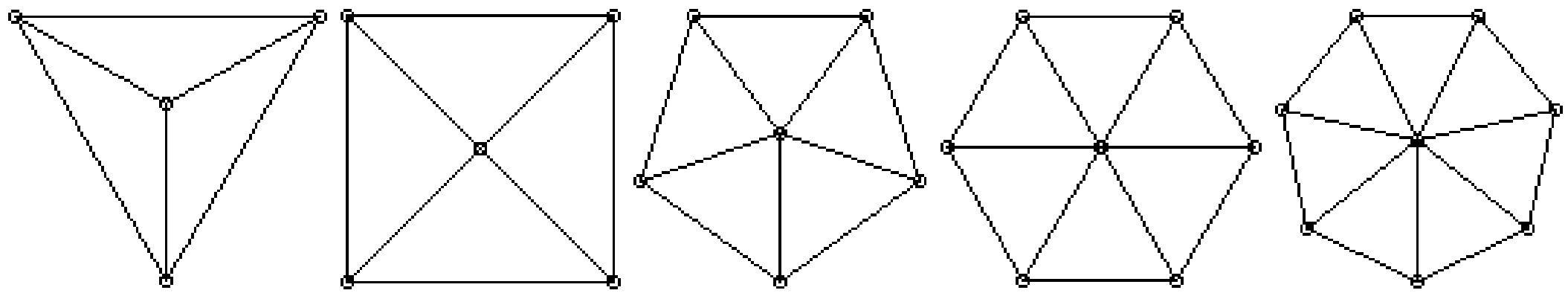